\documentclass[prd,amsmath,amssymb,nofootinbib]{revtex4} 
\usepackage{graphicx} 
\usepackage{amsmath}
\usepackage{amsfonts,amsbsy}
\usepackage{amssymb}
\usepackage{color}
\def\beq{\begin{equation}}
\def\eeq{\end{equation}}
\def\beq{\begin{eqnarray}}
\def\eeq{\end{eqnarray}}

\begin{document}

\title{\bf  Angular correlations in  pA collisions from CGC: multiplicity and mean transverse momentum dependence of $v_2$}

\author{
Tolga Altinoluk$^{a}$, N\'estor Armesto$^{b}$, Alex Kovner$^{c}$, Michael Lublinsky$^{d}$ and Vladimir V. Skokov$^{e}$ 
}

\affiliation{
$^a$National Centre for Nuclear Research, 02-093 Warsaw, Poland \\
$^b$Departamento de F\'{i}sica de Part\'{i}culas and IGFAE, Universidade de Santiago de Compostela, 15782 Santiago de Compostela, Galicia-Spain \\
$^c$Physics Department, University of Connecticut, 2152 Hillside Road, Storrs, CT 06269, USA \\
$^d$Physics Department, Ben-Gurion University of the Negev, Beer Sheva 84105, Israel  \\
$^e$North Carolina State University, Raleigh, NC 27695, USA and \\
RIKEN-BNL Research Center, Brookhaven National Laboratory, Upton, NY 11973, USA
}

\begin{abstract}
%The multiplicity and mean transverse momentum dependence of the second Fourier coefficient $v_2$ in  pA collisions  are explored from the initial state 
%perspective.
 Within the dense-dilute Color Glass Condensate approach, and using the Golec-Biernat-Wuesthoff model for the dipole scattering amplitude, we calculate   $v_2^2$ as well as the correlations between $v_2^2$ and both the total multiplicity and the mean transverse momentum of produced particles.  We find that the correlations are generally very small consistent with the observations. We note an interesting sharp change in the value of $v^2_2$ as well as of its correlations as a function of the width of the transverse momentum bin. This crossover is associated with the change from Bose enhancement dominance of the correlation for narrow bin to HBT dominated correlations for larger bin width.
\end{abstract}

\maketitle

\section{Introduction}
\label{intro}

The observation at the Large Hadron Collider (LHC) of strong  long range rapidity correlations with a characteristic structure in azimuth - "the ridge", in small size collision systems, pp and pA ~\cite{ridge},
 is a very interesting result.  The long range in rapidity implies, by causality arguments, that the correlations must originate in the initial stages of the collision, where the collectivity must emerge from the underlying microscopic dynamics.
 Two very different approaches have been pursued to provide a comprehensive explanation of the observed azimuthal correlations. The first one relies on strong final state interactions between the many produced particles, that lead to a relativistic hydrodynamic description of the evolution of the final state fireball~\cite{bozek}. This approach is quite successful in describing experimental data. It is well justified in nucleus-nucleus collisions where the ridge is also observed~\cite{ridge2}, but the theoretical rationale for its application to small systems (such as the ones produced in  pp collisions) is still wanting.
 An alternative approach stresses the initial state correlations  encoded in the light cone wave-functions  of the incoming hadrons~\cite{gg}, and, in first approximation, 
ignores any final state effects. The present paper is devoted to further exploration of correlations within  the latter framework.

There is a vast literature devoted to initial state correlations, see e.g.~\cite{Altinoluk:2020wpf} for the most recent review  and references therein. The main theoretical framework to compute multiparticle production at high energies is the Color Glass Condensate (CGC) effective theory, see e.g.~\cite{Gelis:2010nm,Kovchegov:2012mbw}.
Building on earlier calculations~\cite{earlyhbt,aklm,Nikolaev:2005zj,double}, most previous work on the subject has been focused on understanding  angular Fourier harmonics $v_n$ (e.g.~\cite{kevin}). 
 Two distinct  quantum interference effects have been identified as giving rise to the emergence of even harmonics: 
 the Bose enhancement of gluons in the incoming projectile wave function  and the Hanbury Brown-Twiss (HBT)  interference  in the emission of gluons in 
 the final state~\cite{us,Kovchegov:2013ewa,urs}~\footnote{Effects due to  anisotropic color domains in the target~\cite{Kovner:2012jm} and density gradients~\cite{Levin:2011fb} have also been addressed in this framework.}. The quantum interference effects between quarks and antiquarks were subsequently studied~\cite{PB,Martinez:2018ygo,Martinez:2018tuf,amir}, and the formalism has been extended to calculate inclusive production of more than two particles in pp~\cite{Ozonder:2014sra,Ozonder:2017wmh} and pA~\cite{amir,Altinoluk:2018hcu,Altinoluk:2018ogz}. Further extensions allowed for the inclusion of odd harmonics via density~\cite{v3} and non-eikonal~\cite{Agostini:2019avp,Agostini:2019hkj} corrections, and considering anisotropies in the target produced via fluctuations~\cite{Dumitru:2014vka}.

In the famous pioneering publication of CMS on the ridge in pp collisions (first paper in~\cite{ridge}), the striking feature was the prominence of the ridge signal in rare high multiplicity events. 
The ridge signal was not observed in events approaching mean multiplicity.  More recent analysis (utilizing modified background subtraction procedures)
revealed azimuthal correlations also in minimal bias events, albeit with a somewhat weaker strength. It has also been observed that the momentum integrated $v_2$, once it can be measured, shows very little dependence on multiplicity in a rather wide multiplicity range (see e.g. the ATLAS analysis in \cite{ridge}). It is clearly important to explore further the multiplicity dependence of the ridge, both experimentally 
and  theoretically.  

 Our work is devoted to this theoretical problem.  Our calculations are performed for large values of transverse momenta, $k^2\gg Q_s^2$ and keeping only leading contributions at large $N_c$. We calculate the correlation between $v_2$ and the multiplicity within the CGC framework. We also study the correlation of $v_2$ 
with the mean  transverse momentum of the particles produced in the collisions, motivated by experimental data~\cite{Aad:2019fgl} and by phenomenological works~\cite{Schenke:2020uqq} where such correlations have been proposed as a tool to constrain the initial conditions for hydrodynamic evolution and to discriminate initial from final state correlations.  Despite being motivated by experimental findings, we do not consider our approach as sufficiently
rigorous from the phenomenological perspective and, hence, we are not making any attempt to compare with data. The main take home conclusion of our work is rather the qualitative features of the above correlations, which show an interesting characteristic dependence on the transverse momentum bin size used to calculate $v_2$. We discuss these features in detail in the text as well as in the discussion section.

The manuscript is structured as follows. In Section~\ref{formalism}, we present the details of dense-dilute CGC approach  as well as the model assumptions that we are using to compute the multiparticle production.  Section~\ref{N_v2} is devoted to the calculation of the second flow harmonic $v_2^2$ as well as correlations of $v^2_2$ with multiplicity, and the correlation of $v^2_2$ with the average transverse momentum of produced particles. Numerical results for these quantities are presented in Section~\ref{numerics}. We conclude with a discussion in Section~\ref{conclu}. Technical details are provided in the Appendices.

\section{Multi gluon production in dilute-dense scattering}
\label{formalism}
\subsection{The basic setup}

Our goal in this paper is to study correlations between the second flow harmonic coefficient $v^2_2$ and the particle multiplicity and average transverse momentum.  We will calculate these correlations in the so called dense-dilute approximation.
Below, we closely follow Ref.~\cite{Altinoluk:2018ogz}, in which  double and triple inclusive gluon production in this approach were calculated in the CGC framework~\cite{Gelis:2010nm,Kovchegov:2012mbw}. 

 In this set up, the number of produced gluons for a given configuration of the projectile (proton) and the target (nucleus) is given by 
\begin{equation}
\left.\frac{dN}{d^{2}kdy}\right|_{\rho_{p},\rho_{ t}}=\frac{2g^{2}}{(2\pi)^{3}}\int\frac{d^{2}q}{(2\pi)^{2}}\frac{d^{2}q'}{(2\pi)^{2}}\Gamma({k},{q},{q}')\rho_{p}^{a}(-{q}')\left[U^{\dagger}({k}-{q}')U({k}-{q})\right]_{ab}\rho_{ p}^{b}({q}).
\end{equation}
Here  the Lipatov vertex $L$ and its square $\Gamma$ are defined as
\begin{eqnarray}
L^i(k,q)&=&\bigg[ \frac{(k-q)^i}{(k-q)^2}-\frac{k^i}{k^2}\bigg] ,\ \\
\Gamma({k},{q},{q}')&=&
L(k,k-q)\cdot L(k,k-q')
 = \left(\frac{{q}}{q^{2}}-\frac{{k}}{k^{2}}\right)
\cdot \left(\frac{{q}'}{q'^{2}}-\frac{{k}}{k^{2}}\right).
\end{eqnarray}
Besides, $\rho_{\rm p}(p)$ is a given configuration of the color charged density in the projectile, and $U(q)$ is the eikonal scattering matrix -- the adjoint Wilson line -- for scattering of a single gluon on a fixed configuration of target fields.
The target Wilson lines depend on the target color sources, $\rho_{\rm t}$; we suppress this in our notation for simplicity.  

The single inclusive and double inclusive gluon production in this approach
are given by 
\begin{equation}
	\frac{dN^{(1)}}{d^{2}kdy}=\left\langle \left\langle  \frac{dN}{d^{2}kdy}|_{\rho_{ p},\rho_{ t}} 
	\right \rangle_{ p} 
	\right\rangle_{ t} 
\end{equation}
and 
\begin{equation}
\frac{d N^{(2)}}{d^{2}k_{1}dy_{1}d^{2}k_{2}dy_{2}}= 
\left\langle \left \langle
\left. \frac{dN}{d^{2}k_{1}dy_{1}}\right|_{\rho_{ p},\rho_{ t}}
\left.\frac{dN}{d^{2}k_{2}dy_{2}} \right|_{\rho_{ p},\rho_{t}}
 \right\rangle_{ p} 
 \right\rangle_{ t} \, ,
\end{equation}
where the averaging is performed over the projectile and target color charge configurations:
\begin{equation}\label{Op}
	\left\langle O(\rho_{\ p})
 	\right\rangle_{ p}  = 
	\frac{1}{Z_{ p}} \int {\cal D} \rho_{ p}\  W_{ p}(\rho_{ p})\  O(\rho_{ p})
\end{equation}
and 
\begin{equation}\label{Ot}
	\left\langle O(\rho_{ t})
 	\right\rangle_{ t}    = 
	\frac{1}{Z_{ t}} \int {\cal D} \rho_{ t}\  W_{ t}(\rho_{ t})\  O(\rho_{ t})\,. 
\end{equation}
The normalization factors, $Z_{ p,t}$, are fixed so that
\begin{equation}
	\left\langle 1 
 	\right\rangle_{ p} 
	=	\left\langle 1 
	\right\rangle_{ t} =1\,.  
\end{equation}
The total multiplicity $N$ (per unit of rapidity) is
\begin{equation} \label{barN}
	 N\,=\, \int_k \frac{dN^{(1)}}{d^{2}kdy}\ .
	\end{equation}
The mean transverse momentum squared per particle $\bar k^2$ is calculated as 
\begin{equation}\label{bark}
	\bar k^2\,=\, \frac{1}{  N}\int_k k^2\,\frac{dN^{(1)}}{d^{2}kdy}\ .
	\end{equation}
	
	Here and below we suppress the rapidity index.
Note that formally both of these quantities may be divergent -- the multiplicity diverges in the infrared if we treat the projectile as translationally invariant in the transverse plane, while the mean momentum diverges in the ultraviolet under fairly general assumptions about the structure of the projectile and target averages. We will regulate these divergencies by introducing physically motivated cutoffs (see below).	

The azimuthal  flow harmonics $v^2_n$ are defined as 
\begin{equation}
	%_
v^2_n(k_1,k_2)\equiv	   \frac{\int d\phi_1 d \phi_2 e^{i n(\phi_1-\phi_2)}  \, \frac { d^2N^{(2)} } { d^2k_1 d^2 k_2 }  }{ \int d\phi_1 d \phi_2   \, \frac {  d^2N^{(2)} } { d^2k_1 d^2 k_2 }  }\ ,
\end{equation}
where $\phi_1$, $\phi_2$ are the azimuthal angles of the corresponding transverse momenta. Below, we focus on $v^2_2$ only.
In this framework, each collision event corresponds to a fixed configuration of $\rho_p$ and $\rho_t$. 
The averaging  introduced in (\ref{Op}) and (\ref{Ot}) is equivalent to averaging over all possible events. 

An ideal way fo studying the dependence of $v_2$ on multiplicity would be to select from the total ensemble only events  with  the total multiplicity $N_i$ in some multiplicity bin (labeled by 
index $i$) and calculate $v_2$ by averaging only over those events. 
Mathematically one would have to introduce a projector $P_i(\rho_p,\rho_t)$ on this subspace of configurations and then modify the averaging procedure as
\beq
W_p[\rho_p]\,W_t[\rho_t]\,\rightarrow \,W_p[\rho_p]\,W_t[\rho_t]\, P_i(\rho_p,\rho_t).
\eeq 
In practice, constructing $P_i$ is a complicated task that so far has not been  accomplished (see, however, \cite{Dumitru:2018iko,Dumitru:2017cwt}). A simpler but nevertheless quite informative observable is a correlation between $v_2$ and  $N$,  i.e. $\langle\langle v_2(k_1,k_2)|_{\rho_p,\rho_t}N|_{\rho_p,\rho_t}\rangle_p\rangle_t$, and similarly between $v_2$ and the transverse momentum per particle. The averaging in these expressions goes over the whole ensemble of events, and thus there is no need to consider particular sub ensembles.

The calculation of these correlations requires us to calculate the
three gluon inclusive production  
\begin{equation}\label{3g}
 \int d^2k_1  d\phi_2 d \phi_3 e^{i 2(\phi_2-\phi_3)} \frac { dN^{(3)} } { d^2k_1 d^2 k_2 d^2k_3 }     =
 \int d^2k_1  d\phi_2 d \phi_3 e^{i 2(\phi_2-\phi_3)}\,\left\langle \left\langle  \frac{dN}{d^{2}k_1dy_1} \frac{dN}{d^{2}k_2dy_2} \frac{dN}{d^{2}k_3dy_3} 
	\right \rangle_{ p} 
	\right\rangle_{ t} 
\end{equation}
and
\begin{equation}
	%_
  \int d^2k_1 k_1^2  d\phi_2 d \phi_3 e^{i 2(\phi_2-\phi_3)} \frac { dN^{(3)} } { d^2k_1 d^2 k_2 d^2k_3 }  \,=\,
\int d^2k_1 k_1^2  d\phi_2 d \phi_3 e^{i 2(\phi_2-\phi_3)}    
\,\left\langle \left\langle  \frac{dN}{d^{2}k_1dy_1} \frac{dN}{d^{2}k_2dy_2} \frac{dN}{d^{2}k_3dy_3} 
	\right \rangle_{ p} 
	\right\rangle_{ t} \ .
\end{equation}
Fortunately,  three gluon inclusive production  at mid rapidity has already been  investigated in Ref.~\cite{Altinoluk:2018ogz} and we will use many of the results of this paper. 

In order to be able to progress with the computations, we need to know the projectile and target averaging functionals. 
Here we are going to use simple models frequently used in a variety of CGC based calculations. For the projectile averaging we will use the simple Gaussian McLerran-Venugopalan (MV) model~\cite{mv} specified by
\begin{equation}
	\langle \rho_{ p}^a({p}) \rho_{ p}^b({k})  \rangle_{ p}  \equiv\, \mu^2(k,p)\,
	= (2\pi)^2  \mu_{} ^2(p)  \delta({p}+{k}) \delta^{ab}\,
	\label{mu2_GBW}
\end{equation}
which corresponds to the weight functional
\begin{equation}
	W_{ p}(\rho_{ p}) = \exp\left( - \int \frac{d^2 q}{(2\pi)^2} 
	\rho_{ p}^a(-{q})
	\frac{1}{2\mu^2(q)}
	\rho_{ p}^a({q})
	\right) \,.	
\end{equation}
Note that this weight functional defines a statistical ensemble that is translationally invariant in the transverse plane. This approximation is only reasonable for momenta of produced particles larger than the inverse radius of the projectile. Thus, in the following we will always assume $k_{\rm min}>1/R$. As mentioned above, for the calculation of total multiplicity we will need to impose an infrared cutoff, which we will choose to be of the order of the transverse radius of the proton. 

We will further simplify our calculations by choosing  $\mu$ to be independent of momentum. This last assumption means that the color charge density is completely uncorrelated in the transverse plane. Although this is clearly not the whole truth, one expects the density-density correlations to be only important at distance scales of the order of the confinement radius and above. Since we  limit ourselves to consider momenta greater than the inverse radius of the proton, the presence of such correlations should not affect our results.

The averaging over the Wilson lines of the target will be performed in the approximation articulated recently in Ref.~\cite{amir}. In this approximation any product of Wilson lines is factored into pairs according to the basic Wick contraction
\begin{equation}\label{wick}
	\left\langle U_{ab}({p})U_{cd}({q}) \right\rangle_{\rm t} = \frac{(2\pi)^2}{N_{c}^{2}-1}\delta_{ac}\delta_{bd}\delta^2({p}+{q}) d({p})\,,
\end{equation}
with the "adjoint dipole" amplitude defined as
\begin{equation}
	d(p) = \frac{1}{N_c^2-1}\int d^2x e^{ix\cdot p}\langle{\rm tr} \ \left[ U^\dagger(x)  U(0) \right] \rangle_{\rm t}\,.
\end{equation}
As explained in Ref.~\cite{amir} this approximation is appropriate for the dense target 
 regime. It collects all terms in the $n$-particle cross section which have the leading dependence on the area of the projectile.  The approximation only includes terms which contain ``small size''  color singlets in the projectile propagating through the target. Any non singlet state that in the transverse plane is removed by more than $1/Q_s$ ( where $Q_s$ is the saturation momentum of the target) from other propagating partons must have a vanishing $S$-matrix on the dense (black disk) target. On the other hand if the singlet state contains more than two partons, one looses a power of the area when integrating over the coordinates of the partons. Thus the leading contribution to the $S$-matrix in the (target) black disk limit is due to  colorless projectile dipoles of the size smaller than the inverse target saturation momentum.  
The same approximation for the quadrupole amplitude has been used previously in Ref.~\cite{Kovchegov:2013ewa} where its consistency with the explicit modelling of the Wilson line correlators via MV model has been verified.
% ~Ref.~\cite{JalilianMarian:2004da}.  

Note that this averaging procedure for the target is formally  equivalent (disregarding subtleties related to the definition of the Haar measure)  to the following form of the weight functional:
\begin{equation}  \label{fac}
W_t[U]=\exp\left\{-\frac{1}{2}\int \frac{d^2q}{(2\pi)^2}\frac{1}{d(q)}{\rm tr}[U^\dagger(q)U(q)]\right\}\,.
\end{equation}

Finally, we will adopt the Golec-Biernat - Wusthoff (GBW)  model~\cite{gbw} for the dipole
\beq\label{gbw}
d(p)= \frac{4\pi}{Q_s^2}\,e^{-p^2/Q_s^2}\,  .
\eeq
This model is known to be a good description for momentum transfer $p$ of order of the saturation momentum and below. Although it does not properly account for the perturbative high momentum tail of the momentum transfer, we believe that it  is quite adequate for the purposes of qualitative understanding of correlations, which is the goal of this paper.

\subsection{From one to three gluons}
\label{gluon_prod}

%In order to study the observables of interest (correlations between multiplicity and $v_2$ in Section \ref{N_v2},  and correlations between transverse momentum and $v_2$ in Section \ref{Pt_v2}), one needs the expressions of the differential multiplicities for single, double and triple inclusive gluon production which have been studied in detail in \cite{Altinoluk:2018ogz} for dilute-dense scattering and the origin of the quantum interference effects have been shown.
 In this subsection, we briefly summarize the results of~\cite{Altinoluk:2018ogz} focusing on the contributions that will be essential for the computation of the observables that we are interested in. 

Let us start with the single inclusive gluon production with transverse momentum $k_1$: 
\beq\label{N1}
\frac{dN^{(1)}}{d^2k_1}=4\pi\,\alpha_s\,  (N_c^2-1)\int \frac{d^2q_1}{(2\pi)^2} \, d(q_1) \,  \mu^2(k_1-q_1, q_1-k_1)\, L^i(k_1,q_1)L^i(k_1,q_1),
\eeq
where
% $\tilde c_0\equiv \alpha_s\, (4\pi)\, (N_c^2-1)$
 $q_1$ is the momentum transfer from the target during the interaction and $(k_1-q_1)$ is the momentum of the gluon in the incoming projectile wave function. 
%Here, $d(q_1)$ is the dipole operator in momentum space for which we will adopt the GBW model 
%
%\beq
%d(p)\equiv \frac{4\pi}{Q_s^2}e^{-p^2/Q_s^2}\,  
%\eeq
%
%for the computations of the observables of interest. Moreover, the function $\mu^2(k,p)$ originates from the projectile averaging of the color charge densities which will be treated within the MV model, ie
%
%\beq
%\label{mu2_GBW}
%\mu^2(k,p)\equiv \mu^2\,  (2\pi)^2\, \delta^{(2)}(k+p)\, 
%\eeq
%
%and the function $L^i(k,q)$ stands for the Lipatov vertex which reads 
%
%\beq
%L^i(k,q)=\bigg[ \frac{(k-q)^i}{(k-q)^2}-\frac{k^i}{k^2}\bigg] \, . 
%\eeq
%
The expression for the double inclusive gluon spectrum can be organized as
\beq
\label{Double_Inc_Gen}
\frac{dN^{(2)}}{d^2k_2d^2k_3}=\frac{dN^{(2)}}{d^2k_2d^2k_3}\bigg|_{dd}+\frac{dN^{(2)}}{d^2k_2d^2k_3}\bigg|_Q+ O\bigg(\frac{1}{(N_c^2-1)^2}\bigg).
\eeq
Here the first term 
\beq\label{uncor}
\frac{dN^{(2)}}{d^2k_2d^2k_3}\bigg|_{dd}=\frac{dN^{(1)}}{d^2k_2}\frac{dN^{(1)}}{d^2k_3}
\eeq
is the square of the single inclusive spectrum and  is the uncorrelated contribution to the double inclusive spectrum which is unimportant for our purposes. On the other hand, the second term in Eq. \eqref{Double_Inc_Gen}, is a quadrupole  contribution which encodes the quantum interference effects. In the approximation of Eq. \eqref{wick}  its explicit expression reads 
\beq\label{N2Q}
\frac{dN^{(2)}}{d^2k_2d^2k_3}\bigg|_Q= (4\pi)^2 \alpha_s^2(N_c^2-1)\, \int \frac{d^2q_2}{(2\pi)^2} \frac{d^2q_3}{(2\pi)^2} \, d(q_2) d(q_3)\, \Big[ I_{Q,1}+I_{Q,2}\Big],
\eeq
where %$\bar c_0\equiv \alpha_s^2 (4\pi)^2 (N_c^2-1)$ and 
\beq
I_{Q,1}&=&\mu^2(k_2-q_2, q_3-k_3)\, \mu^2(k_3-q_3, q_2-k_2)\, L^i(k_2,q_2)L^i(k_2,q_2)\, L^j(k_3,q_3)L^j(k_3,q_3)\, + (k_3\to - k_3),\\
I_{Q,2}&=& \mu^2(k_2-q_2,q_2-k_3)\, \mu^2(k_3-q_3, q_3-k_2) \, L^i(k_2,q_2)L^i(k_2,q_3)\, L^j(k_3,q_2)L^j(k_3,q_3)\, + (k_3\to - k_3).
\eeq
%
%Here the quadrupole was factorized using the mean field approximation  (\ref{fac}).
Our calculations in this paper will be always to leading nontrivial order in $1/N_c^2$, and thus we do not specify the subleading term in Eq. \eqref{N2Q}.

The physical meaning of the two terms in Eq. \eqref{N2Q} has been extensively discussed in~\cite{Altinoluk:2018ogz}. The first term, $I_{Q,1}$, in the translationally invariant approximation contains two contributions. One is proportional to $\delta^{(2)}(k_2-q_2-(k_3-q_3))$. Given that $k_i-q_i$ is the momentum of the $i$-th gluon in the incoming wave function, this contribution clearly is due to the standard Bose enhancement of gluons in the incoming projectile state. The second contribution to $I_{Q,1}$ is proportional to $\delta^{(2)}(k_2-q_2+(k_3-q_3))$. This contribution is due to the gluonic condensate in the projectile wave function, which is equal in magnitude to the Bose enhanced contribution. For simplicity we will refer to it as "backward" Bose enhancement, although one should keep in mind that the physics of this term is distinct from the physics of Bose enhancement.
The second term,  $I_{Q,2}$ contains a delta- function of the final state momenta $\delta^{(2)}(k_2\pm k_3)$.  These correspond to the Hanbury-Brown Twiss correlations between the emitted gluons. The HBT correlations here exist for collinear as well as anti collinear momenta, since the gluons do not carry any physical charges. We will refer to these contributions as forward and backward HBT correlations respectively.

The triple inclusive gluon spectrum can be written as 
\beq
\label{Triple_Spec_Gen}
\frac{dN^{(3)}}{d^2k_1d^2k_2d^2k_3}=\frac{dN^{(3)}}{d^2k_1d^2k_2d^2k_3}\bigg|_{ddd}+\frac{dN^{(3)}}{d^2k_1d^2k_2d^2k_3}\bigg|_{dQ}+\frac{dN^{(3)}}{d^2k_1d^2k_2d^2k_3}\bigg|_{X} \ .
\eeq
 The first two terms in Eq. \eqref{Triple_Spec_Gen} correspond to the case where at least one of the gluons is uncorrelated with the other two. It is easy to see that in order to have a nontrivial correlation of $v_2$ with either multiplicity or average momentum, all three gluons have to be correlated. Therefore the first two terms in  Eq. \eqref{Triple_Spec_Gen} do not contribute to the observables of interest. 
 
 The only nontrivial contributions to the observables of interest arise from the last term in Eq. \eqref{Triple_Spec_Gen} which corresponds to fully correlated part of the triple inclusive gluon spectrum at leading $N_c$.  Its explicit expression reads
\beq
\label{uninteg_3g_N}
\frac{dN^{(3)}}{d^2k_1d^2k_2d^2k_3}\bigg|_X\approx \alpha_s^3(4\pi)^3(N_c^2-1)
\int \frac{d^2q_1}{(2\pi)^2}\frac{d^2q_2}{(2\pi)^2}\frac{d^2q_3}{(2\pi)^2}\, 
d(q_1)d(q_2)d(q_3)\, 
\Big[ I_{X,1}+I_{X,2}+I_{X,3}+I_{X,4}+I_{X,5}\Big].
\eeq
%
%where $c_0=\alpha_s^3(4\pi)^3(N_c^2-1)$ and 
%
The various expressions that enter Eq. \eqref{uninteg_3g_N} at leading order in $1/N_c^2$ are given and discussed below. The first term reads
\beq
\label{X_1_initial}
I_{X,1}=\Big[ {\tilde I}_{X,1}+(k_3\to-k_3)\Big]+\Big[ {\tilde I'}_{X,1}+(k_1\to-k_1)\Big],
\eeq
with ${\tilde I}_{X,1}$ and ${\tilde I'}_{X,1}$  defined as 
\beq
\label{X1}
{\tilde I}_{X,1}&=&
\mu^2(k_2-q_2,q_2-k_1) \, \mu^2(k_1-q_1,q_3-k_3) \, \mu^2(k_3-q_3,q_1-k_2)\nonumber\\
&\times& 
L^i(k_1,q_1) L^i(k_1,q_2) \, L^j(k_2,q_2)L^j(k_2,q_1) \, L^k(k_3,q_3)L^k(k_3,q_3)
\nonumber\\
&+&
\mu^2(k_2+q_2,k_1-q_2) \, \mu^2(k_3-q_3,q_1-k_1) \, \mu^2(q_3-k_3,-q_1-k_2)\nonumber\\
&\times&
L^i(k_1,q_1) L^i(k_1,q_2) \, L^j(k_2,-q_1)L^j(k_2,-q_2) \, L^k(k_3,q_3)L^k(k_3,q_3)
\eeq
and
\beq
\label{X1prime}
{\tilde I'}_{X,1}&=&
\mu^2(k_1-q_2,q_2-k_2) \, \mu^2(k_2-q_1,k_3-q_3) \, \mu^2(q_1-k_1,q_3-k_3)\nonumber\\
&\times& 
L^i(k_1,q_1) L^i(k_1,q_2) \, L^j(k_2,q_1)L^j(k_2,q_2) \, L^k(k_3,q_3)L^k(k_3,q_3) \nonumber\\
&+&
\mu^2(-k_1-q_2,q_2-k_2) \, \mu^2(k_2+q_1,q_3-k_3) \, \mu^2(k_3-q_3,k_1-q_1)\nonumber\\
&\times& 
L^i(k_1,-q_2) L^i(k_1,q_1) \, L^j(k_2,-q_1)L^j(k_2,q_2) \, L^k(k_3,q_3)L^k(k_3,q_3).
\eeq
Assuming translational invariance  (\ref{mu2_GBW})  one can clearly identify the various quantum interference effects contributing to three gluon correlations. Below we briefly review them  following~\cite{Altinoluk:2018ogz}:
\\
$\bullet$ The first term in $I_{X,1}\propto\delta^{(2)}(k_1-k_2)\; \delta^{(2)}\big[ (k_1-q_1)-(k_3-q_3)\big]\; \delta^{(2)}\big[(k_3-q_3)-(k_1-q_1)\big]$ arises due to forward HBT correlation between the gluons $1$ and $2$, and forward Bose enhancement between the gluons $1$ and $3$. 
\\
$\bullet$ The second term in $I_{X,1}\propto \delta^{(2)}(k_1+k_2)\; \delta^{(2)}\big[ (k_3-q_3)-(k_1-q_1)\big] \; \delta^{(2)}\big[ (k_1-q_1)-(k_3-q_3)\big]$ arises due to the backward HBT between gluons $1$ and $2$, and forward Bose correlation between the gluons $1$ and $3$. 
\\
$\bullet$ The third term in $I_{X,1}\propto \delta^{(2)}(k_1-k_2) \; \delta^{(2)}\big[(k_1-q_1)+(k_3-q_3)\big] \; \delta^{(2)}\big[ (k_1-q_1)+(k_3-q_3)\big]$ arises due to the forward HBT of the gluons $1$ and $2$, and backward Bose correlation between the gluons $1$ and $3$.
\\
$\bullet$ The forth term in $I_{X,1}\propto \delta^{(2)}(k_1+k_2)\; \delta^{(2)}\big[ (k_1-q_1)+(k_3-q_3)\big] \; \delta^{(2)}\big[ (k_1-q_1)+(k_3-q_3)\big]$ arises due to the backward HBT correlation between the  gluons $1$ and $2$, and backward Bose correlation between the gluons $1$ and $3$.

The terms $I_{X,2}$ and $I_{X,3}$ are obtained from $I_{X,1}$ by  permutation of the momenta of produced gluons
\beq
\label{X_2_initial}
I_{X,2}&=&
\Big[ {\tilde I}_{X,1}(1\to2,2\to3,3\to1)+(k_1\to-k_1)\Big] + \Big[ {\tilde I'}_{X,1}(1\to2,2\to3,3\to1)+(k_2\to-k_2)\Big], \\
\label{X_3_initial}
I_{X,3}&=&
\Big[ {\tilde I}_{X,1}(1\to3,3\to2,2\to1)+(k_2\to-k_2)\Big]+ \Big[{\tilde I'}_{X,1}(1\to3,3\to2,2\to1)+(k_3\to-k_3)\Big].
\eeq
The identification of various quantum interference  effects in $I_{X,2}$ and $I_{X,3}$ follows straightforwardly from  the earlier discussion using the very same permutation transformation.
Note  that the terms $I_{X,2}$ contain an explicit contribution to 
forward/backward HBT of the gluons $2$ and $3$. These terms will be important later when we calculate $v_2$.

Next is the explicit expressions for $I_{X,4}$:
\beq
\label{X_4_initial}
I_{X,4}&=&
\mu^2(k_2-q_2,q_1-k_1) \, \mu^2(k_1-q_1,q_3-k_3) \, \mu^2(k_3-q_3,q_2-k_2)\nonumber\\
&\times& 
L^i(k_1,q_1)L^i(k_1,q_1)\, L^j(k_2,q_2)L^j(k_2,q_2) \, L^k(k_3,q_3)L^k(k_3,q_3) +(k_3\to-k_3) \nonumber\\
&+&
\mu^2(k_2-q_2,q_3-k_3) \, \mu^2(k_3-q_3,k_1-q_1) \, \mu^2(q_1-k_1,q_2-k_2) \nonumber\\
&\times&
L^i(k_1,q_1)L^i(k_1,q_1)\, L^j(k_2,q_2)L^j(k_2,q_2) \, L^k(k_3,q_3)L^k(k_3,q_3) +(k_1\to-k_1) \nonumber\\
&+&
\mu^2(k_2-q_2,k_1-q_1) \, \mu^2(k_3-q_3,q_1-k_1) \, \mu^2(q_3-k_3,q_2-k_2) \nonumber\\
&\times&
L^i(k_1,q_1)L^i(k_1,q_1)\, L^j(k_2,q_2)L^j(k_2,q_2) \, L^k(k_3,q_3)L^k(k_3,q_3) +(k_3\to-k_3) \nonumber\\
&+&
\mu^2(q_1-k_1,q_3-k_3) \, \mu^2(k_1-q_1,q_2-k_2) \, \mu^2(k_2-q_2,k_3-q_3) \nonumber\\
&\times&
L^i(k_1,q_1)L^i(k_1,q_1)\, L^j(k_2,q_2)L^j(k_2,q_2) \, L^k(k_3,q_3)L^k(k_3,q_3) +(k_1\to-k_1) .
 \eeq
Again, all the physical effects behind each term can be identified:
\\
$\bullet$ The first term in $I_{X,4}\propto \delta^{(2)}\big[ (k_2-q_2)-(k_1-q_1)\big] \delta^{(2)}\big[(k_1-q_1)-(k_3-q_3)\big] \; \delta^{(2)}\big[ (k_3-q_3)-(k_2-q_2)\big]$ --  forward Bose correlation of gluons $1$ and $1$,  forward Bose correlation of gluons $1$ and $1$ and forward Bose correlation of gluons $2$ and $3$. 
\\
$\bullet$ The second term in $I_{X,4}\propto \delta^{(2)}\big[ (k_2-q_2)-(k_3-q_3)\big] \delta^{(2)}\big[(k_3-q_3)+(k_1-q_1)\big] \; \delta^{(2)}\big[ (k_1-q_1)+(k_2-q_2)\big]$ -- the forward Bose correlation of gluons $2$ and $3$, backward Bose correlation of gluons $1$ and $3$ and backward Bose correlation of gluons $2$ and $1$. 
\\
$\bullet$ The third term in $I_{X,4}\propto \delta^{(2)}\big[(k_2-q_2)+(k_1-q_1)\big] \; \delta^{(2)}\big[(k_3-q_3)-(k_1-q_1)\big] \; \delta^{(2)}\big[(k_3-q_3)+(k_2-q_2)\big]$ -- the backward Bose correlation of  gluons $1$ and $2$, forward Bose correlation of gluons $1$ and $3$ and backward Bose correlation of gluons $2$ and $3$.
\\
$\bullet$ The fourth term in $I_{X,4}\propto \delta^{(2)}\big[ (k_1-q_1)+(k_3-q_3)\big]\; \delta^{(2)}\big[ (k_1-q_1)-(k_2-q_2)\big]\; \delta^{(2)}\big[ (k_2-q_2)+(k_3-q_3)\big]$  -- the backward Bose correlation of gluons $1$ and $3$, forward Bose correlation of gluons $1$ and $2$ and backward Bose correlation of gluons $2$ and $3$.
\\
Finally, for $I_{X,5}$ we have
\beq
\label{X5_initial}
I_{X,5}&=&
\mu^2(k_2-q_2,q_2-k_1)\, \mu^2(k_1-q_1,q_1-k_3) \, \mu^2(k_3-q_3,q_3-k_2) \nonumber\\
&\times&
L^i(k_1,q_1)L^i(k_1,q_2) \, L^j(k_2,q_2)L^j(k_2,q_3)\, L^k(k_3,q_3)L^k(k_3,q_1)+(k_3\to-k_3)\nonumber\\
&+&
\mu^2(k_2-q_2,q_2-k_3) \, \mu^2(k_3-q_3,q_3+k_1) \, \mu^2(q_1-k_2,-k_1-q_1) \nonumber\\
&\times&
L^i(k_1,-q_1)L^i(k_1,-q_3)\, L^j(k_2,q_2)L^j(k_2,q_1) \, L^k(k_3,q_3)L^k(k_3,q_2) +(k_1\to-k_1)\nonumber\\
&+&
\mu^2(k_2+q_1,k_1-q_1) \, \mu^2(k_3-q_3, q_3-k_1) \, \mu^2(q_2-k_3, -k_2-q_2) \nonumber\\
&\times&
L^i(k_1,q_1)L^i(k_1,q_3) \, L^j(k_2,-q_2)L^j(k_2,-q_1)\, L^k(k_3,q_3)L^k(k_3,q_2) +(k_3\to-k_3)\nonumber\\
&+&
\mu^2(k_2+q_3,k_3-q_3)\, \mu^2(-k_1-q_1, q_1-k_3)\, \mu^2(k_1-q_2,q_2-k_2) \nonumber\\
&\times&
L^i(k_1,-q_1)L^i(k_1,q_2)\, L^j(k_2,q_2)L^j(k_2,-q_3)\, L^k(k_3,q_3)L^k(k_3,q_1) +(k_1\to-k_1),
\eeq
with the corresponding identification:
\\
$\bullet$ The first term in $I_{X,5} \propto \delta^{(2)}(k_1-k_2)\; \delta^{(2)}(k_1-k_3)\; \delta^{(2)}(k_3-k_2)$ -- the forward HBT correlation of  gluons $1$ and $2$, forward HBT of gluons $1$ and $3$, and forward HBT of gluons $2$ and $3$.
\\
$\bullet$ The second term in $I_{X,5} \propto \delta^{(2)}(k_1+k_2)\;  \delta^{(2)}(k_1+k_3)\; \delta^{(2)}(k_3-k_2)$ -- the forward HBT of  gluons $2$ and $3$, backward HBT of gluons $1$ and $3$, backward HBT of gluons $1$ and $2$.
\\
$\bullet$ Third term in $I_{X,5} \propto \delta^{(2)}(k_1+k_2)\; \delta^{(2)}(k_1-k_3)\; \delta^{(2)}(k_3+k_2)$ -- the backward HBT of  gluons $1$ and $2$, forward HBT of gluons $1$ and $3$, backward HBT of gluons $2$ and $3$.
\\
$\bullet$ The fourth term in $I_{X,5} \propto \delta^{(2)}(k_1-k_2)\; \delta^{(2)}(k_1+k_3)\; \delta^{(2)}(k_2+k_3)$ -- the forward HBT of  gluons $1$ and $2$, backward HBT of gluons $1$ and $3$, backward HBT of gluons $2$ and $3$.

\section{The  $v_2$ and the correlations}
\label{N_v2}

\subsection{Total multiplicity, mean transverse momentum and $v_2$}
We start with calculating the total multiplicity and the second flow harmonic coefficient $v_2$.

 Starting from the expression for the
single inclusive spectrum (\ref{N1}), and carrying out the $q_1$ integral we obtain
\beq\label{N1int}
\frac{dN^{(1)}}{d^2k_1}=\alpha_s\, (4\pi)\, (N_c^2-1)\, \mu^2 \, S_\perp\, e^{-k_1^2/Q_s^2} \bigg\{ \frac{2}{k_1^2}-\frac{1}{k_1^2}e^{k_1^2/Q_s^2}+\frac{1}{Q_s^2}\bigg[{\rm Ei}\bigg(\frac{k_1^2}{Q_s^2}\bigg)-{\rm Ei}\bigg(\frac{k_1^2\, \lambda}{Q_s^2}\bigg)\bigg]\bigg\},
\eeq
where $S_\perp$ is the transverse area of the projectile and $Q_s^2$ is the saturation momentum of the target as defined in Eq.~\eqref{gbw}.
Here we have introduced the infrared cutoff $\lambda\ll 1$ by regulating the integration over the Schwinger parameter, see Appendix A. In terms of physical quantities the value of the IR cutoff is of order $\lambda\sim 1/(S_\perp Q_s^2)$.

As is well known,  the spectrum is divergent in the infrared,
in the limit $k_1^2/Q_s^2\to 0$:
\beq\label{specexp}
\frac{dN^{(1)}}{d^2k_1}(k_1^2/Q_s^2\to 0)\simeq\alpha_s\, (4\pi)\, (N_c^2-1)\, \mu^2 \, S_\perp\, \frac{1}{k_1^2}\bigg\{ 1-\big[2+\ln(\lambda)\big]\frac{k_1^2}{Q_s^2} +\big[ 2+\ln(\lambda)-\lambda\big]\frac{k_1^4}{Q_s^4}\bigg\}.
\eeq
For large momenta $k_1^2/Q_s^2\to \infty$, the spectrum reduces to the usual perturbative one:
\beq
\frac{dN^{(1)}}{d^2k_1}(k_1^2/Q_s^2\to \infty)\simeq \alpha_s\, (4\pi)\, (N_c^2-1)\, \mu^2 S_\perp \frac{Q_s^2}{k_1^4}\bigg[1+\frac{2\, Q_s^2}{k_1^2}+...\bigg].
\eeq
At finite $\lambda$ the IR asymptotics of the expression in Eq. \eqref{specexp} is $\frac{dN^{(1)}}{d^2k_1}(k_1^2/Q_s^2\to 0)\sim \mu^2 \, S_\perp/k_1^2$. Interestingly, the range of momenta in which this asymptotic behavior holds at very small $\lambda$ is narrow, $k^2<Q_s^2/|2+\ln\lambda|$. For very small values of $\lambda$ the total multiplicity is dominated by momenta in the range 
$Q_s^2/|2+\ln\lambda|<k^2<Q_s^2$ where the spectrum is actually, flat $\frac{dN^{(1)}}{d^2k_1}\approx \mu^2 \, S_\perp\big[2+\ln(\lambda)\big]$. This is an interesting feature of the spectrum, but it is only apparent at very small values of $\lambda$. In pA scattering for reasonable values of parameters ($Q_s\sim 1$ GeV, $S_\perp\sim 1/\Lambda_{QCD}^2$) we use  $\lambda\sim 1/S_\perp Q_s^2\sim 1/25$. At this value of $\lambda$ the interval of momenta where the spectrum is flat shrinks almost completely, and the spectrum exhibits a rather sharp crossover from a $1/k^2$ behavior  at $k^2<Q_s^2$ to $1/k^4$ at $k^2>Q_s^2$. 

%Here we encounter the very well known fact that the total multiplicity is dominated by the soft modes, where the spectrum behaves as $1/k_1^2$. 
%\beq
%\bar N=\int d^2k_1 \frac{dN^{(1)}}{d^2k_1} \sim\; \alpha_s\, (4\pi)\, (N_c^2-1)\, \mu^2 S_\perp \, \pi\, \big[ \ln(\Lambda_{\rm max})-\ln(\Lambda_{\rm min}) \big]
%\eeq 

%As one can deduce from (\ref{specexp}), $\Lambda^2_{max}\simeq Q_s^2/(2+\ln(\lambda))$, which is of order $Q_s^2$ for phenomenologically reasonable values of $\lambda$.
In the next section we evaluate the $k_1^2$ integral of (\ref{N1int}) numerically taking $\lambda=1/25$. Since the dependence on $\lambda$ is only logarithmic, the result is quite insensitive to the precise value of the IR cutoff.

The mean transverse momentum $\bar k^2$ is formally defined as the average of $k_1^2$ with the weight Eq. \eqref{N1int}.
From the expression in Eq. \eqref{N1int},
\beq
k_1^2\, \frac{dN^{(1)}}{d^2k_1}=\alpha_s\, (4\pi)\, (N_c^2-1)\, \mu^2 \, S_\perp\, e^{-k_1^2/Q_s^2} \bigg\{ 2 -e^{k_1^2/Q_s^2}+\frac{k_1^2}{Q_s^2}\bigg[{\rm Ei}\bigg(\frac{k_1^2}{Q_s^2}\bigg)-{\rm Ei}\bigg(\frac{k_1^2\, \lambda}{Q_s^2}\bigg)\bigg]\bigg\}.
\eeq
%(i) The IR limit $k_1^2/Q_s^2\to 0$:
%\beq
%k_1^2\, \frac{dN^{(1)}}{d^2k_1}\sim\alpha_s\, (4\pi)\, (N_c^2-1)\, \mu^2 \, S_\perp\,\eeq
%(ii) The UV limit $k_1^2/Q_s^2\to \infty$:
The integral over $k_1$  diverges logarithmically in the  UV
\beq
k_1^2\, \frac{dN^{(1)}}{d^2k_1}\simeq \alpha_s\, (4\pi)\, (N_c^2-1)\, \mu^2 S_\perp \frac{Q_s^2}{k_1^2}\bigg[1+\frac{2\, Q_s^2}{k_1^2}\bigg].
\eeq
%integrating over $k_1$, we get
%\beq
%\int d^2k_1\, k_1^2\, \frac{dN^{(1)}}{d^2k_1}\sim\; \tilde c_0\, \mu^2 S_\perp \, \pi\, \big[ (\Lambda_{\rm max})-(\Lambda_{\rm min}) \big]
%\eeq 
%
%integration over $k_1$ is now divergent in the UV and requires a UV cut-off. 
Being divergent, the average momentum defined this way is not a very useful quantity. Instead, whenever we need a quantity that represents a typical momentum of produced particles we will use
\beq
\bar k^2\rightarrow Q_s^2 \ .
\eeq

The second flow coefficient is evaluated using our expressions for the double inclusive gluon spectrum introduced in (\ref{N2Q}).
\beq
\frac{dN^{(2)}}{d^2k_2\, d^2k_3}\bigg|_Q=Q_1+Q_2\ ,
\eeq
where in the large $N_c$ limit and in the approximation of translationally invariant  projectile (see Appendix A)
\beq
Q_1&=& \alpha_s^2\, (4\pi)^2\, (N_c^2-1) \; \, \mu^4\, S_{\perp}\, \frac{1}{\pi Q_s^2}\; e^{-(k_2-k_3)^2/2Q_s^2} \bigg\{ \bigg[ \frac{1}{2}+\frac{2^2\, Q_s^2}{(k_2+k_3)^2}+\frac{2^4\, Q_s^4}{(k_2+k_3)^4}\bigg] \, \frac{1}{k_2^2k_3^2}\frac{(k_2-k_3)^4}{(k_2+k_3)^4}
\\
&&\hspace{6cm}
+\; Q_s^4\; \frac{2^6}{(k_2+k_3)^8}\bigg[1+(k^i_2-k^i_3)\bigg(\frac{k_2^i}{k_2^2}-\frac{k_3^i}{k_3^2}\bigg)\bigg]\bigg\}+(k_3\to-k_3),\nonumber
\eeq
\beq
Q_2&=& \alpha_s^2\, (4\pi)^2\, (N_c^2-1)\;  \, \mu^4\, S_{\perp}\, (2\pi)^2\, \Big[ \delta^{(2)}(k_2+k_3)+\delta^{(2)}(k_2-k_3)\Big]  \; \frac{1}{2} \; \frac{Q_s^4}{k_2^8}\ .
\eeq
These expressions have been obtained assuming large values of transverse momenta $k_{2,3}^2\gg Q_s^2$.

The momentum dependent second flow coefficient is defined as 
\begin{equation}\label{v2}
v^2_2(k_2,k_3)=  \frac{\int d\phi_2 d \phi_3 e^{i 2(\phi_2-\phi_3)}  \, \frac { d^2N^{(2)} } { d^2k_2 d^2 k_3 }  }{\int d\phi_2 d \phi_3   \, \frac { d^2N^{(2)} } { d^2k_2 d^2 k_3 }  }\ .
\end{equation}
One usually also averages the numerator and the denominator in Eq. \eqref{v2} separately over momentum bins of finite width.

The only contribution to the numerator in Eq. \eqref{v2} comes from the correlated term Eq. \eqref{N2Q} since the uncorrelated term vanishes upon angular integration. The denominator, on the other hand, is dominated by the uncorrelated piece which at large $N_c$ is given by Eq. \eqref{uncor}.

Although the general expressions for the two gluon inclusive spectrum have been known for a while~\cite{Altinoluk:2018ogz,Altinoluk:2018hcu}, we are not aware of the actual calculation of $v_2$ in this simple dense-dilute approach. Here we evaluate Eq.\eqref{v2} numerically, and present the results in the next section.
%with $\bar c_0=\alpha_s^2\, (4\pi)^2\, (N_c^2-1)$.
%space{2cm}

\subsection{$v_2$ vs total multiplicity}

We now turn to our observables of interest. We first aim to study correlations between $v_2$ and multiplicity. The standard measure of correlation between two observables $X$ and $Y$ is the Pearson coefficient $R$
\begin{equation}\label{pierson}
R(X,Y)=\frac{\langle (X-\langle X\rangle)(Y-\langle Y\rangle)\rangle}{\sqrt{\langle X^2-\langle X\rangle^2\rangle}\sqrt{\langle Y^2-\langle Y\rangle^2\rangle}}
\end{equation}
which measures the strength of the correlation between $X$ and $Y$ relative to their autocorrelations. 
This type of observable was studied recently in~\cite{Schenke:2020uqq} in order to flesh out the effects of initial state momentum anisotropies.

In our case however the calculation of the Pearson coefficient would involve the calculation of the four gluon inclusive production (i.e. $\langle (v_2^2)\rangle$) which is relatively complicated. In addition, we are not  interested to compare the correlation with the autocorrelations of $v_2$ and $N$, but rather in  comparing to to the average value of the observables themselves. We will therefore not calculate the Pearson coefficient, but rather define the normalized correlator as
\begin{eqnarray}\label{ONv2}
&&{\cal O}_{ N,v_2}
=\int d\phi_2\, d\phi_3 \, e^{i2(\phi_2-\phi_3)}\int d^2k_1 \frac{dN^{(3)}}{d^2k_1\, d^2k_2\, d^2k_3}\ 
\bigg/ \ 
 \int d\phi_2\, d\phi_3 \, e^{i2(\phi_2-\phi_3)} \frac{dN^{(2)}}{d^2k_2d^2k_3} \int d^2k_1 \frac{dN^{(1)}}{d^2k_1}\nonumber\\
&&=\int d\phi_2\, d\phi_3 \, e^{i2(\phi_2-\phi_3)}\int d^2k_1 \frac{dN^{(3)}}{d^2k_1\, d^2k_2\, d^2k_3}\bigg|_X \ 
\bigg/ \ 
 \int d\phi_2\, d\phi_3 \, e^{i2(\phi_2-\phi_3)} \frac{dN^{(2)}}{d^2k_2d^2k_3}\bigg|_Q \int d^2k_1 \frac{dN^{(1)}}{d^2k_1}\ ,
\end{eqnarray}
where the second equality follows since only the fully correlated part of three (two) gluon inclusive production contributes to the numerator (denominator). The numerator in this definition is precisely the same as the numerator in Eq. \eqref{pierson} with $X=v_2^2$ and $Y=N$, but it is normalized to the product $\langle X\rangle\langle Y\rangle$ rather than to the square root of the product of variances of $X$ and $Y$.

The correlation between  $v_2$  and the total multiplicity of produced particles (per unit rapidity)
is related to the inclusive three gluon production cross section (\ref{3g}).  
%
%
%% 
%%\subsection{The $X$ factor}
%
%%Two things to emphasize:
%
%%Analytical calculations are performed  up to the angular integrations. Angular integrations are performed numerically. 
%
%%Analytical results are not the full results. We only keep the exponentially enhanced terms in the analytical computation. 
%
Starting from Eq. \eqref{uninteg_3g_N}, and integrating over $k_1$,   the result can be split similarly as in Eq. \eqref{uninteg_3g_N}: 
\beq
%\frac{d\bar N^{(3)}}{d^2k_2d^2k_3}\equiv
\int d^2k_1\frac{dN^{(3)}}{d^2k_1d^2k_2d^2k_3}\bigg|_X=X_1+X_2+X_3+X_4+X_5\ .
\eeq
We are able to perform the $k_1$ integration analytically, while the remaining angular integrations are performed numerically. 

Recall that we are only considering large transverse momenta of the observed particles, $|k_{2(3)}|\gg Q_s$. 
This large transverse momentum can be achieved in two distinct ways: either A) the incoming projectile gluons already have large transverse momentum and the momentum transfer in the scattering is relatively small, or B) most of the final state momentum is transferred to a projectile gluon in the scattering. The two contributions have very different behaviors. On the one hand, large transfer momentum is exponentially suppressed in the GBW model as $\exp\{-k^2/Q_s^2\}$, which favors contribution A. On the other hand, the number of gluons in the projectile wave function is strongly peaked at small momentum, so that $N_p(p)/N_p(q)\sim q^2/p^2$. Thus the number of incoming gluons at high transverse momentum is suppressed roughly by a factor $1/(S_\perp k^2)$. For very large transverse area this suppression may be significant enough so that contribution B can become comparable or even larger than contribution A. However for a proton projectile this factor is very unlikely to compete with the exponential suppression due to high momentum transfer. In our calculations, therefore, we only keep the contribution due to small ($\sim \mathcal{O}(Q_s)$)  momentum transfer from the target. 

The calculation is fairly lengthy and the details are given in the Appendix A. The results are presented below.
In general we find two types of terms. 
The one type gives a correlation which in momentum space has width of order $Q_s$. This arises from Bose correlations between the incoming gluons $2$ and $3$ in conjunction with either HBT or Bose correlations of any one of these gluons with gluon $1$. These terms are:\\
\noindent$\bullet$ $X_1$:
\beq
\label{X_1_final}
X_1&=& \frac{1}{2}\, \alpha_s^3(4\pi)^6(N_c^2-1)\; \mu^6\; S_\perp\  e^{-(k_2-k_3)^2/2Q_s^2}\; \frac{1}{k_2^4}
\nonumber\\
&\times&
\bigg\{ \bigg\lgroup \frac{1}{2}+Q_s^2\bigg[\frac{1}{k_2^2}+\frac{2^2}{(k_2+k_3)^2}\bigg]
+Q_s^4\bigg[ \frac{3}{k_2^4}+\frac{2!}{k_2^2}\frac{2^2}{(k_2+k_3)^2}+\frac{2^4}{(k_2+k_3)^4}\bigg]\bigg\rgroup
\frac{1}{k_2^2k_3^2}\frac{(k_2-k_3)^4}{(k_2+k_3)^4}\nonumber\\
&&
\hspace{0.5cm}
+\; Q_s^4\frac{2^6}{(k_2+k_3)^8}\bigg[ 1+(k_2^i-k_3^i)\bigg(\frac{k_2^i}{k_2^2}-\frac{k_3^i}{k_3^2}\bigg)\bigg]\bigg\}.
\eeq
As  explained in the previous section, $X_1$ contributes largely to the forward correlation of the produced gluons. Indeed, as seen from its final expression, $X_1$
is enhanced in the forward region $k_2=k_3$,  where the exponential pre factor is equal to unity,
\beq
X_1(k_2=k_3)&=& \alpha_s^3(4\pi)^6(N_c^2-1)\; \mu^6\; S_\perp\; \frac{1}{8}\; \frac{Q_s^4}{k_2^{12}}\ .
\eeq
The width of the forward region is clearly $|k_2-k_3|\sim Q_s$ , and away from this region this expression is exponentially suppressed.

\noindent$\bullet$ $X_3$:
Comparing Eqs. (\ref{X1_shifted}) and (\ref{X3_shifted}), one notes that  $X_3=X_1(k_2\leftrightarrow k_3)$:
%Therefore, we can immediately write down the result for $X_3$:
% I will use this as a double check of the results, since the final results are quite complicated and it is easy to make algebraic mistakes. I have already spotted a mismatch between the final results (overall factor and explicit expression of the momentum structure that accompanies exponential term).   
\beq
\label{X_3_final}
X_3&=& \frac{1}{2}\, \alpha_s^3(4\pi)^6(N_c^2-1)\; \mu^6\; S_\perp\;  e^{-(k_2-k_3)^2/2Q_s^2}\; \frac{1}{k_3^4}
\nonumber\\
&\times&
\bigg\{ \bigg\lgroup \frac{1}{2}+Q_s^2\bigg[\frac{1}{k_3^2}+\frac{2^2}{(k_2+k_3)^2}\bigg]
+Q_s^4\bigg[ \frac{3}{k_3^4}+\frac{2!}{k_3^2}\frac{2^2}{(k_2+k_3)^2}+\frac{2^4}{(k_2+k_3)^4}\bigg]\bigg\rgroup
\frac{1}{k_2^2k_3^2}\frac{(k_2-k_3)^4}{(k_2+k_3)^4}\nonumber\\
&&
\hspace{0.5cm}
+\; Q_s^4\frac{2^6}{(k_2+k_3)^8}\bigg[ 1+(k_2^i-k_3^i)\bigg(\frac{k_2^i}{k_2^2}-\frac{k_3^i}{k_3^2}\bigg)\bigg]\bigg\}.
\eeq
%In the limit  $k_2=k_3$, we get 
%\beq
%X_3&\sim& c_0\; \mu^6\; S_\perp\; (4\pi)^3\; \frac{1}{8}\; \frac{Q_s^4}{k_3^{12}}
%\eeq
%
%
\noindent$\bullet$ $X_4$:
\beq
\label{X_4_final}
X_4&=&\alpha_s^3(4\pi)^6(N_c^2-1)\, \mu^6\, S_\perp\, e^{-(k_2-k_3)^2/2Q_s^2}
\bigg\{ \bigg[ 1+ 8\, \frac{2^2\, Q_s^2}{(k_2+k_3)^2}+76\, \frac{2^4\, Q_s^4}{(k_2+k_3)^4}\bigg]\frac{2^3}{k_2^2k_3^2}\frac{(k_2-k_3)^4}{(k_2+k_3)^8}
\nonumber\\
&&
\hspace{5cm}
+\; \frac{2^4\, Q_s^4}{(k_2+k_3)^4}\frac{2^2}{(k_2+k_3)^2}\bigg[ \frac{5}{2}\frac{ 2^6}{(k_2+k_3)^6}-\frac{9}{4}\frac{2^2}{k_2^2k_3^2(k_2+k_3)^2}\bigg]\bigg\}.
\eeq
We again notice that $X_4$ is enhanced In the limit $k_2=k_3$:
\beq
X_4(k_2=k_3)&\approx&\alpha_s^3(4\pi)^6(N_c^2-1)\, \mu^6\, S_\perp\, \frac{1}{4}\, \frac{Q_s^4}{k_2^{12}}\ .
\eeq 

The second type of terms is due to HBT correlations between gluons $2$ and $3$. These correlations
in the translationally invariant approximation lead to  $\delta$ functional terms, contributing  when $k_2=\pm k_3$. Accounting for a finite projectile area would regulate the delta functions smearing them on the scale of order $1/S_\perp$. Nevertheless, the correlation due to these terms is very narrow. We will come back to this point in the next section when analyzing our numerical results. The terms of this type are:\\
\noindent$\bullet$ $X_2$:
\beq
\label{X_2_final}
X_2&=& \alpha_s^3\frac{1}{2}(4\pi)^7(N_c^2-1)\, \mu^6\;  S_\perp \big[  \delta^{(2)}(k_2+k_3)+\delta^{(2)}(k_2-k_3)\big]\,  \frac{1}{4} \frac{Q_s^6}{k_2^{12}}
\eeq
\noindent$\bullet$  $X_5$:
\beq
\label{X_5_final}
X_5=  \alpha_s^3\frac{1}{2}(4\pi)^7(N_c^2-1)\, \mu^6\, S_\perp \, \Big[ \delta^{(2)}(k_2+k_3)+\delta^{(2)}(k_2-k_3)\Big] \, \frac{1}{8}\, \frac{Q_s^6}{k_2^{12}}
\eeq

In the next section we present the results of the numerical evaluation of the angular integral of these expressions as defined in Eq. \eqref{ONv2}.
%Like the contribution $X_2$, this term leads to correlation very narrow in momentum space.
%
%
%
%

%
%
%
%
%
%

\subsection{$v_2$ vs mean transverse momentum}
\label{Pt_v2}

The second observable we consider is the correlation between mean transverse momentum and $v_2$  defined as 
\beq\label{Okv2}
{\cal O}_{ k,v_2}=\int d\phi_2\, d\phi_3 \, e^{in(\phi_2-\phi_3)}\int d^2k_1 \, k_1^2\, \frac{dN^{(3)}}{d^2k_1\, d^2k_2\, d^2k_3}\bigg|_X\ 
\bigg/ \ 
 \int d\phi_2\, d\phi_3 \, e^{in(\phi_2-\phi_3)} \frac{dN^{(2)}}{d^2k_2d^2k_3}\bigg|_Q Q_s^2\int d^2k_1\,  \frac{dN^{(1)}}{d^2k_1}\ .
\eeq
In accordance to our discussion earlier, we have substituted $Q_s^2$ for the average transverse momentum in the "normalization" in the denominator.

The computation of this observable proceeds very similarly to the one considered in the previous subsection. Details are given in  Appendix B. Here  we  present the results:
\beq
\int d^2k_1 \, k_1^2\, \frac{dN^{(3)}}{d^2k_1\, d^2k_2\, d^2k_3}\bigg|_X= \bar X_1+\bar X_2+\bar X_3+\bar X_4+\bar X_5\ ,
\eeq
with\\
\noindent$\bullet$ $\bar X_1$: 
%HBT of gluons $k_1$ and $k_2$, Bose enhancement of the gluons $(k_1-q_1)$ and $(k_3-q_3)$. 
%\\
%\\ ${\bar X_1}\propto\delta^{(2)}(k_1\pm k_2)\; \delta^{(2)}\big[ (k_1-q_1)\pm(k_3-q_3)\big]\; \delta^{(2)}\big[(k_3-q_3)\pm(k_1-q_1)\big]$
\beq
\label{barX1_final}
\bar X_1&=& \frac{1}{2}\, \alpha_s^3(4\pi)^6(N_c^2-1)\; \mu^6\; S_\perp\; e^{-(k_2-k_3)^2/2Q_s^2}\; \frac{1}{k_2^2}
\nonumber\\
&\times&
\bigg\{ \bigg\lgroup \frac{1}{2}+Q_s^2\bigg[\frac{1}{k_2^2}+\frac{2^2}{(k_2+k_3)^2}\bigg]
+Q_s^4\bigg[ \frac{3}{k_2^4}+\frac{2!}{k_2^2}\frac{2^2}{(k_2+k_3)^2}+\frac{2^4}{(k_2+k_3)^4}\bigg]\bigg\rgroup
\frac{1}{k_2^2k_3^2}\frac{(k_2-k_3)^4}{(k_2+k_3)^4}\nonumber\\
&&
\hspace{0.5cm}
+\; Q_s^4\frac{2^6}{(k_2+k_3)^8}\bigg[ 1+(k_2^i-k_3^i)\bigg(\frac{k_2^i}{k_2^2}-\frac{k_3^i}{k_3^2}\bigg)\bigg]\bigg\}+(k_3\to-k_3),
\eeq
%In the limit  $k_2=k_3$:
\beq
\bar X_1(k_2=k_3)&=& \alpha_s^3(4\pi)^6(N_c^2-1)\; \mu^6\; S_\perp\ \frac{1}{8}\; \frac{Q_s^4}{k_2^{10}}\times 2.
\eeq

\noindent$\bullet$ $\bar X_3$: 
%HBT of gluons $k_1$ and $k_3$, Bose enhancement of the gluons $(k_2-q_2)$ and $(k_3-q_3)$. 
% ${\bar X_3}\propto\delta^{(2)}(k_1\pm k_3)\; \delta^{(2)}\big[ (k_2-q_2)\pm(k_3-q_3)\big]\; \delta^{(2)}\big[(k_3-q_3)\pm(k_2-q_2)\big]$
\beq
\label{barX3_final}
\bar X_3&=& \frac{1}{2}\, \alpha_s^3(4\pi)^6(N_c^2-1)\; \mu^6\; S_\perp\ e^{-(k_2-k_3)^2/2Q_s^2}\; \frac{1}{k_3^2}
\nonumber\\
&\times&
\bigg\{ \bigg\lgroup \frac{1}{2}+Q_s^2\bigg[\frac{1}{k_3^2}+\frac{2^2}{(k_2+k_3)^2}\bigg]
+Q_s^4\bigg[ \frac{3}{k_3^4}+\frac{2!}{k_3^2}\frac{2^2}{(k_2+k_3)^2}+\frac{2^4}{(k_2+k_3)^4}\bigg]\bigg\rgroup
\frac{1}{k_2^2k_3^2}\frac{(k_2-k_3)^4}{(k_2+k_3)^4}\nonumber\\
&&
\hspace{0.5cm}
+\; Q_s^4\frac{2^6}{(k_2+k_3)^8}\bigg[ 1+(k_2^i-k_3^i)\bigg(\frac{k_2^i}{k_2^2}-\frac{k_3^i}{k_3^2}\bigg)\bigg]\bigg\}+(k_2\to-k_2),
\eeq
%In the limit  $k_2=k_3$, we get 
\beq
\bar X_3(k_2=k_3)&=& \alpha_s^3(4\pi)^6(N_c^2-1)\; \mu^6\; S_\perp\  \frac{1}{8}\; \frac{Q_s^4}{k_3^{10}}\times 2.
\eeq

\noindent$\bullet$ $\bar X_4$: 
%Bose enhancement of the gluons $(k_1-q_1)$, $(k_2-q_2)$ and $(k_3-q_3)$. 
% ${\bar X_4}\propto \delta^{(2)}\big[ (k_2-q_2)\pm(k_1-q_1)\big] \delta^{(2)}\big[(k_1-q_1)\pm(k_3-q_3)\big] \; \delta^{(2)}\big[ (k_3-q_3)\pm(k_2-q_2)\big]$
\beq
\label{barX4_final}
\bar X_4&=& \alpha_s^3(4\pi)^6(N_c^2-1)\; \mu^6\; S_\perp\  e^{-(k_2-k_3)^2/2Q_s^2} \bigg\{ \bigg[ 1+\frac{9}{2}\frac{2^2\, Q_s^2}{(k_2+k_3)^2}+15\frac{2^4\, Q_s^4}{(k_2+k_3)^4}\bigg] \frac{2}{k_2^2\, k_3^2}\frac{(k_2-k_3)^4}{(k_2+k_3)^6}\nonumber\\
&& 
\hspace{5cm}
+\, \frac{2^4 \, Q_s^4}{(k_2+k_3)^4}\frac{2^2}{(k_2+k_3)^2}\bigg[ \frac{3}{2}\frac{2^4}{(k_2+k_3)^4}-\frac{5}{4}\frac{1}{k_2^2\, k_3^2}\bigg]\bigg\} +(k_3\to-k_3),
\eeq
%In the limit $k_2=k_3$, we get
\beq
\bar X_4(k_2=k_3)= \alpha_s^3(4\pi)^6(N_c^2-1)\; \mu^6\, S_\perp\;  \frac{1}{4} \, \frac{Q_s^4}{k_2^{10}}\, \times 2.
\eeq
\noindent$\bullet$ $\bar X_2$: 
%HBT of gluons $k_2$ and $k_3$, Bose enhancement of the gluons $(k_2-q_2)$ and $(k_1-q_1)$. 
%\\
%\\ ${\bar X_2}\propto\delta^{(2)}(k_2\pm k_3)\; \delta^{(2)}\big[ (k_2-q_2)\pm(k_1-q_1)\big]\; \delta^{(2)}\big[(k_1-q_1)\pm(k_2-q_2)\big]$
\beq
\label{barX2_final}
\bar X_2&=& \alpha_s^3(4\pi)^6(N_c^2-1)\, \mu^6\, (2\pi)\;  S_\perp \big[  \delta^{(2)}(k_2+k_3)+\delta^{(2)}(k_2-k_3)\big]\,  \frac{1}{4} \frac{Q_s^6}{k_2^{10}}\ .
\eeq

\noindent $\bullet$ $\bar X_5$: 
%HBT of gluons $k_1$, $k_2$ and $k_3$\\
%$\bar X_5 \propto \delta^{(2)}(k_1-k_2)\; \delta^{(2)}(k_1-k_3)\; \delta^{(2)}(k_3-k_2)$
\beq
\label{barX5_final}
\bar X_5\approx  \alpha_s^3(4\pi)^6(N_c^2-1)\, \mu^6\, (2\pi)\, S_\perp \, \Big[ \delta^{(2)}(k_2+k_3)+\delta^{(2)}(k_2-k_3)\Big] \,  \frac{1}{8}\, \frac{Q_s^6}{k_2^{10}}\ .
\eeq

In the next section we present the results of the numerical evaluation.

%  $\bar\Lambda_{max}$ we get 
%\beq
%\bar k^2= \int d^2k_1\, k_1^2\, \frac{dN^{(1)}}{d^2k_1}\sim\; \tilde c_0\, \mu^2 S_\perp \, \pi Q_s^2\,  \big[ \ln(\Lambda_{\rm max})-\ln(\Lambda_{\rm min}) \big]
%\bar k^2= \int d^2k_1\, k_1^2\, \frac{dN^{(1)}}{d^2k_1}\sim\; \tilde c_0\, \mu^2 S_\perp \, \pi Q_s^2\,  \big[ \ln(\bar\Lambda_{\rm max})-\ln(Q_s) \big]
%\eeq
%{\color{red} there is again some problem with the integration}
%with $\tilde c_0=\alpha_s\, (4\pi)\, (N_c^2-1)$. 

%As the first step let us integrate fully correlated piece of the three gluon spectrum given in Eq. \eqref{uninteg_3g_N} over $k_1$. To perform the computation we will treat the function $\mu^2$ within the MV model, ie
%%
%\beq
%\mu^2(k,p)\equiv \mu^2\,  (2\pi)^2\, \delta^{(2)}(k+p)\, 
%\eeq
%%
%we will use GBW model for the dipole operators
%%
%\beq
%\label{GBW}
%d(p)\equiv \frac{4\pi}{Q_s^2}e^{-p^2/Q_s^2}\,  
%\eeq
%%
%and the Lipatov vertex reads 
%%
%\beq
%\label{Lipatov}
%L^i(k,q)=\bigg[ \frac{(k-q)^i}{(k-q)^2}-\frac{k^i}{k^2}\bigg]
%\eeq

\section{Numerical results}
\label{numerics}
We now turn to numerical evaluation of the correlators discussed above. Here we mainly present the results, keeping their discussion for the next Section.

 Note that in all the figures we plot momentum in units of $Q_s$ and the quantities of interest multiplied by the factor $(N_c^2-1)S_\perp Q_s^2$ in order to exhibit the universal features of the result applicable to any target (any value of $Q_s$) and projectile (any value of $S_\perp$). The ratios we calculate also do not depend on the projectile scale $\mu^2$. To extract a number relevant for p-Pb or p-Au scattering one should take the realistic value $(N_c^2-1)S_\perp Q_s^2\sim 200$.
 
 For the normalization in Eqs.~\eqref{ONv2} and \eqref{Okv2}, the value of the cutoff $\lambda$ has to be specified in the integration Eq.~(\ref{N1int}).  While $\lambda=1/25$ was selected, we have checked that
varying $\lambda$ in reasonable limits does not  appreciably change the results.

We start with calculating $v_2$, Eq.~\eqref{v2}. In addition to the angular integration we also integrate the absolute values of transverse momenta within finite width bins. 
Thus we calculate
\begin{equation}\label{v2bin}
v^2_2(k,k',\Delta)= \frac{ \int_{k-\Delta/2}^{k+\Delta/2} k_2dk_2\int_{k^\prime-\Delta/2}^{k^\prime+\Delta/2} k_3dk_3\int d\phi_2 d \phi_3 e^{i 2(\phi_2-\phi_3)}  \, \frac { d^2N^{(2)} } { d^2k_2 d^2 k_3 }  }{\int_{k-\Delta/2}^{k+\Delta/2} k_2dk_2\int_{k^\prime-\Delta/2}^{k^\prime+\Delta/2} k_3dk_3\int d\phi_2 d \phi_3   \, \frac { d^2N^{(2)} } { d^2k_2 d^2 k_3 }  }\ .
%\frac{1}{\Delta^2} \int_{k-\Delta/2}^{k+\Delta/2} dk_2\int_{k^\prime-\Delta/2}^{k^\prime+\Delta/2} dk_3v_2(k_2,k_3)
\end{equation}
We take $k\gg\Delta$, $k'\gg\Delta$ and $\Delta\sim Q_s$.

We find it interesting to explore the interplay between  the relative position of the centers of the two bins, $k$ and $k'$ and the width of a bin $\Delta$. As discussed above, $v^2_2$ receives contributions form two types of correlations: the Bose and the HBT correlations. While the width of the Bose correlation in momentum space is naturally of order $Q_s$, the HBT correlations have  much shorter range (in our expressions they are formally represented by a delta function). Thus we expect that when $|k- k'|<\Delta$ both, the HBT and Bose effects will contribute to $v^2_2$, however when there is no overlap between the two bins, the HBT correlation should disappear. We thus expect a characteristic dependence of $v^2_2$ on $\Delta$ (at fixed $k-k'$) such that $v^2_2$ should vary steeply when $k-k'\approx \Delta$.

\begin{figure}
 \resizebox{0.49\textwidth}{!}{%
  \includegraphics{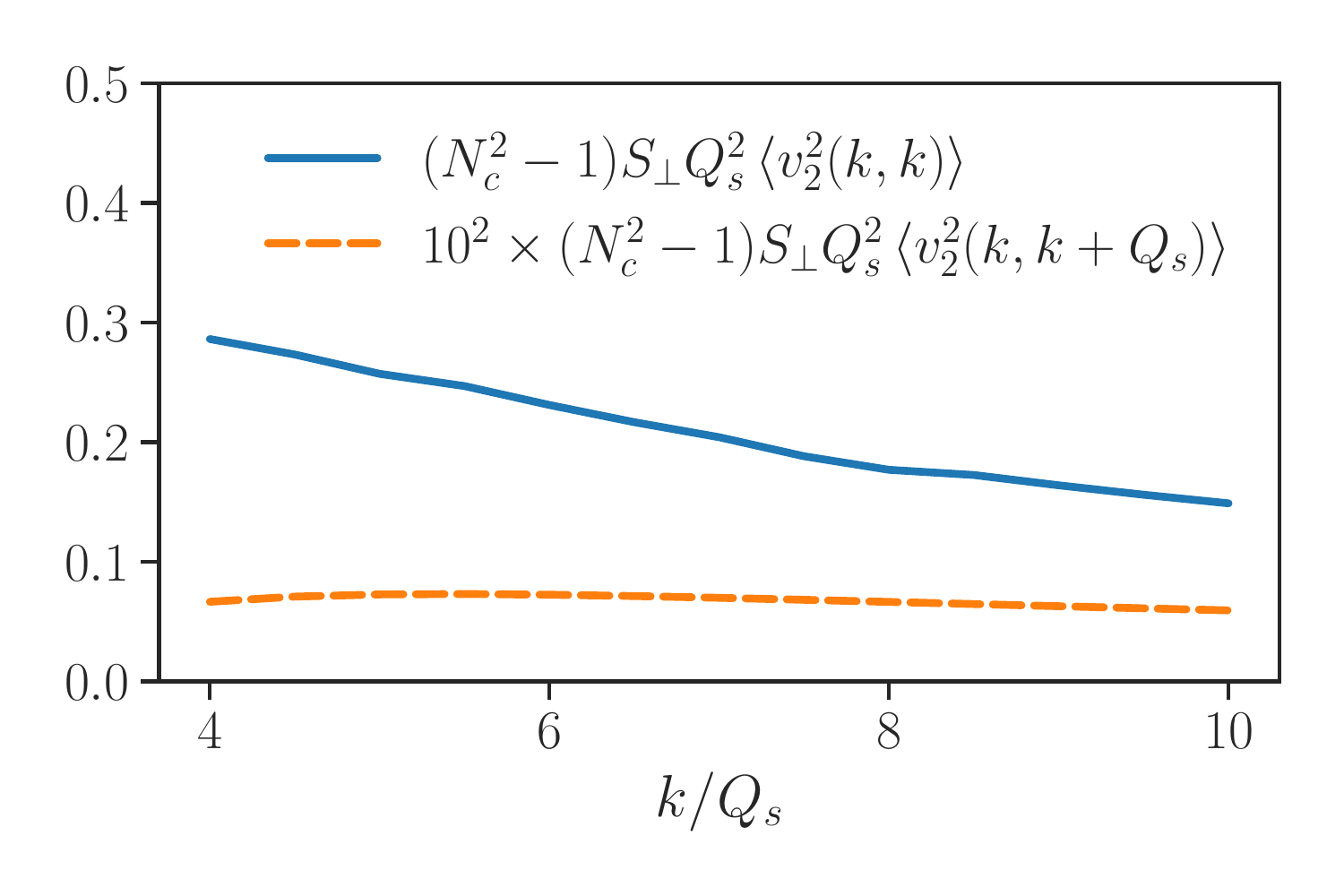}
}
\resizebox{0.49\textwidth}{!}{%
  \includegraphics{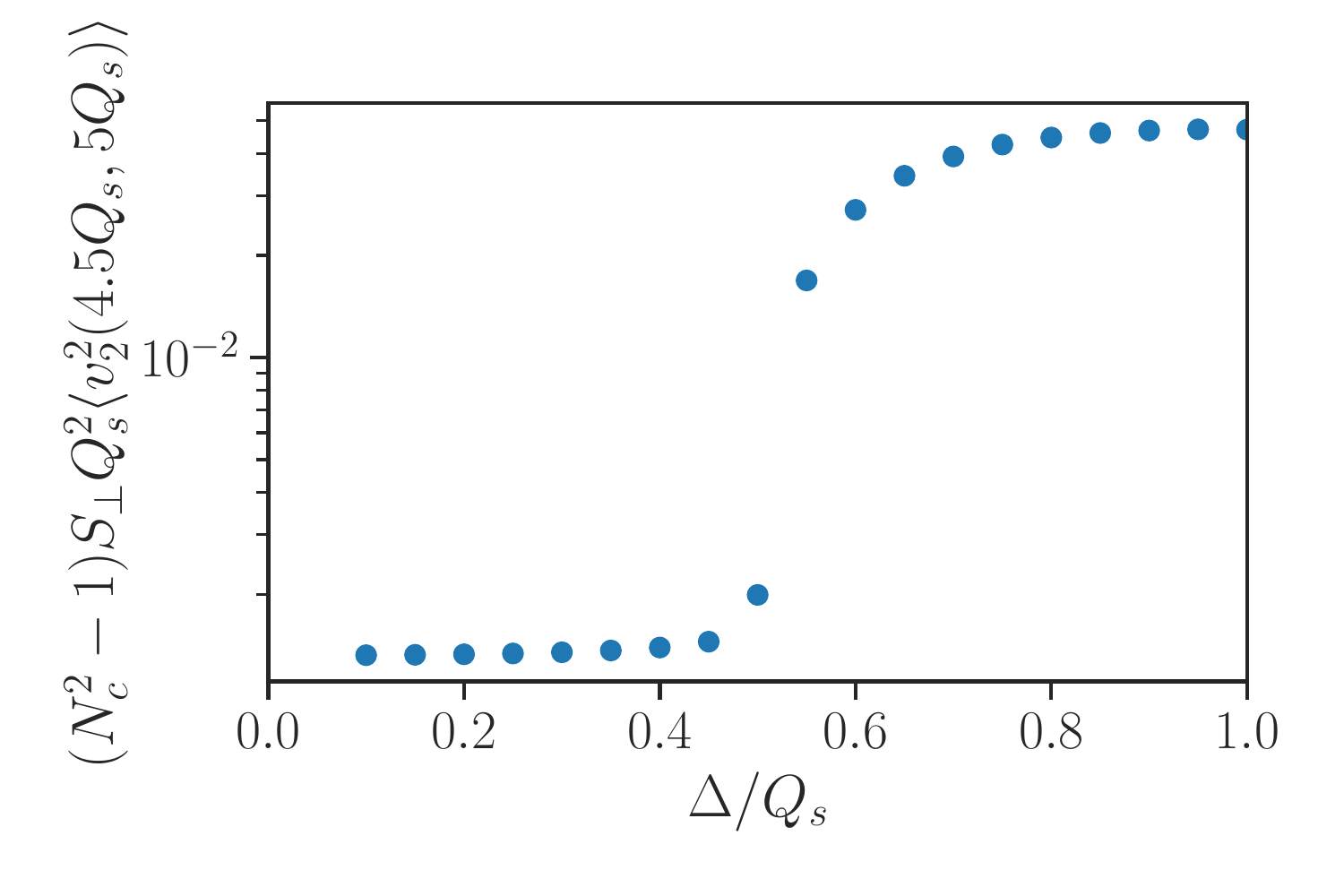}
}
\caption{
Left  panel:
The second flow harmonic, $v_2^2$ as a function of the momentum. 
The calculation of $v^2_2$ is performed for two cases: a)  the same momentum of the pair, b) the momentum of the pair is offset by the saturation momentum of the target in order to avoid the gluon HBT effect. The bin width in both cases is $\Delta=Q_s/2$. \\ 
Right panel:
	The second flow harmonic, $v_2^2$ as a function of the bin width.  The centers of the two bins are chosen at $k=4.5 Q_s$, $k'=5Q_s$.%Two particle correlations in $pp$ and $p$Pb collisions at the LHC measured by the ATLAS Collaboration~\cite{Aaboud:2016yar}, for different energies and particle multiplicities in the event.
}
\label{fig:0}       % Give a unique label
\end{figure}

Fig.~\ref{fig:0} shows our results for $v^2_2$.  In the left panel we see that the dependence of $v_2^2$ on the transverse momentum is rather different for overlapping and non overlapping momentum bins. In the right panel we observe, as expected, a sharp change in $v^2_2$ at the point when the width of the interval equals the distance between the interval midpoints. Interestingly we learn from Fig.~\ref{fig:0} that the contribution of the HBT correlations to $v^2_2$ is overwhelmingly large: it is by about a factor of $\sim 50$ dominates over the contribution of Bose enhancement (right panel of Fig.~\ref{fig:0}).

Next up is the correlation of $v^2_2$ with multiplicity,  Eq.~(\ref{ONv2}).
Again we integrate over bins of width $\Delta$ for the two momenta,
\beq
\frac{dN^{(3)}}{d^2k_1d^2k_2d^2k_3}\bigg|_X \,\rightarrow \, \int_{k-\Delta/2}^{k+\Delta/2} k_2dk_2\int_{k^\prime-\Delta/2}^{k^\prime+\Delta/2} k_3dk_3 \frac{dN^{(3)}}{d^2k_1d^2k_2d^2k_3}\bigg|_X\ .
\eeq

\begin{figure}
\resizebox{0.49\textwidth}{!}{%
  \includegraphics{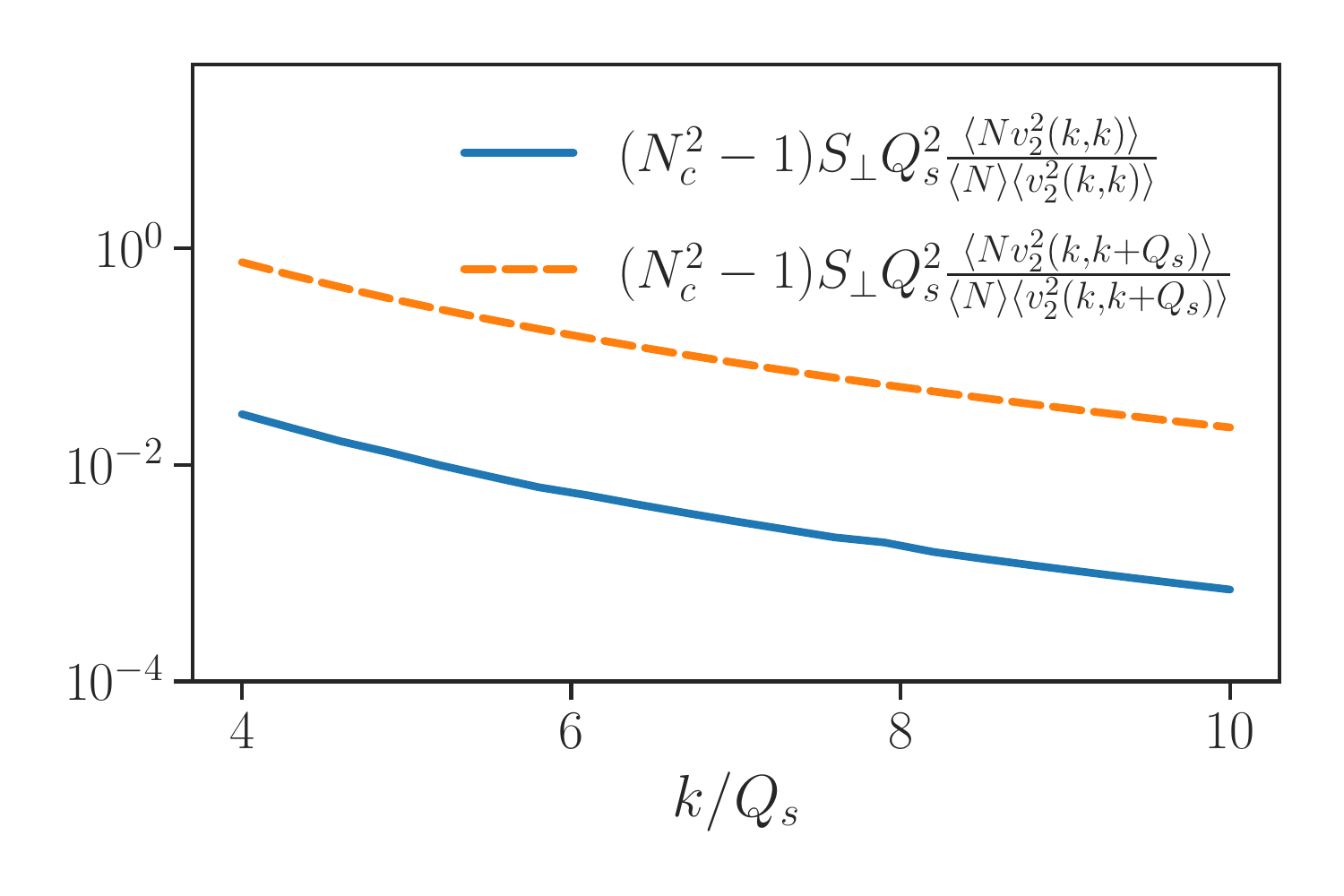}
}
\resizebox{0.49\textwidth}{!}{%
  \includegraphics{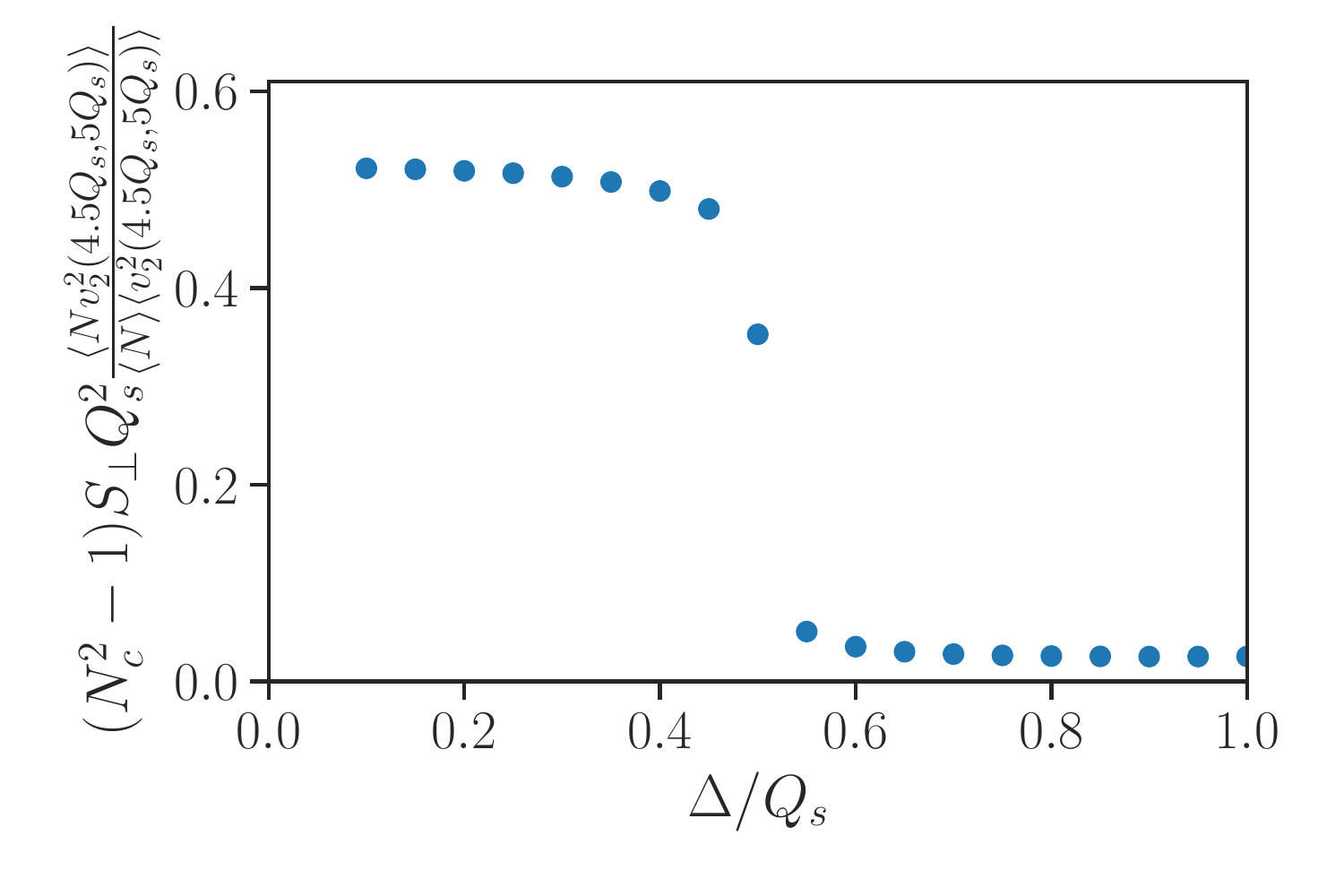}
}
\caption{
Left  panel:
	The three particle correlation function ${\cal O}_{ N,v_2}$ defined by  the normalized correlations between  $v^2_2$  and the total multiplicity of produced particles. The calculation of $v^2_2$ is performed for two cases: a)  the same momentum of the pair, b) the momentum of the pair is offset by the saturation momentum of the target in order to avoid the gluon HBT effect. The bin width in both cases is $\Delta=Q_s/2$.  \\
%Two particle correlations in $pp$ and $p$Pb collisions at the LHC measured by the ATLAS Collaboration~\cite{Aaboud:2016yar}, for different energies and particle multiplicities in the event.
Right panel: The three particle correlation function ${\cal O}_{ N,v_2}$ as a function of the bin width. 
}
\label{fig:1}       % Give a unique label
%\label{fig:2}       % Give a unique label
\end{figure}

Our numerical results for the correlation function between $v^2_2$ and the total multiplicity  are presented in Fig.~\ref{fig:1}.   
We first take coinciding bins, that is $k=k^\prime$  and the bin  width $\Delta=Q_s/2$. In this kinematics $v_2^2$  is dominated by HBT.
The result is the solid (blue) curve  in Fig.~\ref{fig:1}.  The dashed curve in  Fig.~\ref{fig:1} displays the situation when the momenta are offset by $Q_s$, that is $k^\prime=k+Q_s$. 
This choice eliminates the HBT contribution to the azimuthal anisotropy $v^2_2$.
Fig.~\ref{fig:1} shows that the normalized correlation function is strongly suppressed for values of bin width for which $v^2_2$ is sizable, which is when the HBT effect in $v^2_2$ is dominant. 
%we integrated over the momentum bin with the width $Q_s/2$ centred at the specified momentum. 
%Note that we quote the results in dimensionless units and hence do no need to specify $Q_s$. 
%The plotted ratios are also independent of the projectile scale $\mu$, neither on the area $S_\perp$. 

The same effect is also demonstrated in Fig.~\ref{fig:1}, where we show the correlation function as a function of the bin width $\Delta$.  
For illustration, we chose the centers of the bins at $k= 4.5 Q_s$ and $k^\prime=5 Q_s$. 
When $\Delta/Q_s$ is small, the bins are not overlapping and no HBT contribution is present in $v^2_2$. At these values of bin width the correlation between $v^2_2$ and multiplicity is sizable. However for $\Delta>\frac{1}{2}Q_s=|k-k^\prime|$ there is a steep decrease of the correlation and it very sharply drops to negligible values.

\begin{figure}
\resizebox{0.49\textwidth}{!}{%
  \includegraphics{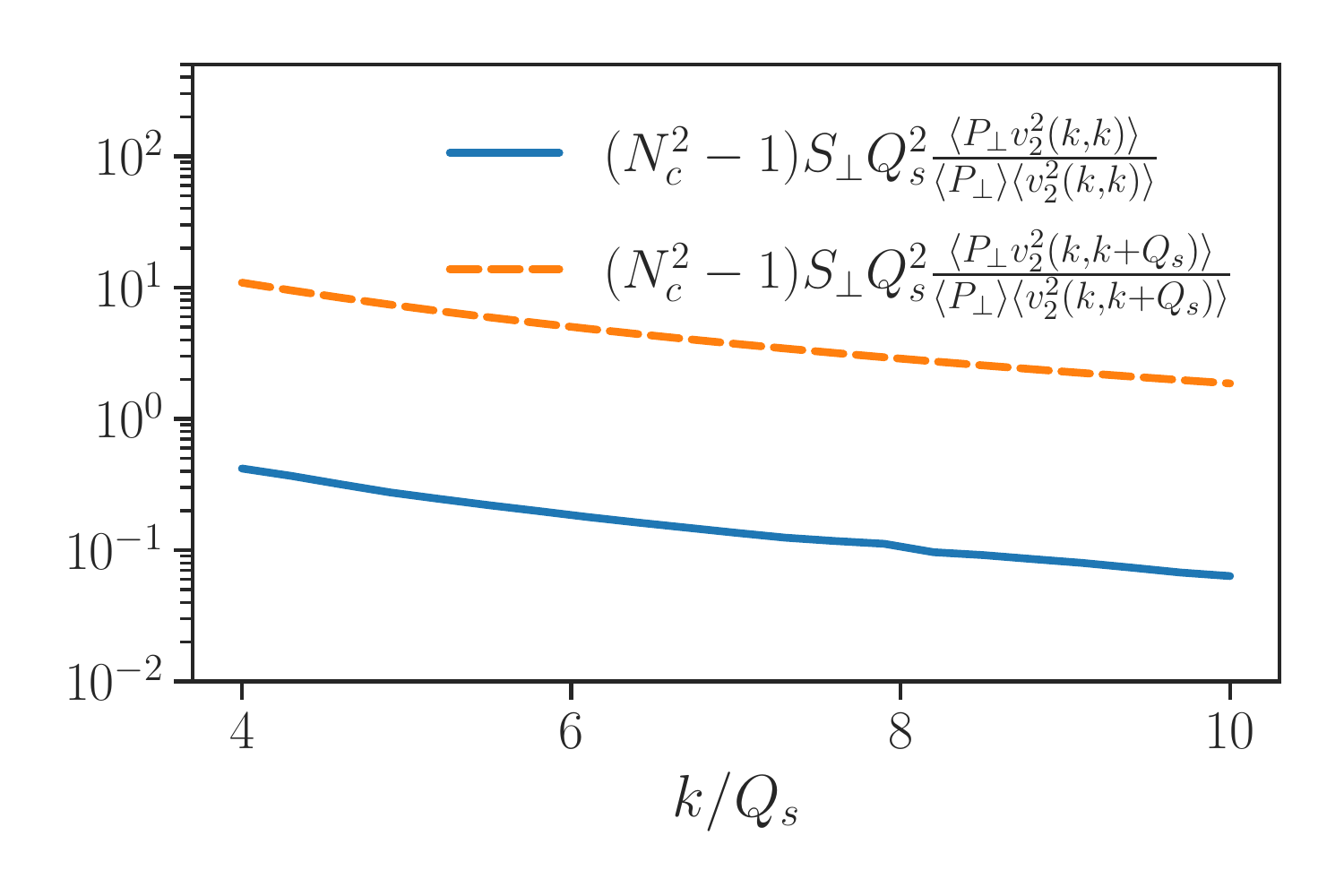}
}
\resizebox{0.49\textwidth}{!}{%
  \includegraphics{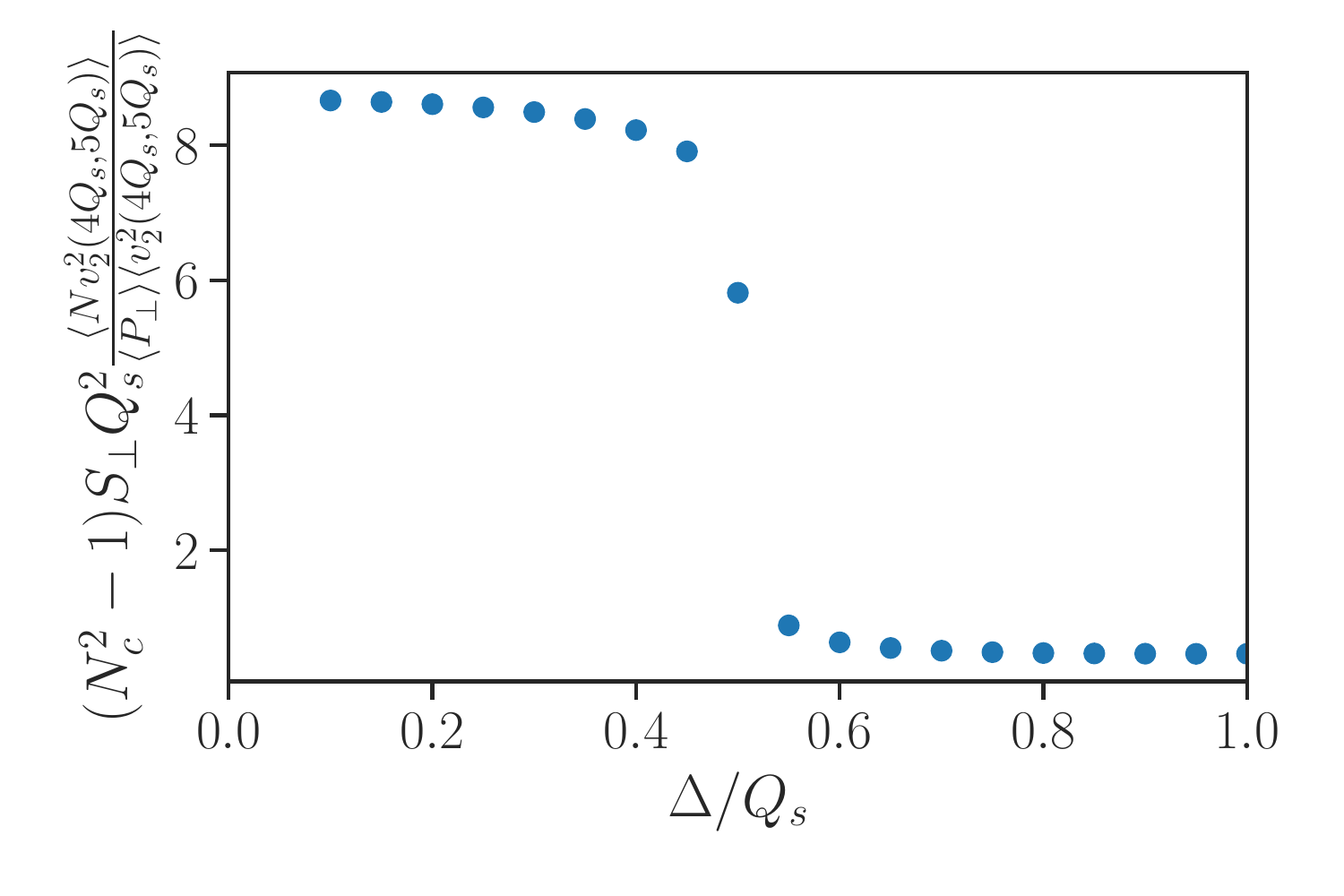}
}
\caption{Left  panel: The three particle correlation function ${\cal O}_{ k,v_2}$. Kinematics is the same as in Fig. \ref{fig:1}.
%Two particle correlations in $pp$ and $p$Pb collisions at the LHC measured by the ATLAS Collaboration~\cite{Aaboud:2016yar}, for different energies and particle multiplicities in the event.
\\ 
Right panel: The three particle correlation ${\cal O}_{ k,v_2}$ as a function of bin width.
}
\label{fig:3}       % Give a unique label
%\label{fig:4}       % Give a unique label
\end{figure}

We observe a similar behavior for the correlation of $v^2_2$ with transverse momentum. Fig.~\ref{fig:3} shows this correlation as a function of transverse momentum and  the same quantity as a function of the bin width.

\begin{figure}
\resizebox{0.5\textwidth}{!}{%
  \includegraphics{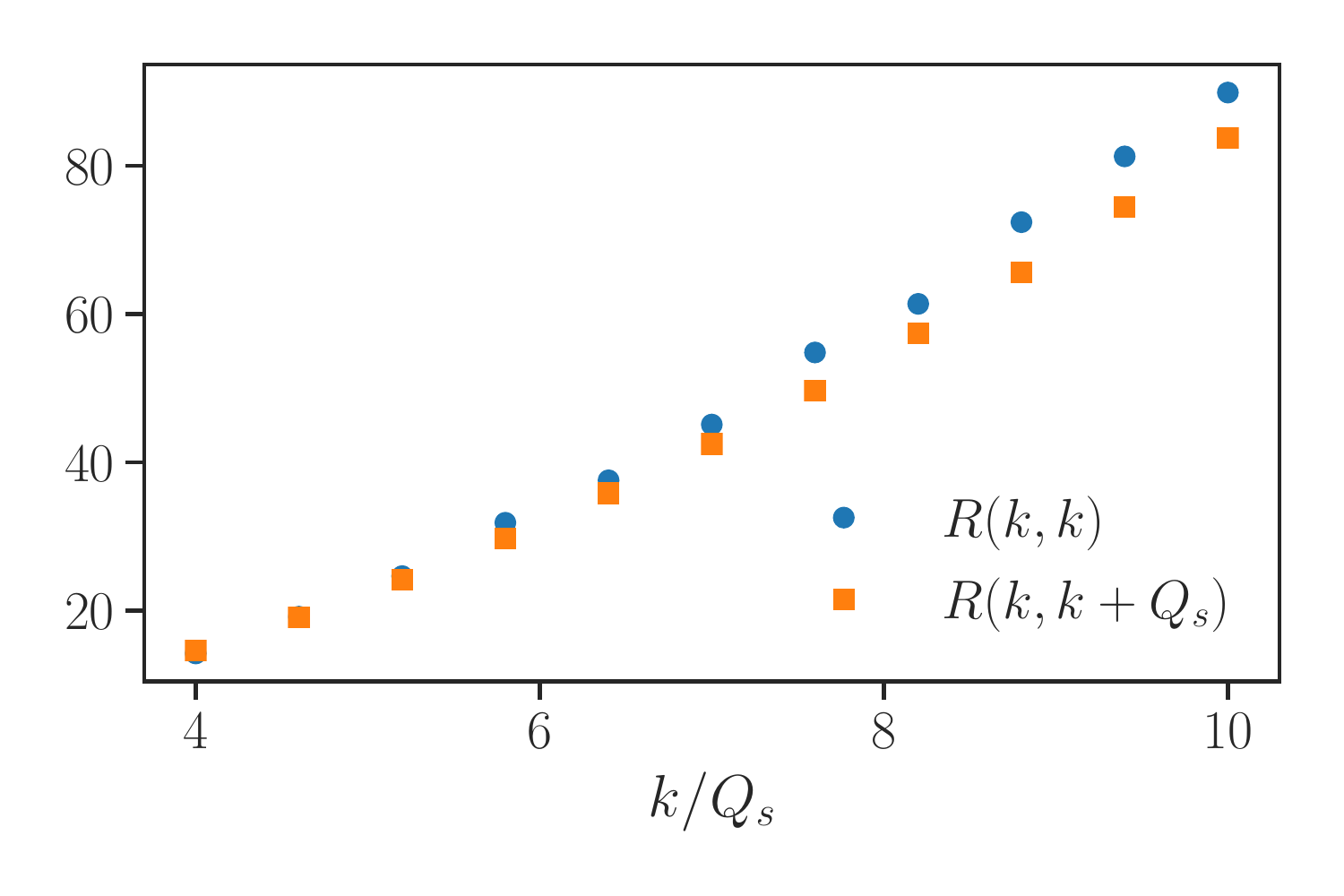}
}
\caption{The ratio $R\equiv {\cal O}_{ k,v_2}/{\cal O}_{ N,v_2}$ as a function of transverse momentum.
%Two particle correlations in $pp$ and $p$Pb collisions at the LHC measured by the ATLAS Collaboration~\cite{Aaboud:2016yar}, for different energies and particle multiplicities in the event.
}
\label{fig:5}       % Give a unique label
\end{figure}

Finally, Fig.~\ref{fig:5} shows the ratio $R\equiv {\cal O}_{ k,v_2}/{\cal O}_{ N,v_2}$ as a function of transverse momentum. The correlation with transverse momentum clearly drops with $k$ slower than the correlation with multiplicity.

 ~
~~
~

\section{Discussion}
\label{conclu}
Our results are quite curious. 

First, regarding  $v^2_2$ we find a very characteristic sharp transition in the value of $v^2_2$ as a function of the bin width. Referring to Fig.~\ref{fig:0} the value of $v_2$ rises sharply from $v_2\approx 2\times 10^{-3}$ to $v_2\approx 1.5\times 10^{-2}$, i.e. almost by an order of magnitude as the bin width is increased from $\Delta<|k-k'|$ to $\Delta>|k-k'|$. This transition is entirely due to the fact that at this value of bin width the HBT correlations start contributing to the second flow harmonic, since the HBT peak is much narrower than the Bose correlation. Although the presence of the transition is expected on these grounds, the fact that the value of $v_2$ rises by such a large amount is worth noting. We conclude that the contribution of the HBT correlations to $v_2$ completely overwhelms the contribution of Bose enhancement.
The actual numerical value for $v_2$ that we get is quite reasonable. For bins centered at $k=4.5 Q_s\approx 4.5$ GeV and $k'=5 Q_s\approx 5$ GeV and $\Delta=Q_s$  (right panel in Fig.~\ref{fig:0}) we find $v_2\approx 1.5\times 10^{-2}$. This is to be compared to typical values of $0.05\ -\ 0.08$ for transverse momentum integrated $v_2$ in p-Pb collisions at LHC~\cite{ridge}. Since the integrated $v_2$ is dominated by the lower momenta, this discrepancy of a factor of $5$ or so may be attributable to the relatively high value of transverse momentum in our calculation. The trend of $v_2$ rising towards low momenta is clearly seen in the left panel of Fig.~\ref{fig:0}. 

 We do reiterate though, that our calculation is not meant as a phenomenological fit in any way, but rather as a qualitative study of the effects of quantum statistics on the correlations.
As such, we note that the sharp rise in $v_2$ with bin width is a very characteristic behavior, and it would be very interesting to explore such dependence experimentally.

We wish to comment on two peculiar features seen on the left panel in Fig.~\ref{fig:0}. First, it is interesting to note that in the regime dominated by Bose correlations (dashed curve), $v^2_2$ is only very weakly dependent on the transverse momentum. Although it does slowly decrease towards large momenta, this decrease cannot be  discerned  for the range of momenta  on the plot. This suggests that the Bose correlated part of the two particle production scales with the same power of momentum as the square of the single particle spectrum. 

Another property to note is that the ratio of the HBT to Bose contributions as seen in the left panel of Fig.~\ref{fig:0} seems to be even greater than on the right panel. The ratio between the solid and dashed curves at $k\sim 4-5 Q_s$ on the left panel is around $\sim 500$, rather than the factor $\sim 50$ that we have inferred from the right panel. Admittedly,  it is not a completely fair comparison, as the solid line on the left panel corresponds to the two momenta sampled from the same bin, while on the right panel the centers of the two bins are displaced by $0.5Q_s$. Still it is a little surprising that a relatively small displacement of the bin centers leads to such a dramatic effect. The reason for this is our treatment of the projectile as translationally invariant, which leads to a delta-function  HBT correlation. In this situation the HBT contribution is essentially given by the overlap area of two rings corresponding to the two momentum bins. Displacing the centers of the bins relative to each other even by a small amount leads to a significant change of the overlap area, and thus the HBT contribution has a sharp peak at zero displacement. As we mentioned before, in a more realistic treatment which takes into account finite transverse size of the proton, the HBT peak should be smeared to have a width $\sim 0.2$ GeV (the inverse radius of the projectile) which should significantly soften the dependence on $|k-k'|$. We have checked numerically that smearing the HBT peak  does indeed have such an effect. For that reason we limit our consideration to kinematic situations where the distance between the bin centers is greater than $0.3- 0.4 Q_s$.

Moving on to the correlation of $v^2_2$ with multiplicity as well as with the transverse momentum, we again observe a very characteristic dependence on the bin width. For small bin width where the HBT does not contribute, the correlation is sizable,  $\sim 3\times 10^{-3}$. However for larger bin width this correlation drops by a factor of about $30$ to $50$, and is negligible. Note that the transition in ${\cal O}_{ N,v_2}$ and ${\cal O}_{ k,v_2}$ is the opposite to that in $v_2$: whereas $v_2$ is smaller at small $\Delta$, the correlations ${\cal O}_{ N,v_2}$ and ${\cal O}_{ k,v_2}$ are larger, and vice versa. This tells us that although the contribution of HBT to $v_2$ is much larger than that of the Bose enhancement, the HBT is much weaker correlated with total multiplicity (and transverse momentum) than the Bose enhancement is. In fact, since the magnitude of the drop in Fig.~\ref{fig:1} is about the same as the magnitude of the rise in Fig.~\ref{fig:0}, we conclude that the numerator in Eq. (\ref{ONv2}) is a rather smooth function of $\Delta$, and the drop in Fig.~\ref{fig:1} is driven entirely by the sharp rise in the denominator in Eq. (\ref{ONv2}) (and the same is true for Eq. \eqref{Okv2}).

The correlation between $v_2$ and $N$  is a decreasing function of momentum. This is easy to understand, since the multiplicity is dominated by soft particles, while the correlation we track originates with gluons that already in the incoming wave function have large transverse momentum.  We thus have no reason to expect a correlation at high values of $k$. Another aspect of this is seen in Fig.~\ref{fig:5} which shows that the correlation of $v_2$  with $k$ remains larger at high momentum than the correlation of $v_2$ with $N$, since particles with higher momentum contribute more significantly to average momentum than to the total multiplicity.

Qualitatively the smallness of the correlations between $v_2$ and $N$ is consistent with the experimental data \cite{ridge}. Experimentally the change in the integrated $v_2$ in p-Pb collisions is at most $40\%$ over an order of magnitude change in $N$. Again, we note that our calculation is valid only for large $k$. The correlation clearly grows towards smaller values of $k$, but within our approximation we cannot push much below $k\sim 4Q_s$.

To conclude, we have calculated $v_2$ and correlations of $v_2$ with the total multiplicity and average transverse momentum per particle in the dense-dilute CGC approach using the GBW model for dipole amplitude. Our results are valid at large $N_c$ and large transverse momentum. We have not made an attempt to include any additional effects beyond multiple scattering in our calculation. We find a reasonable magnitude for $v_2$ and very small correlations with total multiplicity, consistent with data. An interesting observation we make is a characteristic very strong crossover in $v_2(k,k',\Delta)$ as a function of the width of the transverse momentum bin $\Delta$ at fixed $k$ and $k'$, associated with the dominance of the HBT contribution. It would be very interesting to explore such a dependence experimentally.

\section*{Acknowledgements}

NA has received financial support from Xunta de Galicia (Centro singular de investigaci\'on de Galicia accreditation 2019-2022), by European Union ERDF, and by  the ``Mar\'{\i}a  de Maeztu" Units  of  Excellence program  MDM-2016-0692  and  the Spanish Research State Agency under project FPA2017-83814-P.
TA is supported by Grant No. 2018/31/D/ST2/00666 (SONA\-TA 14 - National Science Centre, Poland). 
AK is supported by the NSF Nuclear Theory grants 1614640 and 1913890.
ML was supported by the Israeli Science Foundation (ISF) grant \#1635/16.
ML and AK were also supported by the Binational Science Foundation grant  \#2018722.
VS acknowledges support by the DOE Office of Nuclear Physics through Grant No. DE-SC0020081. VS thanks the ExtreMe Matter Institute EMMI (GSI Helmholtzzentrum f\"ur Schwerionenforschung, Darmstadt, Germany) for partial support and hospitality.
This work has been performed in the framework of COST Action CA 15213 ``Theory of hot matter and relativistic heavy-ion collisions" (THOR), MSCA RISE 823947 ``Heavy ion collisions: collectivity and precision in saturation physics''  (HI\-EIC) and has received funding from the European Un\-ion's Horizon 2020 research and innovation programm under grant agreement No. 824093.

\section{Appendix}

\subsection{Computation of  $X_i$ for the correlations between multiplicity and $v_2$ }

\subsubsection{Computation of $X_1$}
Consider $X_1$ first.
The $k_1$ integral is facilitated by  Eq. \eqref{mu2_GBW}: 
%As the first step let us integrate fully correlated piece of the three gluon spectrum given in Eq. \eqref{uninteg_3g_N} over $k_1$. To perform the computation we will treat the function $\mu^2$ within the MV model, ie
%%
%\beq
%\mu^2(k,p)\equiv \mu^2\,  (2\pi)^2\, \delta^{(2)}(k+p)\, 
%\eeq
%%
%we will use GBW model for the dipole operators
%%
%\beq
%d(p)\equiv \frac{4\pi}{Q_s^2}e^{-p^2/Q_s^2}\,  
%\eeq
%%
%and the Lipatov vertex reads 
%%
%\beq
%L^i(k,q)=\bigg[ \frac{(k-q)^i}{(k-q)^2}-\frac{k^i}{k^2}\bigg]
%\eeq
%
%
%In order to understand how the computation works, let us take a closer look at the $I_{X,1}$ term: 
%
\beq
%\frac{d\bar N}{d^2k_2\, d^2k_3}\bigg|_{X,1}\equiv 
\int d^2k_1 \frac{dN^{(3)}}{d^2k_1\, d^2k_2\, d^2k_3}\bigg|_{X,1}
=\big[(X_{1,a}+X_{1,b})+(k_3\to-k_3)\big]+\big[ (X_{1,c}+X_{1,d})+(k_1\to-k_1)\big],
\eeq
where $X_{1,a}$ is the first term in Eq. \eqref{X1}, $X_{1,b}$ is the second term in Eq. \eqref{X1}, $X_{1,c}$ is the first term in Eq. \eqref{X1prime} and $X_{1,d}$ is the second term in Eq. \eqref{X1prime}, all integrated over $k_1$.
The explicit expression of $X_{1,a}$ reads 
\beq
X_{1,a}&=&\alpha_s^3(4\pi)^3(N_c^2-1)\, \mu^6 \, (2\pi)^6 \int d^2k_1\, \frac{d^2q_1}{(2\pi)^2} \, \frac{d^2q_2}{(2\pi)^2} \, \frac{d^2q_3}{(2\pi)^2} \, d(q_1)\, d(q_2)\, d(q_3)\nonumber\\
&\times&
\delta^{(2)}(k_1-k_2)\, \delta^{(2)}(k_1-q_1+q_3-k_3)\, \delta^{(2)}(k_3-q_3+q_1-k_2)\nonumber\\
&\times&
L^i(k_1,q_1)L^i(k_1,q_2) \, L^j(k_2,q_2)L^j(k_2,q_1)\, L^k(k_3,q_3)L^k(k_3,q_3).
\eeq
We use the first $\delta$-function to integrate over $k_1$, the second $\delta$-function to integrate over $q_1$, and the third $\delta$-function becomes $\delta(0)$ which, as usual,
is regulated by the transverse area $S_\perp$ of the projectile.
The result reads 
\beq
X_{1,a}&=&\alpha_s^3(4\pi)^3(N_c^2-1)\, \mu^6 \, (2\pi)^2\, S_\perp \int  \frac{d^2q_2}{(2\pi)^2} \, \frac{d^2q_3}{(2\pi)^2} \, d\big[ k_2-(k_3-q_3)\big] \, d(q_2)\, d(q_3)\, 
\bigg[ \frac{(k_3-q_3)^i}{(k_3-q_3)^2}-\frac{k_2^i}{k_2^2}\bigg] 
\nonumber\\
&\times&
\bigg[ \frac{(k_2-q_2)^i}{(k_2-q_2)^2}-\frac{k_2^i}{k_2^2}\bigg]
\bigg[ \frac{(k_2-q_2)^j}{(k_2-q_2)^2}-\frac{k_2^j}{k_2^2}\bigg] \bigg[ \frac{(k_3-q_3)^j}{(k_3-q_3)^2}-\frac{k_2^j}{k^2_2}\bigg]\; 
\bigg[ \frac{(k_3-q_3)^k}{(k_3-q_3)^2}-\frac{k_3^k}{k_3^2}\bigg] \bigg[ \frac{(k_3-q_3)^k}{(k_3-q_3)^2}-\frac{k_3^k}{k_3^2}\bigg].
\eeq
Note that the dipole $d\big[ k_2-(k_3-q_3)\big]$ is enhanced when its argument is small. 
Hence $X_{1,a}$ is a contribution to the forward ($k_2\simeq k_3$) correlation of  gluons $2$ and $3$. \\

The rest of the terms in $X_1$ can be computed in a similar way: 
\beq
X_{1,b}&=&\alpha_s^3(4\pi)^3(N_c^2-1)\, \mu^6 \, (2\pi)^2\, S_\perp \int  \frac{d^2q_2}{(2\pi)^2} \, \frac{d^2q_3}{(2\pi)^2} \, 
d\big[ k_2+(k_3-q_3)\big] \, d(q_2)\, d(q_3)\, 
\bigg[ \frac{(k_3-q_3)^i}{(k_3-q_3)^2}+\frac{k_2^i}{k_2^2}\bigg] 
\nonumber\\
&\times&
\bigg[ \frac{(k_2+q_2)^i}{(k_2+q_2)^2}-\frac{k_2^i}{k_2^2}\bigg]
\bigg[ \frac{(k_3-q_3)^j}{(k_2-q_2)^2}+\frac{k_2^j}{k_2^2}\bigg] \bigg[ \frac{(k_2+q_2)^j}{(k_2+q_2)^2}-\frac{k_2^j}{k^2_2}\bigg]\; 
\bigg[ \frac{(k_3-q_3)^k}{(k_3-q_3)^2}-\frac{k_3^k}{k_3^2}\bigg] \bigg[ \frac{(k_3-q_3)^k}{(k_3-q_3)^2}-\frac{k_3^k}{k_3^2}\bigg].
\eeq
$X_{1,b}$  is a contribution to the backward correlation of  gluons $2$ and $3$. By renaming $q_2\to-q_2$ and $q_3\to-q_3$, it is easy to see that  $X_{1,b}=X_{1,a}(k_3\to-k_3)$. 
\beq
X_{1,c}&=&\alpha_s^3(4\pi)^3(N_c^2-1)\, \mu^6 \, (2\pi)^2\, S_\perp \int  \frac{d^2q_2}{(2\pi)^2} \, \frac{d^2q_3}{(2\pi)^2} \, 
d\big[ k_2+(k_3-q_3)\big] \, d(q_2)\, d(q_3) \, 
\bigg[ \frac{(k_3-q_3)^i}{(k_3-q_3)^2}+\frac{k_2^i}{k_2^2}\bigg]  \nonumber\\
&\times&
\bigg[ \frac{(k_2-q_2)^i}{(k_2-q_2)^2}-\frac{k_2^i}{k_2^2}\bigg]
\bigg[ \frac{(k_3-q_3)^j}{(k_3-q_3)^2}+\frac{k_2^j}{k_2^2}\bigg] \bigg[ \frac{(k_2-q_2)^j}{(k_2-q_2)^2}-\frac{k_2^j}{k_2^2}\bigg] \; 
\bigg[ \frac{(k_3-q_3)^k}{(k_3-q_3)^2}-\frac{k_3^k}{k_3^2}\bigg] \bigg[ \frac{(k_3-q_3)^k}{(k_3-q_3)^2}-\frac{k_3^k}{k_3^2}\bigg].
\eeq
$X_{1,c}$  is a contribution to the backward correlation of  gluons $2$ and $3$. By renaming $q_3\to-q_3$, it is easy to show that $X_{1,c}=X_{1,a}(k_3\to-k_3)$. 
\beq
X_{1,d}&=&\alpha_s^3(4\pi)^3(N_c^2-1)\, \mu^6 \, (2\pi)^2\, S_\perp \int  \frac{d^2q_2}{(2\pi)^2} \, \frac{d^2q_3}{(2\pi)^2} \, 
d\big[ k_2-(k_3-q_3)\big] \, d(q_2)\, d(q_3) \, 
\bigg[ \frac{(k_3-q_3)^i}{(k_3-q_3)^2}-\frac{k_2^i}{k_2^2}\bigg]  \nonumber\\
&\times&
\bigg[ \frac{(k_2-q_2)^i}{(k_2-q_2)^2}-\frac{k_2^i}{k_2^2}\bigg]
\bigg[ \frac{(k_3-q_3)^j}{(k_3-q_3)^2}-\frac{k_2^j}{k_2^2}\bigg] \bigg[ \frac{(k_2-q_2)^j}{(k_2-q_2)^2}-\frac{k_2^j}{k_2^2}\bigg] \; 
\bigg[ \frac{(k_3-q_3)^k}{(k_3-q_3)^2}-\frac{k_3^k}{k_3^2}\bigg] \bigg[ \frac{(k_3-q_3)^k}{(k_3-q_3)^2}-\frac{k_3^k}{k_3^2}\bigg].
\eeq
$X_{1,d}$  is a contribution to the forward correlation of gluons $2$ and $3$. $X_{1,d}=X_{1,a}$. 

Combining all the terms, we get
\beq
&&
\hspace{-0.8cm}
X_{1}= 4\, \alpha_s^3(4\pi)^3(N_c^2-1)\, \mu^6 \, (2\pi)^2\, S_\perp \int  \frac{d^2q_2}{(2\pi)^2} \, \frac{d^2q_3}{(2\pi)^2} \bigg\{ 
d\big[ k_2-(k_3-q_3)\big] \, d(q_2)\, d(q_3) \, 
%\nonumber\\
%&\times&
\bigg[ \frac{(k_3-q_3)^i}{(k_3-q_3)^2}-\frac{k_2^i}{k_2^2}\bigg] 
\\
&\times&
\bigg[ \frac{(k_2-q_2)^i}{(k_2-q_2)^2}-\frac{k_2^i}{k_2^2}\bigg] 
\bigg[ \frac{(k_3-q_3)^j}{(k_3-q_3)^2}-\frac{k_2^j}{k_2^2}\bigg] \bigg[ \frac{(k_2-q_2)^j}{(k_2-q_2)^2}-\frac{k_2^j}{k_2^2}\bigg] \; 
\bigg[ \frac{(k_3-q_3)^k}{(k_3-q_3)^2}-\frac{k_3^k}{k_3^2}\bigg] \bigg[ \frac{(k_3-q_3)^k}{(k_3-q_3)^2}-\frac{k_3^k}{k_3^2}\bigg] +(k_3\to-k_3)\bigg\}. \nonumber
\eeq
In $X_{1}$, the first term is a contribution to the forward correlation of the gluons 2 and 3. The second term with $(k_3\to-k_3)$
is a contribution to the backward correlation of the gluons 2 and 3. Shifting and renaming the variables  the expression can be brought to a more compact form:
\beq
\label{X1_shifted}
X_1&\equiv& 4\, \alpha_s^3(4\pi)^3(N_c^2-1)\, \mu^6(2\pi)^2\, S_\perp\, 
\int \frac{d^2q_2}{(2\pi)^2}\frac{d^2q_3}{(2\pi)^2}\, \bigg\{
d(q_2+k_2) \, d(q_3+k_2) \, d(q_3+k_3)\\
&&
\hspace{2.5cm}
\times
\bigg[ \frac{q^i_2}{q_2^2}+\frac{k^i_2}{k_2^2}\bigg] \bigg[ \frac{q^j_2}{q_2^2}+\frac{k^j_2}{k_2^2}\bigg]
\bigg[ \frac{q^i_3}{q_3^2}+\frac{k^i_2}{k_2^2}\bigg] \bigg[ \frac{q^j_3}{q_3^2}+\frac{k^j_2}{k_2^2}\bigg]
\bigg[ \frac{q^k_3}{q_3^2}+\frac{k^k_3}{k_3^2}\bigg] \bigg[ \frac{q^k_3}{q_3^2}+\frac{k^k_3}{k_3^2}\bigg] +(k_3\to-k_3)\bigg\}. \nonumber
\eeq
%
%
%
%Alternative way to continue computation is to adopt: 
%\beq
%\mu^2(k,q)= e^{-\frac{(k+q)^2}{4B^{-1}}}
%\eeq 
% with $B$ being the gluonic area of the projectile and 
% \beq
% L^i(k,q_1)L^i(k,q_2)\to\frac{(2\pi)^2}{4\mu^2}\exp\bigg\{ -\frac{1}{16\mu^2}\big[ (k-q_1)+(k-q_2)\big]^2\bigg\}
% \eeq 
%%
%with $\mu^2=B^{-1}$ we get the Wigner function description used by Raju et al + Cyrille et al. We can also take $4\mu^2=Q_s^2$. 
%\subsection{Computation of $X_1$ and $X_3$}
%
Using the GBW model (\ref{gbw}) for the dipoles $d$,  the $X_1$ contribution can be written as 
\beq
X_1&=&4\, \alpha_s^3(4\pi)^3(N_c^2-1)\, \mu^6\, (2\pi)^2\, S_{\perp}\, \frac{1}{(2\pi)^4}\, \bigg(\frac{4\pi}{Q_s^2}\bigg)^3 \, e^{-(2k_2^2+k_3^2)/Q_s^2} \nonumber\\
&\times&
\int d^2q_2\, e^{-(q_2^2+2\, q_2\cdot k_2)/Q_s^2}
\bigg[ \frac{k_2^i}{k_2^2}\frac{k_2^j}{k_2^2}+\frac{k_2^i}{k_2^2}\frac{q_2^j}{q_2^2}+\frac{q_2^i}{q_2^2}\frac{k_2^j}{k_2^2}+\frac{q_2^i}{q_2^2}\frac{q_2^j}{q_2^2}\bigg]\nonumber\\
&\times&
\int d^2q_3\, e^{-[2q_3^2+2q_3\cdot(k_2+k_3)]/Q_s^2}
\bigg[ \frac{k_2^i}{k_2^2}\frac{k_2^j}{k_2^2}+\frac{k_2^i}{k_2^2}\frac{q_3^j}{q_3^2}+\frac{q_3^i}{q_3^2}\frac{k_2^j}{k_2^2}+\frac{q_3^i}{q_3^2}\frac{q_3^j}{q_3^2}\bigg]
\bigg[ \frac{1}{q_3^2}+2\, \frac{q_3^k}{q_3^2}\frac{k_3^k}{k_3^2}+\frac{1}{k_3^2}\bigg].
\eeq
Note that $X_1 \propto e^{-(2k_2^2+k_3^2)/Q_s^2}$ which provides an exponential suppression for produced gluon momenta much larger than the target saturation scale $Q_s$. However we still have  the $q_2$ and $q_3$ integrations to perform, and those may bring a compensating exponential enhancement.  Next, we will perform these integrations concentrating on such possible enhancing factors.
Let us first focus on the $q_2$ integration (the $q_3$ integration is done similarly and will follow). There are three types of Gaussian integrals to consider: 
\beq\label{3int}
I_{0,0}&=&\int d^2q_2 \, e^{-(q_2^2+2\, q_2\cdot k_2)/Q_s^2}\ ,\\
I^i_{1,2}&=& \int d^2q_2 \, e^{-(q_2^2+2\, q_2\cdot k_2)/Q_s^2} \, \frac{q_2^i}{q_2^2}\ ,\nonumber\\
I_{2,4}^{ij}&=&\int d^2q_2 \, e^{-(q_2^2+2\, q_2\cdot k_2)/Q_s^2} \, \frac{q_2^i}{q_2^2}\frac{q_2^j}{q_2^2}\ . \nonumber
\eeq
The first integral is trivial,
% to perform. After completing the square and renaming the integration variable, one gets a Gaussian integral and the result reads 
%
\beq
\label{1_Type_0}
I_{0,0}=\pi Q_s^2\;  e^{k_2^2/Q_s^2}.
\eeq
The second and the third integrals are performed with the help of  a Schwinger parameter $t$ introduced for the $q_2^2$ factor in the denominator:
%Then, the second type of integral can be written as 
\beq
I_{1,2}^i&=&\int_0^{+\infty} dt \exp\bigg[ \frac{k_2^2}{Q_s^2(1+Q_s^2t)}\bigg]\int d^2q \exp\bigg[ -\frac{(1+Q_s^2t)}{Q_s^2}q^2\bigg] \bigg[ q^i-\frac{k_2^i}{(1+Q_s^2t)}\bigg]\nonumber\\
&=&-\pi Q_s^2\, k_2^i \int_0^{+\infty} dt \frac{1}{(1+Q_s^2t)^2}\, \exp\bigg[\frac{k_2^2}{Q_s^2(1+Q_s^2t)}\bigg].
\eeq 
Changing variables to  $t'=1/(1+Q_s^2t)$, the result reads 
\beq
\label{1_Type_1}
I_{1,2}^i&=&-\pi Q_s^2\, k_2^i\frac{1}{Q_s^2}\int_0^1 dt' \, e^{k_2^2t'/Q_s^2}=\pi Q_s^2\, \frac{k_2^i}{k_2^2}\Big(1-e^{k_2^2/Q_s^2}\Big).
\eeq
For the last  integration, $I_{2,4}$, a similar procedure is followed, but with two  Schwinger parameters $t_1$ and $t_2$ for each $1/q_2^2$ factor. This makes it possible to perform the integration over $q_2$ and one of the Schwinger parameters. However, the remaining integral over the second Schwinger parameter is divergent, reflecting the original
IR divergence of the $q_2$ integration.  The origin of this IR divergence is pretty clear - it is an artefact of the approximation  (\ref{mu2_GBW}) with momentum independent $\mu$.
In fact, color neutrality should be imposed at some non-perturbative IR scale $\Lambda_{min}$ below which $\mu$ must vanish, $\mu(p<\Lambda_{min})=0$.  
Roughly  $\Lambda^2_{min}\sim 1/S_\perp$. We find it more convenient introducing a  cutoff $\lambda\rightarrow 0$ on the Schwinger parameter than the sharp IR cutoff $\Lambda_{min}$ to 
regulate this divergence. The relation between the two is
\beq
\lambda\simeq \Lambda^2_{min}/Q_s^2\simeq 1/(S_\perp Q_s^2).
\eeq
The final result reads
\beq
\label{1_Type_2}
I_{2,4}^{ij}&=&\pi Q_s^2\, \bigg\{ \frac{\delta^{ij}}{2}\bigg\lgroup \frac{1}{Q_s^2}+\frac{1}{k_2^2}\Big(1-e^{k^2_2/Q_s^2}\Big)+\frac{1}{Q_s^2}\bigg[ {\rm Ei}\bigg(\frac{k_2^2}{Q_s^2}\bigg)-{\rm Ei}\bigg(\frac{k_2^2\, \lambda}{Q_s^2}\bigg)\bigg]\bigg\rgroup
-\frac{k_2^i}{k_2^2}\frac{k_2^j}{k_2^2}
\bigg\lgroup \frac{k_2^2}{Q_s^2}+
\Big(1-e^{k_2^2/Q_s^2}\Big)\bigg\rgroup\bigg\},
\eeq
where ${\rm Ei}$ is an exponential integral special function. Combining the results given in Eqs. (\ref{1_Type_0}), (\ref{1_Type_1}) and (\ref{1_Type_2}), 
the overall result of the $q_2$ integration reads 
\beq
\label{q_2_final_result}
&&
\int d^2q_2\, e^{-(q_2^2+2\, q_2\cdot k_2)/Q_s^2}
\bigg[ \frac{k_2^i}{k_2^2}\frac{k_2^j}{k_2^2}+\frac{k_2^i}{k_2^2}\frac{q_2^j}{q_2^2}+\frac{q_2^i}{q_2^2}\frac{k_2^j}{k_2^2}+\frac{q_2^i}{q_2^2}\frac{q_2^j}{q_2^2}\bigg]
\nonumber\\
&&
\hspace{2cm}
=\pi Q_s^2\, \bigg\{\frac{k_2^i}{k_2^2}\frac{k_2^j}{k_2^2}\bigg(1-\frac{k_2^2}{Q_s^2}\bigg)
+\frac{\delta^{ij}}{2}\frac{1}{k_2^2}\bigg\lgroup\frac{k_2^2}{Q_s^2}+\Big(1-e^{k_2^2/Q_s^2}\Big)+\frac{k_2^2}{Q_s^2}\bigg[ {\rm Ei}\bigg(\frac{k_2^2}{Q_s^2}\bigg)- {\rm Ei}\bigg(\frac{k_2^2 \, \lambda}{Q_s^2}\bigg)\bigg]\bigg\rgroup\bigg\}.
\eeq
So far the integration over $q_2$ was computed without any approximation. Yet, as we have mentioned above, we are interested only in terms in $X_1$ 
that are not exponentially suppressed.  In other words, only exponentially enhanced terms in (\ref{q_2_final_result}) are of interest. The $\lambda$-dependent terms are not of that type
and can be neglected.
%we are interested in computing the terms without exponential suppression in the final result after integration over $q_3$.  In this case, we can neglect the exponential integral function that depends on the cutoff $\lambda$. 
For $k_2^2\gg Q_s^2$, the result can be simplified using the large argument asymptotic  expansion of the exponential integral  function, 
\beq
{\rm Ei}(x)\approx \frac{e^x}{x}\bigg( 1+\frac{1!}{x}+\frac{2!}{x^2}+\frac{3!}{x^3}+ \cdots\bigg),  \ \  x\rightarrow \infty.
\eeq
 %Then, we can write the exponentially enhanced contribution of each the three types of integrals as
%\beq
%\label{exp_I_00}
%I_{0,0}&=& \pi Q_s^2\;  e^{k_2^2/Q_s^2}\\
%\label{exp_I_12}
%I^i_{1,2}&\sim& -\pi Q_s^2\; e^{k_2^2/Q_s^2} \: \frac{k_2^i}{k_2^2} \\
%I^{ij}_{2,4}&\sim& \pi Q_s^2 \; e^{k_2^2/Q_s^2} \bigg[ \frac{\delta^{ij}}{2}\frac{Q_s^2}{k_2^4}+\frac{k_2^i}{k_2^2}\frac{k_2^j}{k_2^2}\bigg]
%\eeq 
Thus the exponentially enhanced contribution to the total $q_2$ integral reads 
\beq
\label{Exp_Type_2}
\int d^2q_2\, e^{-(q_2^2+2\, q_2\cdot k_2)/Q_s^2}
\bigg[ \frac{k_2^i}{k_2^2}\frac{k_2^j}{k_2^2}+\frac{k_2^i}{k_2^2}\frac{q_2^j}{q_2^2}+\frac{q_2^i}{q_2^2}\frac{k_2^j}{k_2^2}+\frac{q_2^i}{q_2^2}\frac{q_2^j}{q_2^2}\bigg]\approx \pi Q_s^2 \;  e^{k_2^2/Q_s^2}\; \frac{\delta^{ij}}{2}\frac{Q_s^2}{k_2^4}\; \bigg(1+\frac{2!\,Q_s^2}{k_2^2}+\frac{3!\, Q_s^4}{k_2^4} \bigg) .
\eeq 
We have kept sub-leading terms of the order $Q_s^4/k_2^4$ since, as we will see later, those will turn out to be the first non-vanishing terms when $k_2=k_3$.
%Note that this is the leading term in the expansion of the special function. We know that the first non-vanishing term in the $k_2=k_3$ limit is of order $Q_s^4$ for $X_1$ contribution. Therefore, we keep the terms of $Q_s^2$ and $Q_s^4$ in the expansion, which allows us to organize the 
After the $q_2$ integration has been evaluated, $X_1$  still contains a $q_3$ integral, which is our next target:
\beq
\label{X_1_before_q3}
X_1&\approx& \alpha_s^3(4\pi)^3(N_c^2-1)\; \mu^6\; S_\perp\; \frac{1}{(2\pi)}\,  \frac{(4\pi)^3}{Q_s^2}\; e^{-(k_2^2+k_3^2)/Q_s^2}\, \frac{1}{k_2^4}\; \bigg(1+\frac{2!\,Q_s^2}{k_2^2}+\frac{3!\, Q_s^4}{k_2^4} \bigg) \nonumber\\
&\times&
\int d^2q_3\, e^{-[2q_3^2+2q_3\cdot(k_2+k_3)]/Q_s^2}
\bigg[ \frac{1}{k_2^2}+2\frac{k_2^i}{k_2^2}\frac{q_3^i}{q_3^2}+\frac{1}{q_3^2}\bigg]
\bigg[ \frac{1}{q_3^2}+2\, \frac{q_3^j}{q_3^2}\frac{k_3^j}{k_3^2}+\frac{1}{k_3^2}\bigg] \\
&\approx& \alpha_s^3(4\pi)^3(N_c^2-1)\; \mu^6\; S_\perp\; \frac{1}{(2\pi)}\,  \frac{(4\pi)^3}{Q_s^2}\; e^{-(k_2^2+k_3^2)/Q_s^2}\, \frac{1}{k_2^4}\; \bigg(1+\frac{2!\, Q_s^2}{k_2^2}+\frac{3!\, Q_s^4}{k_2^4} \bigg) \nonumber\\
% c_0\; \mu^6\; S_\perp\; \frac{(4\pi)^2}{Q_s^2}\; \frac{2}{k_2^4} \; e^{-(k_2^2+k_3^2)/Q_s^2}\; \nonumber\\
&\times& \; 
\bigg\{ \frac{1}{k_2^2k_3^2}T_{0,0}+\bigg(\frac{1}{k_2^2}+\frac{1}{k_3^2}\bigg)T_{0,2}+T_{0,4}+\frac{2(k_2+k_3)^i}{k_2^2k_3^2}T^i_{1,2}+2\bigg(\frac{k_2^i}{k_2^2}+\frac{k_3^i}{k_3^2}\bigg)T^i_{1,4}+4\frac{k_2^i}{k_2^2}\frac{k_3^i}{k_3^2}T^{ij}_{2,4} \bigg\}, \nonumber
\eeq
where 
\beq
T_{0,0}&=&\int d^2q_3\, e^{-[2q_3^2+2q_3\cdot(k_2+k_3)]/Q_s^2}, \\
\label{T02_def}
T_{0,2}&=& \int d^2q_3\, e^{-[2q_3^2+2q_3\cdot(k_2+k_3)]/Q_s^2} \frac{1}{q_3^2}\ ,\\
\label{T_04_def}
T_{0,4}&=& \int d^2q_3\, e^{-[2q_3^2+2q_3\cdot(k_2+k_3)]/Q_s^2} \frac{1}{q_3^4}\ ,\\
T^i_{1,2}&=& \int d^2q_3\, e^{-[2q_3^2+2q_3\cdot(k_2+k_3)]/Q_s^2} \frac{q_3^i}{q_3^2}\ , \\
\label{T14_def}
T^i_{1,4}&=& \int d^2q_3\, e^{-[2q_3^2+2q_3\cdot(k_2+k_3)]/Q_s^2} \frac{q_3^i}{q_3^4} \ ,\\
T^{ij}_{2,4}&=& \int d^2q_3\, e^{-[2q_3^2+2q_3\cdot(k_2+k_3)]/Q_s^2} \frac{q_3^iq_3^j}{q_3^4}\ .
\eeq
%
% T_04_def, T_04_soln
%
The structure of these integrals over $q_3$ are very similar to the ones encountered in the $q_2$ integrations (\ref{3int}). Therefore, the integration over $q_3$ can be performed in a similar manner. Again, keeping only the exponentially enhanced terms, the results read: 
%can be performed in the same way as before. 
%So, we can immediately present the final results, again keeping the exponentially enhanced terms only:
\beq
T_{0,0}&=&\pi Q_s^2\; e^{(k_2+k_3)^2/2Q_s^2} \; \frac{1}{2}\ ,\\
\label{T02_soln}
T_{0,2}&\approx&\pi Q_s^2\; e^{(k_2+k_3)^2/2Q_s^2} \; \frac{2^2}{(k_2+k_3)^2}\bigg[ \frac{1}{2}+\frac{1}{4}\frac{2^2\, Q_s^2}{(k_2+k_3)^2}+\frac{1}{4}\frac{2^4\, Q_s^4}{(k_2+k_3)^4}\bigg],\\
%\frac{2}{(k_2+k_3)^2}+\frac{4\, Q_s^2}{(k_2+k_3)^4}+\frac{16\, Q_s^4}{(k_2+k_3)^6}\bigg]\\
\label{T_04_soln}
T_{0,4}&\approx& \pi Q_s^2\; e^{(k_2+k_3)^2/2Q_s^2} \; \frac{2^4}{(k_2+k_3)^4} \, \bigg[ \frac{1}{2}+\frac{2^2\, Q_s^2}{(k_2+k_3)^2}+\frac{9}{4}\frac{2^4\, Q_s^4}{(k_2+k_3)^4}\bigg],\\
%\bigg[ \frac{2^3}{(k_2+k_3)^4} +\frac{2^6\, Q_s^2}{(k_2+k_3)^6}+9\frac{ 2^6\, Q_s^4}{(k_2+k_3)^8}\bigg]\\
T^i_{1,2}&\approx&-\pi Q_s^2\:  e^{(k_2+k_3)^2/2Q_s^2} \; \frac{(k_2+k_3)^i}{(k_2+k_3)^2} \ ,\\
\label{T14_soln}
T^i_{1,4}&\approx& -\pi Q_s^2\:  e^{(k_2+k_3)^2/2Q_s^2} \; \frac{(k_2+k_3)^i}{(k_2+k_3)^2}
\, \frac{2^2}{(k_2+k_3)^2}\, \bigg[ 1+\frac{2^2\, Q_s^2}{(k_2+k_3)^2}+\frac{3}{2}\frac{2^4\, Q_s^4}{(k_2+k_3)^4}\bigg],\\
%\bigg[ \frac{4}{(k_2+k_3)^2}+\frac{16\; Q_s^2}{(k_2+k_3)^4}+3\frac{2^5\; Q_s^4}{(k_2+k_3)^6}\bigg] \\
T^{ij}_{2,4}&\approx& \pi Q_s^2\:  e^{(k_2+k_3)^2/2Q_s^2} \; \frac{2^2}{(k_2+k_3)^2}\bigg[ \frac{1}{2}\frac{(k_2+k_3)^i(k_2+k_3)^j}{(k_2+k_3)^2}
+\frac{\delta^{ij}}{8} \frac{2^2\, Q_s^2}{(k_2+k_3)^2}\bigg( 1+ \frac{2^2\, Q_s^2}{(k_2+k_3)^2}\bigg)\bigg].
% \bigg\{ \delta^{ij}\bigg[ \frac{2\, Q_s^2}{(k_2+k_3)^4}+\frac{8\, Q_s^4}{(k_2+k_3)^6}\bigg]
%+\frac{2\, (k_2+k_3)^i(k_2+k_3)^j}{(k_2+k_3)^4}\bigg\}
\eeq 
Finally, using these results in Eq. \eqref{X_1_before_q3}, $X_1$ can be written as  
\beq
X_1&\approx& \frac{1}{2}\, \alpha_s^3(4\pi)^6(N_c^2-1)\; \mu^6\; S_\perp\ e^{-(k_2-k_3)^2/2Q_s^2}\; \frac{1}{k_2^4}\bigg[ 1+\frac{2!\, Q_s^2}{k_2^2}+\frac{3!\, Q_s^4}{k_2^4}\bigg] 
\bigg\{ \frac{1}{2}\frac{1}{k_2^2k_3^2}\frac{(k_2-k_3)^4}{(k_2+k_3)^4}\\
&+&
\frac{2^2\, Q_s^2}{(k_2+k_3)^2}\frac{1}{k_2^2k_3^2}\frac{(k_2-k_3)^4}{(k_2+k_3)^4}
%\nonumber\\
%&+&
+ \frac{2^4\, Q_s^4}{(k_2+k_3)^4}\bigg\lgroup \frac{1}{k_2^2k_3^2}\frac{(k_2-k_3)^4}{(k_2+k_3)^4}+\frac{2^2}{(k_2+k_3)^4}\bigg[ 1+(k^i_2-k^i_3)\bigg(\frac{k_2^i}{k_2^2}-\frac{k_3^i}{k_3^2}\bigg)\bigg]\bigg\rgroup\bigg\},\nonumber
\eeq
%
%\frac{1}{2}\, \frac{1}{k_2^6k_3^2} \bigg\{ \bigg[\frac{1}{2}-\frac{4\, (k_2\cdot k_3)}{(k_2+k_3)^2}+\frac{8\, (k_2\cdot k_3)^2}{(k_2+k_3)^4}\bigg]+\frac{4\, Q_s^2}{(k_2+k_3)^2}\bigg[ 1-\frac{8\, (k_2\cdot k_3)}{(k_2+k_3)^4}(k_2^2+k_3^2)\bigg]\nonumber\\
%&&+\frac{2^4\, Q_s^4}{(k_2+k_3)^4}\bigg[ 1+ \frac{12\,  k_2^2 k_3^2}{(k_2+k_3)^4}-\frac{12 \,(k_2\cdot k_3)}{(k_2+k_3)^4}(k_2^2+k_3^2)\bigg]\bigg\}
which can be organized differently and rewritten as in Eq. \eqref{X_1_final}. 
%\beq
%X_1&\approx& \frac{1}{2}\, \alpha_s^3(4\pi)^6(N_c^2-1)\; \mu^6\; S_\perp\ e^{-(k_2-k_3)^2/2Q_s^2}\; \frac{1}{k_2^4}
%\nonumber\\
%&\times&
%\bigg\{ \bigg\lgroup \frac{1}{2}+Q_s^2\bigg[\frac{1}{k_2^2}+\frac{2^2}{(k_2+k_3)^2}\bigg]
%+Q_s^4\bigg[ \frac{3}{k_2^4}+\frac{2!}{k_2^2}\frac{2^2}{(k_2+k_3)^2}+\frac{2^4}{(k_2+k_3)^4}\bigg]\bigg\rgroup
%\frac{1}{k_2^2k_3^2}\frac{(k_2-k_3)^4}{(k_2+k_3)^4}\nonumber\\
%&&
%\hspace{0.5cm}
%+\; Q_s^4\frac{2^6}{(k_2+k_3)^8}\bigg[ 1+(k_2^i-k_3^i)\bigg(\frac{k_2^i}{k_2^2}-\frac{k_3^i}{k_3^2}\bigg)\bigg]\bigg\}
%\eeq
%It has been explained at the beginning of this calculation, $X_1$ contributes largely to forward correlation of the produced gluons. Indeed, as seen from its final expression, $X_1$
%is enhanced in the limit  $k_2=k_3$,  
%\beq
%X_1(k_2=k_3)&\approx& \alpha_s^3(4\pi)^6(N_c^2-1)\; \mu^6\; S_\perp\  \frac{1}{8}\; \frac{Q_s^4}{k_2^{12}}
%\eeq
This completes our analytical analysis of $X_1$.

%Similar analysis can be done for the other terms as well. %Here, we just list the final result: 

\subsubsection{Computation of $X_2$}
After performing the trivial integration over $k_1$,  $X_2$  reads
\beq
&&
X_2=4 \, \alpha_s^3(4\pi)^3(N_c^2-1)\, \mu^6 \, (2\pi)^4\, S_\perp\, \big[ \delta^{(2)}(k_2-k_3)+\delta^{(2)}(k_2+k_3) \big] \, \int \frac{d^2q_1}{(2\pi)^2} \, \frac{d^2q_2}{(2\pi)^2} \, \frac{d^2q_3}{(2\pi)^2} \, d(q_1) \, d(q_2) \, d(q_3)\nonumber\\
&&
\times\, 
\bigg[ \frac{(k_2-q_2)^i}{(k_2-q_2)^2}-\frac{k_2^i}{k_2^2}\bigg] \bigg[ \frac{(k_2-q_3)^i}{(k_2-q_3)^2}-\frac{k_2^i}{k_2^2}\bigg] \; 
\bigg[ \frac{(k_2-q_2)^j}{(k_2-q_2)^2}-\frac{k_2^j}{k_2^2}\bigg] \bigg[ \frac{(k_2-q_3)^j}{(k_2-q_3)^2}-\frac{k_2^j}{k_2^2}\bigg] \nonumber\\
&&
\times\,  
\bigg[ \frac{(k_2-q_2)^k}{(k_2-q_2)^2}-\frac{(k_2-q_2+q_1)^k}{(k_2-q_2+q_1)^2}\bigg]
\bigg[ \frac{(k_2-q_2)^k}{(k_2-q_2)^2}-\frac{(k_2-q_2+q_1)^k}{(k_2-q_2+q_1)^2}\bigg].
\eeq
The first term in $X_2$ is an explicit contribution to the forward HBT of  gluons 2 and 3. The second term, which is indeed a mirror image $(k_3\to-k_3)$, is an explicit contribution to the backward HBT of  gluons 2 and 3. 
By shifting the variables and renaming them, $X_2$ contribution can be put into a more compact and convenient form:
\beq
\label{X2_shifted}
X_2&\equiv&4\, \alpha_s^3(4\pi)^3(N_c^2-1)\, \mu^6 \, (2\pi)^4 \, S_\perp \Big[ \delta^{(2)}(k_2+k_3)+\delta^{(2)}(k_2-k_3) \Big] 
\int \frac{d^2q_1}{(2\pi)^2}\frac{d^2q_2}{(2\pi)^2}\frac{d^2q_3}{(2\pi)^2} \, 
\\
&\times&
d(q_1+q_2) \, d(q_2+k_2)\, d(q_3+k_2) 
\bigg[ \frac{q^k_1}{q^2_1}+\frac{q^k_2}{q_2^2}\bigg]\bigg[ \frac{q^k_1}{q^2_1}+\frac{q^k_2}{q_2^2}\bigg] \, 
\bigg[ \frac{q_2^i}{q_2^2}+\frac{k_2^i}{k_2^2}\bigg] \bigg[ \frac{q_2^j}{q_2^2}+\frac{k_2^j}{k_2^2}\bigg] \, 
\bigg[ \frac{q_3^i}{q_3^2}+\frac{k_2^i}{k_2^2}\bigg] \bigg[ \frac{q_3^j}{q_3^2}+\frac{k_2^j}{k_2^2}\bigg] . \nonumber
\eeq
After we use the GBW model given in Eq. \eqref{gbw} for the dipole operators, $X_2$  can be organized in the following way: 
\beq
X_2&=&4\, \alpha_s^3(4\pi)^3(N_c^2-1)\, \mu^6\, (2\pi)^4\, S_\perp \, \big[ \delta^{(2)}(k_2+k_3)+\delta^{(2)}(k_2-k_3)\big]\, \frac{1}{(2\pi)^6}\, \frac{(4\pi)^3}{Q_s^6} \, e^{-2k_2^2/Q_s^2} \nonumber\\
&\times&\int d^2q_2 \; e^{-(2q_2^2+2q_2\cdot k_2)/Q_s^2} \bigg[ \frac{q_2^i}{q_2^2}\frac{q_2^j}{q_2^2}+\frac{k_2^i}{k_2^2}\frac{q_2^j}{q^2_2}+\frac{k_2^j}{k_2^2}\frac{q_2^i}{q^2_2}+\frac{k_2^i}{k_2^2}\frac{k_2^j}{k_2^2}\bigg]
\int d^2q_1\; e^{-(q_1^2+2q_1\cdot q_2)/Q_s^2} \bigg[ \frac{1}{q_1^2}+2\frac{q_1^k}{q_1^2}\frac{q_2^k}{q_2^2}+\frac{1}{q_2^2}\bigg]
\nonumber\\
&\times& 
\int d^2q_3\; e^{-(q_3^2+2q_3\cdot k_2)/Q_s^2}\bigg[ \frac{q_3^i}{q_3^2}\frac{q_3^j}{q_3^2}+\frac{q_3^i}{q_3^2}\frac{k_2^j}{k_2^2}+\frac{q_3^j}{q_3^2}\frac{k_2^i}{k_2^2}+\frac{k_2^i}{k_2^2}\frac{k_2^j}{k_2^2}\bigg].
\eeq
Let us start from the integration over $q_3$. This integral has been computed and the exact result is given in Eq. (\ref{q_2_final_result}). Since we are interested in computing the exponentially enhanced contributions from the $q_3$ integration, the approximate result is given in Eq. \eqref{Exp_Type_2}. Using this result in $X_2$, we get
\beq
X_2&\approx&4\, \alpha_s^3(4\pi)^3(N_c^2-1)\, \mu^6\, (2\pi)^4\, S_\perp \, \big[ \delta^{(2)}(k_2+k_3)+\delta^{(2)}(k_2-k_3)\big]\, \frac{1}{(2\pi)^6}\, \frac{(4\pi)^3}{Q_s^6} \, \pi Q_s^2 \; e^{-k_2^2/Q_s^2}
\nonumber\\
&\times&
 \frac{1}{2k_2^2}\bigg[ \frac{Q_s^2}{k_2^2}+2!\frac{Q_s^4}{k_2^4}+3!\frac{Q_s^6}{k_2^6}\bigg]
%\nonumber\\
%&\times& 
\int d^2q_2 \; e^{-(2q_2^2+2q_2\cdot k_2)/Q_s^2}\bigg[ \frac{1}{q_2^2}+2\frac{q_2^i}{q_2^2}\frac{k_2^i}{k_2^2}+\frac{1}{k_2^2}\bigg]
\nonumber\\
&\times&
\int d^2q_1\; e^{-(q_1^2+2q_1\cdot q_2)/Q_s^2}\bigg[ \frac{1}{q_1^2}+2\frac{q_1^j}{q_1^2}\frac{q_2^j}{q_2^2}+\frac{1}{q_2^2}\bigg].
\eeq
Now, let us consider the integration over $q_1$ which can be organized as follows:
\beq
\int d^2q_1\; e^{-(q_1^2+2q_1\cdot q_2)/Q_s^2}\bigg[ \frac{1}{q_1^2}+2\frac{q_1^j}{q_1^2}\frac{q_2^j}{q_2^2}+\frac{1}{q_2^2}\bigg]=\frac{1}{q_2^2}I_{0,0}+2\frac{q_2^i}{q_2^2}I^i_{1,2}+I_{0,2}\ .
\eeq
$I_{0,0}$ and $I^i_{1,2}$  were computed previously and the results are given in Eqs. (\ref{1_Type_0}) and (\ref{1_Type_1}) respectively. The new integral $I_{0,2}$ can be computed in a similar manner, and the exponentially enhanced piece reads 
\beq
\label{I_02}
I_{0,2}=\int d^2q_1 \; e^{-(q_1^2+2 q_1\cdot q_2)}\frac{1}{q_1^2}\approx \pi Q_s^2\; e^{q_2^2/Q_s^2}\bigg[\frac{1}{q_2^2}+\frac{Q_s^2}{q_2^4}+2!\frac{Q_s^4}{q_2^6}+3!\frac{Q_s^6}{q_2^8}\bigg].
\eeq
Putting everything together, $X_2$, upon integration over $q_1$, reads
%The exponentially enhanced piece of the above integral then reads 
%\beq
%I_{0,2}\approx \pi Q_s^2\; e^{q_2^2/Q_s^2}\bigg[\frac{1}{q_2^2}+\frac{Q_s^2}{q_2^4}+2!\frac{Q_s^4}{q_2^6}+3!\frac{Q_s^6}{q_2^8}\bigg]
%\eeq
%Then, $q_1$ integration can be written as 
%\beq
%\int d^2q_1\; e^{-(q_1^2+2q_1\cdot q_2)/Q_s^2}\bigg[ \frac{1}{q_1^2}+2\frac{q_1^j}{q_1^2}\frac{q_2^j}{q_2^2}+\frac{1}{q_2^2}\bigg]\approx \pi Q_s^2\; e^{q_2^2/Q_s^2}\frac{1}{q_2^2}\bigg[ \frac{Q_s^2}{q_2^2}+2! \frac{Q_s^4}{q_2^4}+3!\frac{Q_s^6}{q_2^6}\bigg]
%\eeq
%Upon integration over $q_1$, $X_2$ contribution reads 
\beq
X_2&\approx&\alpha_s^3(4\pi)^3(N_c^2-1)\, \mu^6\, S_\perp \big[  \delta^{(2)}(k_2+k_3)+\delta^{(2)}(k_2-k_3)\big]\, (4\pi)^3 \, Q_s^2 \, e^{-k_2^2/Q_s^2}\frac{1}{2}\, \frac{1}{k_2^4}\bigg[ 1+ 2!\frac{Q_s^2}{k_2^2}\bigg] \nonumber\\
&\times&
\int d^2q_2\; e^{-(q_2^2+2q_2\cdot k_2)/Q_s^2}\bigg[ \frac{1}{q_2^6}+2\frac{k_2^i}{k_2^2}\frac{q_2^i}{q_2^2}+\frac{1}{k_2^2}\frac{1}{q_2^4}\bigg]
\bigg[ 1+2!\frac{Q_s^2}{q_2^2}\bigg],
\eeq
where we have neglected the $O(Q_s^6)$ terms which originate from $q_1$ and $q_3$ integrals since they will contribute to higher order terms upon integration over $q_2$.  Let us organize the $X_2$ contribution in the following way: 
\beq
\label{X_2_next-to-final}
X_2&\approx& \alpha_s^3(4\pi)^3(N_c^2-1)\, \mu^6\, S_\perp \big[  \delta^{(2)}(k_2+k_3)+\delta^{(2)}(k_2-k_3)\big]\, (4\pi)^3 \, Q_s^2 \, e^{-k_2^2/Q_s^2}\frac{1}{2}\, \frac{1}{k_2^4}\bigg[ 1+ 2!\frac{Q_s^2}{k_2^2}\bigg] \nonumber\\
&\times&\bigg\{ \bigg[ \frac{1}{k_2^2}I_{0,4}+2\frac{k_2^i}{k_2^2}I^i_{1,6} +I_{0,6}\bigg]+2!Q_s^2\bigg[ \frac{1}{k_2^2}I_{0,6}+2\frac{k_2^i}{k_2^2}I^i_{1,8}+I_{0,8}\bigg]\bigg\},
\eeq
where the new type of integrals are defined, computed and expanded to the appropriate order as  
\beq
\label{I_04}
I_{0,4}&=&\int d^2q_2\; e^{-(q_2^2+2q_2\cdot k_2)/Q_s^2}\frac{1}{q_2^4} \approx \pi Q_s^2\; e^{k_2^2/Q_s^2}\frac{1}{k_2^4}\bigg[1+4\frac{Q_s^2}{k_2^2}\bigg],\\
I_{0,6}&=&\int d^2q_2\; e^{-(q_2^2+2q_2\cdot k_2)/Q_s^2}\frac{1}{q_2^6} \approx \pi Q_s^2\; e^{k_2^2/Q_s^2}\frac{1}{k_2^6}\bigg[1+9\frac{Q_s^2}{k_2^2}\bigg], \\
I_{0,8}&=&\int d^2q_2\; e^{-(q_2^2+2q_2\cdot k_2)/Q_s^2}\frac{1}{q_2^8} \approx \pi Q_s^2\; e^{k_2^2/Q_s^2}\frac{1}{k_2^8}\ ,\\
I^i_{1,6}&=&\int d^2q_2\; e^{-(q_2^2+2q_2\cdot k_2)/Q_s^2}\frac{q_2^i}{q_2^6} \approx -\pi Q_s^2\; e^{k_2^2/Q_s^2}\;\frac{k_2^i}{k_2^2}\frac{1}{k_2^4}\bigg[1+6\frac{Q_s^2}{k_2^2}\bigg],\\
I^i_{1,8}&=&\int d^2q_2\; e^{-(q_2^2+2q_2\cdot k_2)/Q_s^2}\frac{q_2^i}{q_2^8} \approx -\pi Q_s^2\; e^{k_2^2/Q_s^2}\;  \frac{k_2^i}{k_2^2}\frac{1}{k_2^6}\ .
\eeq
%Exponentially enhanced contributions to each of the above integrals after expanded to the appropriate order I_02, I_04
%\beq
%I_{0,4}&\approx& \pi Q_s^2\; e^{k_2^2/Q_s^2}\frac{1}{k_2^4}\bigg[1+4\frac{Q_s^2}{k_2^2}\bigg]\\
%I_{0,6}&\approx& \pi Q_s^2\; e^{k_2^2/Q_s^2}\frac{1}{k_2^6}\bigg[1+9\frac{Q_s^2}{k_2^2}\bigg]\\
%I_{0,8}&\approx& \pi Q_s^2\; e^{k_2^2/Q_s^2}\frac{1}{k_2^8}\\
%I^i_{1,6}&\approx& -\pi Q_s^2\; e^{k_2^2/Q_s^2}\;\frac{k_2^i}{k_2^2}\frac{1}{k_2^4}\bigg[1+6\frac{Q_s^2}{k_2^2}\bigg]\\
%I^i_{1,8}&\approx& -\pi Q_s^2\; e^{k_2^2/Q_s^2}\;  \frac{k_2^i}{k_2^2}\frac{1}{k_2^6}
%\eeq
Using these results, it is straightforward to realize that the combinations appearing in Eq. \eqref{X_2_next-to-final} are  
\beq
\frac{1}{k_2^2}I_{0,4}+2\frac{k_2^i}{k_2^2}I^i_{1,6} +I_{0,6}&\approx&\pi Q_s^2\; e^{k_2^2/Q_s^2} \; \frac{Q_s^2}{k_2^8}+\mathcal{O}(Q_s^6),\\
\frac{1}{k_2^2}I_{0,6}+2\frac{k_2^i}{k_2^2}I^i_{1,8}+I_{0,8}&\approx& \mathcal{O}(Q_s^4).
\eeq
Thus, the final expression for $X_2$  can be written as in Eq. \eqref{X_2_final}. 

%Therefore, we get 
%\beq
%X_2&\approx& \alpha_s^3(4\pi)^3(N_c^2-1)\, \mu^6\, (2\pi)\;  S_\perp \big[  \delta^{(2)}(k_2+k_3)+\delta^{(2)}(k_2-k_3)\big]\, (4\pi)^3 \frac{1}{4} \frac{Q_s^6}{k_2^{12}}
%\eeq

\subsubsection{Computation of $X_3$}
For the computation of $X_3$, we start from Eq. \eqref{X_3_initial} and use the definitions given in Eqs. \eqref{X1} and \eqref{X1prime}. We  perform the integrals over $k_1$ and $q_2$ by using the $\delta$-functions. Then, $X_3$ can be written as  
\beq
X_3&=&4\,  \alpha_s^3(4\pi)^3(N_c^2-1)\, \mu^6 \, (2\pi)^2\, S_\perp\, \int \frac{d^2q_1}{(2\pi)^2}\, \frac{d^2q_3}{(2\pi)^2} \bigg\{ d(q_1)\, d\big[ (k_3-q_3)-k_2\big]\, d(q_3)\, \nonumber\\
&\times&
\bigg[ \frac{(k_3-q_3)^i}{(k_3-q_3)^2}-\frac{k_3^i}{k_3^2}\bigg] \bigg[ \frac{(k_3-q_1)^i}{(k_3-q_1)^2}-\frac{k_3^i}{k_3^2}\bigg]
%\nonumber\\
%&\times&
\bigg[ \frac{(k_3-q_3)^j}{(k_3-q_3)^2}-\frac{k_3^j}{k_3^2}\bigg] \bigg[ \frac{(k_3-q_1)^j}{(k_3-q_1)^2}-\frac{k_3^j}{k_3^2}\bigg]\; 
\nonumber\\
&\times&
\bigg[ \frac{(k_3-q_3)^k}{(k_3-q_3)^2}-\frac{k_2^k}{k_2^2}\bigg] \bigg[ \frac{(k_3-q_3)^k}{(k_3-q_3)^2}-\frac{k_2^k}{k_2^2}\bigg] +(k_2\to-k_2)\bigg\}.
\eeq
The first term in $X_3$ is a contribution to the forward correlation of  gluons 2 and 3. Its mirror image given by $(k_2\to-k_2)$ is a contribution to the backward correlation of  gluons 2 and 3. Again by shifting and renaming the integration variables, $X_3$ contribution can be organized as follows: 
\beq
\label{X3_shifted}
X_3&\equiv&  4\, \alpha_s^3(4\pi)^3(N_c^2-1)\, \mu^6(2\pi)^2\, S_\perp\, 
\int \frac{d^2q_2}{(2\pi)^2} \frac{d^2q_3}{(2\pi)^2} \bigg\{
d(q_2+k_2) \, d(q_2+k_3)\, d(q_3+k_3) \nonumber\\
&&
\hspace{1.2cm}
\times
\bigg[ \frac{q_2^i}{q_2^2}+{\frac{k_3^i}{k_3^2}}\bigg] \bigg[ \frac{q_2^j}{q_2^2}+{\frac{k_3^j}{k_3^2}}\bigg]\, 
\bigg[ \frac{q_3^i}{q_3^2}+{\frac{k_3^i}{k_3^2}}\bigg] \bigg[ \frac{q_3^j}{q_3^2}+{\frac{k_3^j}{k_3^2}}\bigg] \, 
\bigg[ \frac{q_2^k}{q_2^2}+\frac{k_2^k}{k_2^2}\bigg] \bigg[ \frac{q_2^k}{q_2^2}+\frac{k_2^k}{k_2^2}\bigg] +(k_2\to-k_2)\bigg\}.
\eeq
By comparing Eqs. (\ref{X1_shifted}) and (\ref{X3_shifted}), it is straight forward to realize that  $X_1=X_3(k_2\leftrightarrow k_3)$. Therefore, one can immediately write down the result for $X_3$ as in Eq. \eqref{X_3_final}. 
% I will use this as a double check of the results, since the final results are quite complicated and it is easy to make algebraic mistakes. I have already spotted a mismatch between the final results (overall factor and explicit expression of the momentum structure that accompanies exponential term).   
%\beq
%X_3&\approx& \frac{1}{2}\, \alpha_s^3(4\pi)^6(N_c^2-1)\; \mu^6\; S_\perp\  e^{-(k_2-k_3)^2/2Q_s^2}\; \frac{1}{k_3^4}
%\nonumber\\
%&\times&
%\bigg\{ \bigg\lgroup \frac{1}{2}+Q_s^2\bigg[\frac{1}{k_3^2}+\frac{2^2}{(k_2+k_3)^2}\bigg]
%+Q_s^4\bigg[ \frac{3}{k_3^4}+\frac{2!}{k_3^2}\frac{2^2}{(k_2+k_3)^2}+\frac{2^4}{(k_2+k_3)^4}\bigg]\bigg\rgroup
%\frac{1}{k_2^2k_3^2}\frac{(k_2-k_3)^4}{(k_2+k_3)^4}\nonumber\\
%&&
%\hspace{0.5cm}
%+\; Q_s^4\frac{2^6}{(k_2+k_3)^8}\bigg[ 1+(k_2^i-k_3^i)\bigg(\frac{k_2^i}{k_2^2}-\frac{k_3^i}{k_3^2}\bigg)\bigg]\bigg\}
%\eeq
%In the limit  $k_2=k_3$, we get 
%\beq
%X_3&\approx& \alpha_s^3(4\pi)^6(N_c^2-1)\; \mu^6\; S_\perp\  \frac{1}{8}\; \frac{Q_s^4}{k_3^{12}}
%\eeq

\subsubsection{Computation of $X_4$}
We start from Eq. \eqref{X_4_initial}, use the definitions for the Lipatov vertices and the MV model for $\mu^2$, and integrate over $q_2$ and $k_1$ to write  $X_4$  in the following way: 
\beq
X_4&=&4\,  \alpha_s^3(4\pi)^3(N_c^2-1)\, \mu^6 \, (2\pi)^2\, S_\perp\, \int \frac{d^2q_1}{(2\pi)^2}\, \frac{d^2q_3}{(2\pi)^2} \bigg\{ d(q_1)\, d\big[ (k_3-q_3)-k_2\big]\, d(q_3)
\nonumber\\
&\times&
\bigg[ \frac{(k_3-q_3)^i}{(k_3-q_3)^2}-\frac{(k_3-q_3+q_1)^i}{(k_3-q_3+q_1)^2}\bigg]
\bigg[ \frac{(k_3-q_3)^i}{(k_3-q_3)^2}-\frac{(k_3-q_3+q_1)^i}{(k_3-q_3+q_1)^2}\bigg]\; 
\bigg[  \frac{(k_3-q_3)^j}{(k_3-q_3)^2}-\frac{k_2^j}{k_2^2}\bigg] \bigg[  \frac{(k_3-q_3)^j}{(k_3-q_3)^2}-\frac{k_2^j}{k_2^2}\bigg]\nonumber\\
&\times&
\bigg[ \frac{(k_3-q_3)^k}{(k_3-q_3)^2} -\frac{k_3^k}{k_3^2}\bigg] \bigg[ \frac{(k_3-q_3)^k}{(k_3-q_3)^2} -\frac{k_3^k}{k_3^2}\bigg]+(k_3\to-k_3)\bigg\}.
\eeq
The first term in $X_4$ is a contribution to the forward correlation of  gluons 2 and 3. The mirror image is a contribution the backward correlation of  gluons 2 and 3. We again shift and rename the integration variables to rewrite  $X_4$ as follows:
\beq
\label{X4_shifted}
X_4&=&  4\, \alpha_s^3(4\pi)^3(N_c^2-1)\, \mu^6(2\pi)^2\, S_\perp\, 
\int \frac{d^2q_2}{(2\pi)^2} \frac{d^2q_3}{(2\pi)^2} \bigg\{
d(q_2+q_3)\, d(q_3+k_2)\, d(q_3+k_3)\nonumber\\
&&
\hspace{1cm}
\times
\bigg[ \frac{q_2^i}{q_2^2}+\frac{q_3^i}{q_3^2}\bigg] \bigg[ \frac{q_2^i}{q_2^2}+\frac{q_3^i}{q_3^2}\bigg]\, 
\bigg[ \frac{q_3^j}{q_3^2}+\frac{k_2^j}{k_2^2}\bigg] \bigg[ \frac{q_3^j}{q_3^2}+\frac{k_2^j}{k_2^2}\bigg] \, 
\bigg[ \frac{q_3^k}{q_3^2}+\frac{k_3^k}{k_3^2}\bigg] \bigg[ \frac{q_3^k}{q_3^2}+\frac{k_3^k}{k_3^2}\bigg]+ (k_3\to-k_3) \bigg\}.
\eeq
Now, we use the GBW model for the dipole operators, and organize  $X_4$  as 
\beq
X_4&= &4\, \alpha_s^3(4\pi)^3(N_c^2-1)\, \mu^6\, (2\pi)^2\, S_\perp\, \frac{1}{(2\pi)^4} \, \bigg(\frac{4\pi}{Q_s^2}\bigg)^3\, e^{-(k_2^2+k_3^2)/Q_s^2}
\int d^2q_3 \, e^{-[3q_3^2+2q_3\cdot(k_2+k_3)]/Q_s^2}
\nonumber\\
&\times&
\bigg[ 
\frac{1}{k_2^2k_3^2} 
+\bigg(\frac{1}{k_2^2}+\frac{1}{k_3^2}\bigg)\frac{1}{q_3^2} 
%\nonumber\\
%&&
+2\bigg(\frac{k_2^i}{k_2^2k_3^2}+\frac{k_3^i}{k_2^2k_3^2}\bigg)\frac{q_3^i}{q_3^2} 
+2\bigg(\frac{k_2^i}{k_2^2}+\frac{k_3^i}{k_3^2}\bigg)\frac{q_3^i}{q_3^2}\frac{1}{q_3^2}
+4\frac{k_2^i}{k_2^2}\frac{k_3^j}{k_3^2}\frac{q_3^i}{q_3^2}\frac{q_3^j}{q_3^2}
+\frac{1}{q_3^2}\frac{1}{q_3^2}\bigg]
\nonumber\\
&\times&
\int d^2q_2 \, e^{-(q_2^2+2q_2\cdot q_3)/Q_s^2} \bigg[ \frac{1}{q_3^2}+\frac{1}{q_2^2}+2\frac{q_3^k}{q_3^2}\frac{q_2^k}{q_2^2}\bigg].
\eeq
The integration over $q_2$ can be performed trivially as before. Indeed, the types of  integrals that appear are $I_{0,0}$, $I^k_{1,2}$ and $I_{0,2}$  which were computed previously and the results  given in Eqs. \eqref{1_Type_0}, \eqref{1_Type_1}  and \eqref{I_02} respectively. Using these results, we can write  $X_4$  as 
\beq
X_4&\approx& 4\, \alpha_s^3(4\pi)^3(N_c^2-1)\, \mu^6\, S_\perp\, \frac{1}{(2\pi)^2} \frac{(4\pi)^3}{Q_s^6} \, \pi Q_s^2\, e^{-(k_2^2+k_3^2)/Q_s^2}\int d^2q_3\, e^{-[2q_3^2+2q_3\cdot(k_2+k_3)]/Q_s^2} \bigg[\frac{Q_s^2}{q_3^4}+\frac{2!\, Q_s^4}{q_3^6}+\frac{3!\, Q_s^6}{q_3^8}\bigg]
\nonumber\\
&\times& 
\bigg[ 
\frac{1}{k_2^2k_3^2} 
+\bigg(\frac{1}{k_2^2}+\frac{1}{k_3^2}\bigg)\frac{1}{q_3^2} 
+2\frac{(k_2+k_3)^i}{k_2^2k_3^2}\frac{q_3^i}{q_3^2} 
+2\bigg(\frac{k_2^i}{k_2^2}+\frac{k_3^i}{k_3^2}\bigg)\frac{q_3^i}{q_3^2}\frac{1}{q_3^2}
+4\frac{k_2^i}{k_2^2}\frac{k_3^j}{k_3^2}\frac{q_3^i}{q_3^2}\frac{q_3^j}{q_3^2}
+\frac{1}{q_3^2}\frac{1}{q_3^2}\bigg],
\eeq
which can be organized as
\beq
X_4&\approx& 4\, \alpha_s^3(4\pi)^3(N_c^2-1)\, \mu^6\, S_\perp\, \frac{1}{(2\pi)^2} \frac{(4\pi)^3}{Q_s^6} \, \pi Q_s^4\, e^{-(k_2^2+k_3^2)/Q_s^2}\\
&\times&
\bigg\{ \;\;\;\;
\bigg[ \frac{1}{k_2^2k_3^2}T_{0,4}+\bigg(\frac{1}{k_2^2}+\frac{1}{k_3^2}\bigg)T_{0,6}+T_{0,8}+2\frac{(k_2+k_3)^i}{(k_2+k_3)^2}T^i_{1,6}+2\bigg(\frac{k_2^i}{k_2^2}+\frac{k_3^i}{k_3^2}\bigg)T^i_{1,8}+4\frac{k_2^i}{k_2^2}\frac{k_3^j}{k_3^2}T^{ij}_{2,8} \bigg]
\nonumber\\
&+& 
2\, Q_s^2\bigg[\frac{1}{k_2^2k_3^2}T_{0,6}+\bigg(\frac{1}{k_2^2}+\frac{1}{k_3^2}\bigg)T_{0,8}+T_{0,10}+2\frac{(k_2+k_3)^i}{(k_2+k_3)^2}T^i_{1,8}+2\bigg(\frac{k_2^i}{k_2^2}+\frac{k_3^i}{k_3^2}\bigg)T^i_{1,10}+4\frac{k_2^i}{k_2^2}\frac{k_3^j}{k_3^2}T^{ij}_{2,10} \bigg]
\nonumber\\
&+& \!
3!\, Q_s^4\bigg[\frac{1}{k_2^2k_3^2}T_{0,8}+\bigg(\frac{1}{k_2^2}+\frac{1}{k_3^2}\bigg)T_{0,10}+T_{0,12}+2\frac{(k_2+k_3)^i}{(k_2+k_3)^2}T^i_{1,10}+2\bigg(\frac{k_2^i}{k_2^2}+\frac{k_3^i}{k_3^2}\bigg)T^i_{1,12}+4\frac{k_2^i}{k_2^2}\frac{k_3^j}{k_3^2}T^{ij}_{2,12} \bigg]\bigg\}.\nonumber
\eeq
The new types of integrals that appear above are defined as  
\beq
%T_{0,4}&=& \int d^2q_3\, e^{-[2q_3^2+2q_3\cdot(k_2+k_3)]/Q_s^2} \frac{1}{q_3^4}\\
\label{T06_def}
T_{0,6}&=& \int d^2q_3\, e^{-[2q_3^2+2q_3\cdot(k_2+k_3)]/Q_s^2} \frac{1}{q_3^6}\ ,\\
T_{0,8}&=& \int d^2q_3\, e^{-[2q_3^2+2q_3\cdot(k_2+k_3)]/Q_s^2} \frac{1}{q_3^8}\ ,\\
T_{0,10}&=& \int d^2q_3\, e^{-[2q_3^2+2q_3\cdot(k_2+k_3)]/Q_s^2} \frac{1}{q_3^{10}}\ ,\\
T_{0,12}&=& \int d^2q_3\, e^{-[2q_3^2+2q_3\cdot(k_2+k_3)]/Q_s^2} \frac{1}{q_3^{12}}\ ,\\
\label{T16_def}
T^i_{1,6}&=& \int d^2q_3\, e^{-[2q_3^2+2q_3\cdot(k_2+k_3)]/Q_s^2} \frac{q_3^i}{q_3^6} \ ,\\
T^i_{1,8}&=& \int d^2q_3\, e^{-[2q_3^2+2q_3\cdot(k_2+k_3)]/Q_s^2} \frac{q_3^i}{q_3^8}\ , \\
T^i_{1,10}&=& \int d^2q_3\, e^{-[2q_3^2+2q_3\cdot(k_2+k_3)]/Q_s^2} \frac{q_3^i}{q_3^{10}}\ , \\
T^i_{1,12}&=& \int d^2q_3\, e^{-[2q_3^2+2q_3\cdot(k_2+k_3)]/Q_s^2} \frac{q_3^i}{q_3^{12}}\ , \\
T^{ij}_{2,8}&=& \int d^2q_3\, e^{-[2q_3^2+2q_3\cdot(k_2+k_3)]/Q_s^2} \frac{q_3^iq_3^j}{q_3^8}\ ,\\
T^{ij}_{2,10}&=& \int d^2q_3\, e^{-[2q_3^2+2q_3\cdot(k_2+k_3)]/Q_s^2} \frac{q_3^iq_3^j}{q_3^{10}}\ ,\\
T^{ij}_{2,12}&=& \int d^2q_3\, e^{-[2q_3^2+2q_3\cdot(k_2+k_3)]/Q_s^2} \frac{q_3^iq_3^j}{q_3^{12}}\ ,
\eeq
and $T_{0,4}$ was defined in Eq. \eqref{T_04_def} and its solution  given in Eq. \eqref{T_04_soln}. The new integrals that are defined above can be computed in the same way as before, and the exponentially enhanced contributions to these integrals read
\beq
%T_{0,4}&\approx& \pi Q_s^2\; e^{(k_2+k_3)^2/2Q_s^2}\, \frac{2^4}{(k_2+k_3)^4} \bigg[ \frac{1}{2}+  \frac{2^2\, Q_s^2}{(k_2+k_3)^2}+ \frac{9}{4}\,  \frac{ 2^4\, Q_s^4}{(k_2+k_3)^4} \bigg]\\
\label{T06_soln}
T_{0,6}&\approx& \pi Q_s^2\; e^{(k_2+k_3)^2/2Q_s^2}\, \frac{2^6}{(k_2+k_3)^6} \bigg[ \frac{1}{2}+ \frac{9}{4}\, \frac{2^2\, Q_s^2}{(k_2+k_3)^2}+9\,  \frac{ 2^4\, Q_s^4}{(k_2+k_3)^4} \bigg],\\
T_{0,8}&\approx& \pi Q_s^2\; e^{(k_2+k_3)^2/2Q_s^2}\, \frac{2^8}{(k_2+k_3)^8} \bigg[ \frac{1}{2}+ 4\, \frac{2^2\, Q_s^2}{(k_2+k_3)^2}+25\,  \frac{ 2^4\, Q_s^4}{(k_2+k_3)^4} \bigg],\\
T_{0,10}&\approx& \pi Q_s^2\; e^{(k_2+k_3)^2/2Q_s^2}\, \frac{2^{10}}{(k_2+k_3)^{10}} \bigg[ \frac{1}{2}+\frac{25}{4}\, \frac{2^2\, Q_s^2}{(k_2+k_3)^2}\bigg], \\
T_{0,12}&\approx& \pi Q_s^2\; e^{(k_2+k_3)^2/2Q_s^2}\, \frac{2^{12}}{(k_2+k_3)^{12}}\,  \frac{1}{2}\ ,
\eeq
\beq
\label{T16_soln}
T^i_{1,6}&\approx&-\;  \pi Q_s^2\; e^{(k_2+k_3)^2/2Q_s^2}\, \frac{(k_2+k_3)^i}{(k_2+k_3)^2}\; \frac{2^4}{(k_2+k_3)^4}\bigg[ 1+ 3\; \frac{2^2\, Q_s^2}{(k_2+k_3)^2}+9\; \frac{2^4\, Q_s^4}{(k_2+k_3)^4}\bigg],\\
T^i_{1,8}&\approx&-\;  \pi Q_s^2\; e^{(k_2+k_3)^2/2Q_s^2}\, \frac{(k_2+k_3)^i}{(k_2+k_3)^2}\; \frac{2^6}{(k_2+k_3)^6}\bigg[ 1 + 6\; \frac{2^2\, Q_s^2}{(k_2+k_3)^2}+30\; \frac{2^4\, Q_s^4}{(k_2+k_3)^4}\bigg],\\
T^i_{1,10}&\approx&-\;  \pi Q_s^2\; e^{(k_2+k_3)^2/2Q_s^2}\, \frac{(k_2+k_3)^i}{(k_2+k_3)^2}\; \frac{2^8}{(k_2+k_3)^8}\bigg[ 1 + 10\; \frac{2^2\, Q_s^2}{(k_2+k_3)^2}\bigg],\\
T^i_{1,12}&\approx&-\;  \pi Q_s^2\; e^{(k_2+k_3)^2/2Q_s^2}\, \frac{(k_2+k_3)^i}{(k_2+k_3)^2}\; \frac{2^{10}}{(k_2+k_3)^{10}}\ ,
\eeq
\beq
T^{ij}_{2,8}&\approx& \pi Q_s^2\; e^{(k_2+k_3)^2/2Q_s^2}\; \frac{2^6}{(k_2+k_3)^6}
\bigg\{ \frac{1}{2} \, \frac{(k_2+k_3)^i(k_2+k_3)^j}{(k_2+k_3)^2}+\frac{2^2\, Q_s^2}{(k_2+k_3)^2}\bigg[ \frac{\delta^{ij}}{8}+ 2\, \frac{(k_2+k_3)^i(k_2+k_3)^j}{(k_2+k_3)^2}\bigg]\nonumber\\
&&
\hspace{4.7cm}
+\; \frac{2^4\, Q_s^4}{(k_2+k_3)^4}\bigg[ 3\, \frac{\delta^{ij}}{4}+\frac{15}{2}\, \frac{(k_2+k_3)^i(k_2+k_3)^j}{(k_2+k_3)^2}\bigg]\bigg\},
\\
T^{ij}_{2,10}&\approx& \pi Q_s^2\; e^{(k_2+k_3)^2/2Q_s^2}\; \frac{2^8}{(k_2+k_3)^8}\bigg[ \frac{1}{2}\, \frac{(k_2+k_3)^i(k_2+k_3)^j}{(k_2+k_3)^2}
+\frac{2^2\, Q_s^2}{(k_2+k_3)^2}\bigg[ \frac{\delta^{ij}}{8}+\frac{15}{4}\, \frac{(k_2+k_3)^i(k_2+k_3)^j}{(k_2+k_3)^2}\bigg]\bigg\}, \\
T^{ij}_{2,12}&\approx& \pi Q_s^2\; e^{(k_2+k_3)^2/2Q_s^2}\; \frac{2^{10}}{(k_2+k_3)^{10}}\, \frac{1}{2}\, \frac{(k_2+k_3)^i(k_2+k_3)^j}{(k_2+k_3)^2}\ .
\eeq
Using these results we can finally write the final expression for the exponentially enhanced contributions to $X_4$ as in Eq. \eqref{X_4_final}. 
%
%By using these results, we can write the exponentially enhanced contribution to $X_4$ as
%\beq
%X_4&\approx&\alpha_s^3(4\pi)^3(N_c^2-1)\, \mu^6\, S_\perp\, (4\pi)^3\, e^{-(k_2-k_3)^2/2Q_s^2}
%\bigg\{ \bigg[ 1+ 8\, \frac{2^2\, Q_s^2}{(k_2+k_3)^2}+76\, \frac{2^4\, Q_s^4}{(k_2+k_3)^4}\bigg]\frac{2^3}{k_2^2k_3^2}\frac{(k_2-k_3)^4}{(k_2+k_3)^8}
%\nonumber\\
%&&
%\hspace{5cm}
%+\; \frac{2^4\, Q_s^4}{(k_2+k_3)^4}\frac{2^2}{(k_2+k_3)^2}\bigg[ \frac{5}{2}\frac{ 2^6}{(k_2+k_3)^6}-\frac{9}{4}\frac{2^2}{k_2^2k_3^2(k_2+k_3)^2}\bigg]\bigg\}
%\eeq
%In the limit $k_2=k_3$, we get 
%\beq
%X_4&\approx&\alpha_s^3(4\pi)^3(N_c^2-1)\, \mu^6\, S_\perp\, (4\pi)^3\, \frac{1}{4}\, \frac{Q_s^4}{k_2^{12}}
%\eeq
%%

\subsubsection{Computation of $X_5$}
We start from Eq. \eqref{X5_initial}, use the explicit expressions for the Lipatov vertices and the MV model for $\mu^2$, integrate over $k_1$, and write the $X_5$ contribution as  
\beq
X_5&=&4 \, \alpha_s^3(4\pi)^3(N_c^2-1)\, \mu^6 \, (2\pi)^4\, S_\perp\, \big[ \delta^{(2)}(k_2-k_3)+\delta^{(2)}(k_2+k_3) \big] \, \int \frac{d^2q_1}{(2\pi)^2} \, \frac{d^2q_2}{(2\pi)^2} \, \frac{d^2q_3}{(2\pi)^2} \, d(q_1) \, d(q_2) \, d(q_3)\, 
\nonumber\\
&\times&
 \bigg[ \frac{(k_2-q_1)^i}{(k_2-q_1)^2}-\frac{k_2^i}{k_2^2}\bigg]  \bigg[ \frac{(k_2-q_2)^i}{(k_2-q_2)^2}-\frac{k_2^i}{k_2^2}\bigg] \; 
  \bigg[ \frac{(k_2-q_2)^j}{(k_2-q_1)^2}-\frac{k_2^j}{k_2^2}\bigg]  \bigg[ \frac{(k_2-q_3)^j}{(k_2-q_2)^2}-\frac{k_2^j}{k_2^2}\bigg] \nonumber\\
  &\times&
  \bigg[ \frac{(k_2-q_3)^k}{(k_2-q_3)^2}-\frac{k_2^k}{k_2^2}\bigg]  \bigg[ \frac{(k_2-q_1)^k}{(k_2-q_1)^2}-\frac{k_2^k}{k_2^2}\bigg] .
\eeq
The first term in $X_5$ is an explicit contribution to the forward HBT of  gluons 2 and 3. The second term, which is indeed a mirror image $(k_3\to-k_3)$, is an explicit contribution to the backward HBT of  gluons 2 and 3. Again by shifting and renaming the integration variables, we obtain
\beq
\label{X5_shifted}
X_5&\equiv&4\, \alpha_s^3(4\pi)^3(N_c^2-1)\, \mu^6 \, (2\pi)^4 \, S_\perp \Big[ \delta^{(2)}(k_2+k_3)+\delta^{(2)}(k_2-k_3) \Big] 
\int \frac{d^2q_1}{(2\pi)^2}\frac{d^2q_2}{(2\pi)^2}\frac{d^2q_3}{(2\pi)^2} \, 
\\
&&
%\hspace{2.5cm}
\times\; 
d(q_1+k_2) \, d(q_2+k_2)\, d(q_3+k_2)
\bigg[ \frac{q_1^i}{q_1^2}+\frac{k_2^i}{k_2^2}\bigg] \bigg[ \frac{q_1^k}{q_1^2}+\frac{k_2^k}{k_2^2}\bigg] \, 
\bigg[ \frac{q_2^i}{q_2^2}+\frac{k_2^i}{k_2^2}\bigg] \bigg[ \frac{q_2^j}{q_2^2}+\frac{k_2^j}{k_2^2}\bigg] \, 
\bigg[ \frac{q_3^j}{q_3^2}+\frac{k_2^j}{k_2^2}\bigg] \bigg[ \frac{q_3^k}{q_3^2}+\frac{k_2^k}{k_2^2}\bigg].\nonumber
\eeq
After using the GBW model for the dipole operators, the computation of $X_5$ becomes straightforward since the integrations over $q_1$, $q_2$ and $q_3$ are factorized and it can be written as
\beq
X_5&=&4\, \alpha_s^3(4\pi)^3(N_c^2-1)\, \mu^6\, (2\pi)^4\, S_\perp \, \Big[ \delta^{(2)}(k_2+k_3)+\delta^{(2)}(k_2-k_3)\Big] \, \frac{1}{(2\pi)^6}\, \frac{(4\pi)^3}{Q_s^6} e^{-3k_2^2/Q_s^2} \nonumber\\
&\times&
\int d^2q_1\, e^{-(q_1^2+2q_1\cdot k_2)/Q_s^2 }
\bigg[ \frac{q_1^i}{q_1^2}+\frac{k_2^i}{k_2^2}\bigg]\bigg[ \frac{q_1^k}{q_1^2}+\frac{k_2^k}{k_2^2}\bigg]
%\nonumber\\
%&\times&
\int d^2q_2\, e^{-(q_2^2+2q_2\cdot k_2)/Q_s^2 }
\bigg[ \frac{q_2^i}{q_2^2}+\frac{k_2^i}{k_2^2}\bigg]\bigg[ \frac{q_2^j}{q_2^2}+\frac{k_2^j}{k_2^2}\bigg]
\nonumber\\
&\times&
\int d^2q_3\, e^{-(q_3^2+2q_3\cdot k_2)/Q_s^2 }
\bigg[ \frac{q_3^j}{q_3^2}+\frac{k_2^j}{k_2^2}\bigg]\bigg[ \frac{q_3^k}{q_3^2}+\frac{k_2^k}{k_2^2}\bigg].
\eeq
The exponentially enhanced contributions to each of the above integrals can be read off from Eq.~(\ref{Exp_Type_2}) with the following final result:
\beq
X_5&\approx&4\, \alpha_s^3(4\pi)^3(N_c^2-1)\, \mu^6\, (2\pi)^4\, S_\perp \, \Big[ \delta^{(2)}(k_2+k_3)+\delta^{(2)}(k_2-k_3)\Big] \, \frac{1}{(2\pi)^6}\, \frac{(4\pi)^3}{Q_s^6} e^{-3k_2^2/Q_s^2}
\nonumber\\
&\times&
\bigg[ \pi Q_s^2\; e^{k_2^2/Q_s^2}\; \frac{\delta^{ik}}{2}\frac{Q_s^2}{k_2^4}\bigg]
\bigg[ \pi Q_s^2\; e^{k_2^2/Q_s^2}\; \frac{\delta^{ij}}{2}\frac{Q_s^2}{k_2^4}\bigg]
\bigg[ \pi Q_s^2\; e^{k_2^2/Q_s^2}\; \frac{\delta^{jk}}{2}\frac{Q_s^2}{k_2^4}\bigg],
\eeq
which can be further simplified and the final result can be written as in Eq. \eqref{X_5_final}. 
%\beq
%X_5\approx  \alpha_s^3(4\pi)^3(N_c^2-1)\, \mu^6\, (2\pi)\, S_\perp \, \Big[ \delta^{(2)}(k_2+k_3)+\delta^{(2)}(k_2-k_3)\Big] \, (4\pi)^3\, \frac{1}{8}\, \frac{Q_s^6}{k_2^{12}}
%\eeq

\subsection{Computation of $\bar X_i$ for the correlations between mean transverse momentum and $v_2$}
The computation for $\bar X_i$ is very similar to the one that we performed in the previous section. The only difference is that each contribution listed in Eqs. \eqref{X_1_initial} - \eqref{X5_initial} should be multiplied with $k_1^2$ before performing the integral over $k_1$. Following this procedure and using the explicit expressions for the Lipatov vertices and the MV model for $\mu^2$, we get the following expressions for $\bar X_i$:
\beq
\bar X_{1}&=& 4\, \alpha_s^3(4\pi)^3(N_c^2-1)\, \mu^6 \, (2\pi)^2\, S_\perp \int  \frac{d^2q_2}{(2\pi)^2} \, \frac{d^2q_3}{(2\pi)^2} \bigg\{ 
d\big[ k_2-(k_3-q_3)\big] \, d(q_2)\, d(q_3) \, k_2^2\, 
%\nonumber\\
%&\times&
\bigg[ \frac{(k_3-q_3)^i}{(k_3-q_3)^2}-\frac{k_2^i}{k_2^2}\bigg] 
\\
&\times&
\bigg[ \frac{(k_2-q_2)^i}{(k_2-q_2)^2}-\frac{k_2^i}{k_2^2}\bigg] 
\bigg[ \frac{(k_3-q_3)^j}{(k_3-q_3)^2}-\frac{k_2^j}{k_2^2}\bigg] \bigg[ \frac{(k_2-q_2)^j}{(k_2-q_2)^2}-\frac{k_2^j}{k_2^2}\bigg] \; 
\bigg[ \frac{(k_3-q_3)^k}{(k_3-q_3)^2}-\frac{k_3^k}{k_3^2}\bigg] \bigg[ \frac{(k_3-q_3)^k}{(k_3-q_3)^2}-\frac{k_3^k}{k_3^2}\bigg] +(k_3\to-k_3)\bigg\}, \nonumber
\eeq
\beq
\bar X_2&=&4 \, \alpha_s^3(4\pi)^3(N_c^2-1)\, \mu^6 \, (2\pi)^4\, S_\perp\, \big[ \delta^{(2)}(k_2-k_3)+\delta^{(2)}(k_2+k_3) \big] \, \int \frac{d^2q_1}{(2\pi)^2} \, \frac{d^2q_2}{(2\pi)^2} \, \frac{d^2q_3}{(2\pi)^2} \, d(q_1) \, d(q_2) \, d(q_3)\,\nonumber\\
&&
\times\,  (k_2-q_2+q_1)^2\, 
\bigg[ \frac{(k_2-q_2)^i}{(k_2-q_2)^2}-\frac{k_2^i}{k_2^2}\bigg] \bigg[ \frac{(k_2-q_3)^i}{(k_2-q_3)^2}-\frac{k_2^i}{k_2^2}\bigg] \; 
\bigg[ \frac{(k_2-q_2)^j}{(k_2-q_2)^2}-\frac{k_2^j}{k_2^2}\bigg] \bigg[ \frac{(k_2-q_3)^j}{(k_2-q_3)^2}-\frac{k_2^j}{k_2^2}\bigg] \nonumber\\
&&
\times\,  
\bigg[ \frac{(k_2-q_2)^k}{(k_2-q_2)^2}-\frac{(k_2-q_2+q_1)^k}{(k_2-q_2+q_1)^2}\bigg]
\bigg[ \frac{(k_2-q_2)^k}{(k_2-q_2)^2}-\frac{(k_2-q_2+q_1)^k}{(k_2-q_2+q_1)^2}\bigg],
\eeq
\beq
\bar X_3&=&4\,  \alpha_s^3(4\pi)^3(N_c^2-1)\, \mu^6 \, (2\pi)^2\, S_\perp\, \int \frac{d^2q_1}{(2\pi)^2}\, \frac{d^2q_3}{(2\pi)^2} \bigg\{ d(q_1)\, d\big[ (k_3-q_3)-k_2\big]\, d(q_3)\,
k_3^2\,  
\bigg[ \frac{(k_3-q_3)^i}{(k_3-q_3)^2}-\frac{k_3^i}{k_3^2}\bigg] \\
&\times& \bigg[ \frac{(k_3-q_1)^i}{(k_3-q_1)^2}-\frac{k_3^i}{k_3^2}\bigg]
\bigg[ \frac{(k_3-q_3)^j}{(k_3-q_3)^2}-\frac{k_3^j}{k_3^2}\bigg] \bigg[ \frac{(k_3-q_1)^j}{(k_3-q_1)^2}-\frac{k_3^j}{k_3^2}\bigg]\; 
\bigg[ \frac{(k_3-q_3)^k}{(k_3-q_3)^2}-\frac{k_2^k}{k_2^2}\bigg] \bigg[ \frac{(k_3-q_3)^k}{(k_3-q_3)^2}-\frac{k_2^k}{k_2^2}\bigg] +(k_2\to-k_2)\bigg\} ,\nonumber
\eeq
\beq
\bar X_4&=&4\,  \alpha_s^3(4\pi)^3(N_c^2-1)\, \mu^6 \, (2\pi)^2\, S_\perp\, \int \frac{d^2q_1}{(2\pi)^2}\, \frac{d^2q_3}{(2\pi)^2} \bigg\{ d(q_1)\, d\big[ (k_3-q_3)-k_2\big]\, d(q_3)
\, (k_3-q_3+q_1)^2\, 
\nonumber\\
&\times&
\bigg[ \frac{(k_3-q_3)^i}{(k_3-q_3)^2}-\frac{(k_3-q_3+q_1)^i}{(k_3-q_3+q_1)^2}\bigg]
\bigg[ \frac{(k_3-q_3)^i}{(k_3-q_3)^2}-\frac{(k_3-q_3+q_1)^i}{(k_3-q_3+q_1)^2}\bigg]\; 
\bigg[  \frac{(k_3-q_3)^j}{(k_3-q_3)^2}-\frac{k_2^j}{k_2^2}\bigg] \bigg[  \frac{(k_3-q_3)^j}{(k_3-q_3)^2}-\frac{k_2^j}{k_2^2}\bigg]\nonumber\\
&\times&
\bigg[ \frac{(k_3-q_3)^k}{(k_3-q_3)^2} -\frac{k_3^k}{k_3^2}\bigg] \bigg[ \frac{(k_3-q_3)^k}{(k_3-q_3)^2} -\frac{k_3^k}{k_3^2}\bigg]+(k_3\to-k_3)\bigg\},
\eeq
\beq
\bar X_5&=&4 \, \alpha_s^3(4\pi)^3(N_c^2-1)\, \mu^6 \, (2\pi)^4\, S_\perp\, \big[ \delta^{(2)}(k_2-k_3)+\delta^{(2)}(k_2+k_3) \big] \, \int \frac{d^2q_1}{(2\pi)^2} \, \frac{d^2q_2}{(2\pi)^2} \, \frac{d^2q_3}{(2\pi)^2} \, d(q_1) \, d(q_2) \, d(q_3)\, k_2^2\, 
\nonumber\\
&\times&
 \bigg[ \frac{(k_2-q_1)^i}{(k_2-q_1)^2}-\frac{k_2^i}{k_2^2}\bigg]  \bigg[ \frac{(k_2-q_2)^i}{(k_2-q_2)^2}-\frac{k_2^i}{k_2^2}\bigg] \; 
  \bigg[ \frac{(k_2-q_2)^j}{(k_2-q_1)^2}-\frac{k_2^j}{k_2^2}\bigg]  \bigg[ \frac{(k_2-q_3)^j}{(k_2-q_2)^2}-\frac{k_2^j}{k_2^2}\bigg] \nonumber\\
  &\times&
  \bigg[ \frac{(k_2-q_3)^k}{(k_2-q_3)^2}-\frac{k_2^k}{k_2^2}\bigg]  \bigg[ \frac{(k_2-q_1)^k}{(k_2-q_1)^2}-\frac{k_2^k}{k_2^2}\bigg] .
\eeq
Upon performing the appropriate shifts of the integration variables and renaming them, the total result can be organized as follows:
\beq
%\frac{d\bar N}{d^2k_2d^2k_3}\equiv
\int d^2k_1\, k_1^2\, \frac{dN^{(3)}}{d^2k_1d^2k_2d^2k_3}=\bar X_1+\bar X_2+\bar X_3+\bar X_4+\bar X_5\ ,
\eeq
where 
\beq
\label{barX1_shifted}
\bar X_1&\equiv& 4\, \alpha_s^3(4\pi)^3(N_c^2-1)\, \mu^6(2\pi)^2\, S_\perp\, 
\int \frac{d^2q_2}{(2\pi)^2}\frac{d^2q_3}{(2\pi)^2}\, \bigg\{
d(q_2+k_2) \, d(q_3+k_2) \, d(q_3+k_3) \\
&&
\hspace{2cm}
\times \, k_2^2\, 
\bigg[ \frac{q^i_2}{q_2^2}+\frac{k^i_2}{k_2^2}\bigg] \bigg[ \frac{q^j_2}{q_2^2}+\frac{k^j_2}{k_2^2}\bigg]
\bigg[ \frac{q^i_3}{q_3^2}+\frac{k^i_2}{k_2^2}\bigg] \bigg[ \frac{q^j_3}{q_3^2}+\frac{k^j_2}{k_2^2}\bigg]
\bigg[ \frac{q^k_3}{q_3^2}+\frac{k^k_3}{k_3^2}\bigg] \bigg[ \frac{q^k_3}{q_3^2}+\frac{k^k_3}{k_3^2}\bigg] +(k_3\to-k_3)\bigg\},\nonumber
\eeq
\beq
\label{barX2_shifted}
\bar X_2&\equiv&4\, \alpha_s^3(4\pi)^3(N_c^2-1)\, \mu^6 \, (2\pi)^4 \, S_\perp \Big[ \delta^{(2)}(k_2+k_3)+\delta^{(2)}(k_2-k_3) \Big] 
\int \frac{d^2q_1}{(2\pi)^2}\frac{d^2q_2}{(2\pi)^2}\frac{d^2q_3}{(2\pi)^2} \, 
d(q_1+q_2) \, d(q_2+k_2)\,  
\nonumber\\
&&
\hspace{2cm}
\times\, d(q_3+k_2)\,  q_1^2\, 
\bigg[ \frac{q^k_1}{q^2_1}+\frac{q^k_2}{q_2^2}\bigg]\bigg[ \frac{q^k_1}{q^2_1}+\frac{q^k_2}{q_2^2}\bigg] \, 
\bigg[ \frac{q_2^i}{q_2^2}+\frac{k_2^i}{k_2^2}\bigg] \bigg[ \frac{q_2^j}{q_2^2}+\frac{k_2^j}{k_2^2}\bigg] \, 
\bigg[ \frac{q_3^i}{q_3^2}+\frac{k_2^i}{k_2^2}\bigg] \bigg[ \frac{q_3^j}{q_3^2}+\frac{k_2^j}{k_2^2}\bigg] ,
\eeq
\beq
\label{barX3_shifted}
\bar X_3&\equiv&  4\, \alpha_s^3(4\pi)^3(N_c^2-1)\, \mu^6(2\pi)^2\, S_\perp\, 
\int \frac{d^2q_2}{(2\pi)^2} \frac{d^2q_3}{(2\pi)^2} \bigg\{
d(q_2+k_2) \, d(q_2+k_3)\, d(q_3+k_3) \\
&&
\hspace{2cm}
\times \, k_3^2\, 
\bigg[ \frac{q_2^i}{q_2^2}+{\frac{k_3^i}{k_3^2}}\bigg] \bigg[ \frac{q_2^j}{q_2^2}+{\frac{k_3^j}{k_3^2}}\bigg]\, 
\bigg[ \frac{q_3^i}{q_3^2}+{\frac{k_3^i}{k_3^2}}\bigg] \bigg[ \frac{q_3^j}{q_3^2}+{\frac{k_3^j}{k_3^2}}\bigg] \, 
\bigg[ \frac{q_2^k}{q_2^2}+\frac{k_2^k}{k_2^2}\bigg] \bigg[ \frac{q_2^k}{q_2^2}+\frac{k_2^k}{k_2^2}\bigg] +(k_2\to-k_2)\bigg\},\nonumber
\eeq
\beq
\label{barX4_shifted}
\bar X_4&\equiv&  4\, \alpha_s^3(4\pi)^3(N_c^2-1)\, \mu^6(2\pi)^2\, S_\perp\, 
\int \frac{d^2q_2}{(2\pi)^2} \frac{d^2q_3}{(2\pi)^2} \bigg\{
d(q_2+q_3)\, d(q_3+k_2)\, d(q_3+k_3)\\
&&
\hspace{2cm}
\times \, q_2^2\, 
\bigg[ \frac{q_2^i}{q_2^2}+\frac{q_3^i}{q_3^2}\bigg] \bigg[ \frac{q_2^i}{q_2^2}+\frac{q_3^i}{q_3^2}\bigg]\, 
\bigg[ \frac{q_3^j}{q_3^2}+\frac{k_2^j}{k_2^2}\bigg] \bigg[ \frac{q_3^j}{q_3^2}+\frac{k_2^j}{k_2^2}\bigg] \, 
\bigg[ \frac{q_3^k}{q_3^2}+\frac{k_3^k}{k_3^2}\bigg] \bigg[ \frac{q_3^k}{q_3^2}+\frac{k_3^k}{k_3^2}\bigg]+ (k_3\to-k_3) \bigg\},\nonumber
\eeq
\beq
\label{barX5_shifted}
\bar X_5&\equiv&4\, \alpha_s^3(4\pi)^3(N_c^2-1)\, \mu^6 \, (2\pi)^4 \, S_\perp \Big[ \delta^{(2)}(k_2+k_3)+\delta^{(2)}(k_2-k_3) \Big] 
\int \frac{d^2q_1}{(2\pi)^2}\frac{d^2q_2}{(2\pi)^2}\frac{d^2q_3}{(2\pi)^2} \, 
d(q_1+k_2) \, d(q_2+k_2)\, 
\nonumber\\
&&
\hspace{2.5cm}
\times \, d(q_3+k_2)\, k_2^2\, 
\bigg[ \frac{q_1^i}{q_1^2}+\frac{k_2^i}{k_2^2}\bigg] \bigg[ \frac{q_1^k}{q_1^2}+\frac{k_2^k}{k_2^2}\bigg] \, 
\bigg[ \frac{q_2^i}{q_2^2}+\frac{k_2^i}{k_2^2}\bigg] \bigg[ \frac{q_2^j}{q_2^2}+\frac{k_2^j}{k_2^2}\bigg] \, 
\bigg[ \frac{q_3^j}{q_3^2}+\frac{k_2^j}{k_2^2}\bigg] \bigg[ \frac{q_3^k}{q_3^2}+\frac{k_2^k}{k_2^2}\bigg].
\eeq
\subsubsection{Computation of $\bar X_1$, $\bar X_3$ and $\bar X_5$}
By comparing the Eqs. \eqref{X1_shifted} and \eqref{barX1_shifted} for $\bar X_1$, Eqs. \eqref{X3_shifted} and \eqref{barX3_shifted} for $\bar X_3$ and Eqs.\eqref{X5_shifted} and \eqref{barX5_shifted} for $\bar X_5$, one can immediately realize that the structures of the $q_2$ and $q_3$ dependence of these terms are exactly the same, therefore the results are the same  -- up to the extra factor of $k_2^2$ in the numerator of $\bar X_1$ and $\bar X_5$, and $k_3^2$ in the numerator of $\bar X_3$ when compared to the expressions of $X_1$, $X_3$ and $X_5$.  Thus, without doing any computations, we can conclude that 
\beq
\bar X_1&=& k_2^2 \; X_1\ ,\\
\bar X_3 &=& k_3^2 \; X_3\ ,\\
\bar X_5 &=& k_2^2 \; X_5\ ,
\eeq
and the explicit expressions for $\bar X_1$, $\bar X_3$ and $\bar X_5$ can be written as in Eqs. \eqref{barX1_final}, \eqref{barX3_final} and \eqref{barX5_final} respectively. 

\subsubsection{Computation of $\bar X_2$}
Starting from Eq. \eqref{barX2_shifted} and using the GBW model for the dipole operators, $\bar X_2$ can be organized as follows: 
\beq
\bar X_2 &=& 4\, \alpha_s^3(4\pi)^3(N_c^2-1)\, \mu^6 \, (2\pi)^4 S_\perp \Big[ \delta^{(2)}(k_2+k_3)+\delta^{(2)}(k_2-k_3)\Big] \, \frac{1}{(2\pi)^6}\, \frac{(4\pi)^3}{Q_s^6}\, e^{-2k_2^2/Q_s^2} \nonumber\\
&&\times\, 
\int d^2q_2 \, e^{-(2q_2^2+2q_2\cdot k_2)/Q_s^2} 
\bigg[ \frac{q_2^i}{q_2^2}\frac{q_2^j}{q_2^2} +\frac{q_2^i}{q_2^2}\frac{k_2^j}{k_2^2}+\frac{q_2^j}{q_2^2}\frac{k_2^i}{k_2^2}+\frac{k_2^i}{k_2^2}\frac{k_2^j}{k_2^2}\bigg]
\int d^2q_1\, e^{-(q_1^2+2q_1\cdot q_2)/Q_s^2} \bigg[ 1+2\frac{q_1^k \, q_2^k}{q_2^2}+\frac{q_1^2}{q_2^2}\bigg]
\nonumber\\
&&\times \, 
\int d^2q_3 \,  e^{-(q_3^2+2q_3\cdot k_2)/Q_s^2} \bigg[ \frac{q_3^i}{q_3^2}\frac{q_3^j}{q_3^2} + \frac{q_3^i}{q_3^2}\frac{k_2^j}{k_2^2} +\frac{q_3^j}{q_3^2}\frac{k_2^i}{k_2^2}+ \frac{k_2^i}{k_2^2}\frac{k_2^j}{k_2^2} \bigg].
\eeq
We first perform the integration over $q_3$. The structure of this integral is exactly the same as the one given in Eq. \eqref{Exp_Type_2}. By using this result,  $\bar X_2$  can be written as 
%%
%\beq
%\label{I_q_3}
%\int d^2q_3 \,  e^{-(q_3^2+2q_3\cdot k_2)/Q_s^2} \bigg[ \frac{q_3^i}{q_3^2}\frac{q_3^j}{q_3^2} + \frac{q_3^i}{q_3^2}\frac{k_2^j}{k_2^2} +\frac{q_3^j}{q_3^2}\frac{k_2^i}{k_2^2}+ \frac{k_2^i}{k_2^2}\frac{k_2^j}{k_2^2} \bigg] 
%\approx 
%\pi Q_s^2\;  e^{k_2^2/Q_s^2} \,  \frac{\delta^{ij}}{2} \frac{Q_s^2}{k_2^4}\bigg( 1+ 2\frac{Q_s^2}{k_2^2}+6\frac{Q_s^4}{k_2^4}\bigg)
%\eeq
%%
%Note that each $q_i$ integration brings at least a factor of $Q_s^2$ which will cancel the factor  $Q_s^6$ in the denominator. Since our aim is to compute $\bar X_2$ at $O(Q_s^6)$, for now we keep $O(Q_s^2)$ and $O(Q_s^4)$ in Eq. \eqref{I_q_3}. Upon integration over $q_3$, exponentially enhanced contribution to $X_2$ reads 
%
\beq
\label{barX2_int}
\bar X_2 &\approx&  4\, \alpha_s^3(4\pi)^3(N_c^2-1)\, \mu^6 \, (2\pi)^4 S_\perp \Big[ \delta^{(2)}(k_2+k_3)+\delta^{(2)}(k_2-k_3)\Big] \, \frac{1}{(2\pi)^6}\, \frac{(4\pi)^3}{Q_s^6}\, \pi Q_s^2\, e^{-k_2^2/Q_s^2} \frac{1}{2} \, \frac{Q_s^2}{k_2^4} \bigg( 1+2\frac{Q_s^2}{k_2^2}+6\frac{Q_s^4}{k_2^4}\bigg) \nonumber\\
&&\times \, \int d^2q_2 \, e^{-(2q_2^2+2q_2\cdot k_2)/Q_s^2}\bigg[ \frac{1}{q_2^2}+2\frac{k_2^i}{k_2^2}\frac{q_2^i}{q_2^2}+\frac{1}{k_2^2}\bigg] \, 
\int d^2q_1\, e^{-(q_1^2+2q_1\cdot q_2)/Q_s^2}\bigg[ 1+2 \frac{q_1^k \, q_2^k}{q_2^2}+\frac{q_1^2}{q_2^2}\bigg].
\eeq
Let us now consider the integration over $q_1$ that can be organized as follows:
% which can be written as 
%
\beq
\int d^2q_1\, e^{-(q_1^2+2q_1\cdot q_2)/Q_s^2}\bigg[ 1+2 \frac{q_1^k \, q_2^k}{q_2^2}+\frac{q_1^2}{q_2^2}\bigg]=I_{0,0}+2\frac{q_2^k}{q_2^2}\bar I^k_{1,0}+ \frac{1}{q_2^2}\bar I_{2,0}\ ,
\eeq
%3int, 1_Type_0
with $I_{0,0}$ being defined in Eq. \eqref{3int} and its solution  given in Eq. \eqref{1_Type_0}. The new type of integrals that appear above can be computed trivially: 
\beq
%I_{0,0}&=& \int d^2q_1\; e^{-(q_1^2+2q_1\cdot q_2)/Q_s^2} \, = \pi Q_s^2\:  e^{q_2^2/Q_s^2} \\
\bar I^k_{1,0}&=& \int d^2q_1\; e^{-(q_1^2+2q_1\cdot q_2)/Q_s^2} \, q_1^k \; = - \pi Q_s^2\:  e^{q_2^2/Q_s^2}\, q_2^k\ ,  \\
\bar I_{2,0}&=& \int d^2q_1\; e^{-(q_1^2+2q_1\cdot q_2)/Q_s^2} \, q_1^2 \; = \pi Q_s^2\:  e^{q_2^2/Q_s^2} \big( Q_s^2+q_2^2\big).
\eeq
By using these results the integration over $q_1$ reads
\beq
\label{Integral_q_1}
\int d^2q_1\, e^{-(q_1^2+2q_1\cdot q_2)/Q_s^2}\bigg[ 1+2 \frac{q_1^k \, q_2^k}{q_2^2}+\frac{q_1^2}{q_2^2}\bigg]= \pi Q_s^2\:  e^{q_2^2/Q_s^2} \; \frac{Q_s^2}{q_2^2}\ .
\eeq
Inserting these results into Eq. \eqref{barX2_int}, for the exponentially enhanced contribution in $\bar X_2$ we get  
\beq
\bar X_2 &\approx&  4\, \alpha_s^3(4\pi)^3(N_c^2-1)\, \mu^6 \, (2\pi)^4 S_\perp \Big[ \delta^{(2)}(k_2+k_3)+\delta^{(2)}(k_2-k_3)\Big] \, \frac{1}{(2\pi)^6}\, \frac{(4\pi)^3}{Q_s^6}\,
( \pi Q_s^2)^2\, e^{-k_2^2/Q_s^2} \frac{1}{2} \, \frac{Q_s^4}{k_2^4} \bigg( 1+2\frac{Q_s^2}{k_2^2}\bigg) \nonumber\\
&& \times\, 
\int d^2q_2 \, e^{-(q_2^2+2q_2\cdot k_2)/Q_s^2}\bigg[ \frac{1}{q_2^4}+2\frac{k_2^i}{k_2^2}\frac{q_2^i}{q_2^4}+\frac{1}{q_2^2} \frac{1} {k_2^2}\bigg] .
\eeq
The remaining integral is over $q_2$ which reads  
\beq
\label{New_q2}
\int d^2q_2 \, e^{-(q_2^2+2q_2\cdot k_2)/Q_s^2}\bigg[ \frac{1}{q_2^4}+2\frac{k_2^i}{k_2^2}\frac{q_2^i}{q_2^4}+\frac{1}{q_2^2} \frac{1} {k_2^2}\bigg] = I_{0,4} +2\frac{k_2^i}{k_2^2} I^i_{1,4}+\frac{1}{k_2^2}I_{0,2}\ ,
\eeq
where the integrals $I_{0,2}$ and $I_{0,4}$ were computed previously and the results are given in Eqs. \eqref{I_02}. and \eqref{I_04}. The new type of integral that appears in Eq. \eqref{New_q2} can be computed in the same way and the results reads
\beq
%I_{0,2}&=&  \int d^2q_2 \, e^{-(q_2^2+2q_2\cdot k_2)/Q_s^2} \, \frac{1}{q_2^2} \approx \pi Q_s^2\; e^{k_2^2/Q_s^2}\, \frac{1}{k_2^2} \bigg( 1+ \frac{Q_s^2}{k_2^2}\bigg)
%\\
%%
%I_{0,4}&=& \int d^2q_2 \, e^{-(q_2^2+2q_2\cdot k_2)/Q_s^2} \, \frac{1}{q_2^4} \approx \pi Q_s^2\; e^{k_2^2/Q_s^2}\, \frac{1}{k_2^4} \bigg( 1+4 \frac{Q_s^2}{k_2^2}\bigg) \\
%
I^i_{1,4}&=& \int d^2q_2 \, e^{-(q_2^2+2q_2\cdot k_2)/Q_s^2} \, \frac{q_2^i}{q_2^4} \approx - \pi Q_s^2\; e^{k_2^2/Q_s^2}\, \frac{k_2^i}{k_2^2} \, \frac{Q_s^2}{k_2^2}\, \bigg( 1+ 2\frac{Q_s^2}{k_2^2}\bigg) .
\eeq
Then
\beq
\int d^2q_2 \, e^{-(q_2^2+2q_2\cdot k_2)/Q_s^2}\bigg[ \frac{1}{q_2^4}+2\frac{k_2^i}{k_2^2}\frac{q_2^i}{q_2^4}+\frac{1}{q_2^2} \frac{1} {k_2^2}\bigg] \approx \pi Q_s^2\; e^{k_2/Q_s^2}\, \frac{Q_s^2}{k_2^6}
\eeq
and, finally, we can write the exponentially enhanced contribution to $\bar X_2$ as it is given in Eq. \eqref{barX2_final}.
\subsubsection{Computation of $\bar X_4$}
The last contribution that we consider is $\bar X_4$. Starting from Eq. (\ref{barX4_shifted}), this contribution can be organized as
\beq
\bar X_4&=& 4\, \alpha_s^3(4\pi)^3(N_c^2-1)\, \mu^6\, (2\pi)^2\, S_\perp\, \frac{1}{(2\pi)^4} \, \bigg(\frac{4\pi}{Q_s^2}\bigg)^3\, e^{-(k_2^2+k_3^2)/Q_s^2}
%\nonumber\\
%&\times&
\int d^2q_3 \, e^{-[3q_3^2+2q_3\cdot(k_2+k_3)]/Q_s^2} \nonumber\\
&&\times
\bigg[ 
\frac{1}{k_2^2k_3^2} 
+\bigg(\frac{1}{k_2^2}+\frac{1}{k_3^2}\bigg)\frac{1}{q_3^2} 
%\nonumber\\
%&&
+2\bigg(\frac{k_2^i}{k_2^2k_3^2}+\frac{k_3^i}{k_2^2k_3^2}\bigg)\frac{q_3^i}{q_3^2} 
+2\bigg(\frac{k_2^i}{k_2^2}+\frac{k_3^i}{k_3^2}\bigg)\frac{q_3^i}{q_3^2}\frac{1}{q_3^2}
+4\frac{k_2^i}{k_2^2}\frac{k_3^j}{k_3^2}\frac{q_3^i}{q_3^2}\frac{q_3^j}{q_3^2}
+\frac{1}{q_3^2}\frac{1}{q_3^2}\bigg] \nonumber\\
&&\times
\int d^2q_2 \, e^{-(q_2^2+2q_2\cdot q_3)/Q_s^2} \bigg[ 1+ 2\frac{q_2^k\, q_3^k}{q_3^2}+\frac{q_2^2}{q_3^2}\bigg].
\eeq
The structure of the $q_2$ integral is exactly the same as in Eq. \eqref{Integral_q_1}. Using this result, we can write $\bar X_4$ as
\beq
\bar X_4&=& 4\, \alpha_s^3(4\pi)^3(N_c^2-1)\, \mu^6\, (2\pi)^2\, S_\perp\, \frac{1}{(2\pi)^4} \, \bigg(\frac{4\pi}{Q_s^2}\bigg)^3\, (\pi Q_s^2)\, Q_s^2\, e^{-(k_2^2+k_3^2)/Q_s^2}
%\nonumber\\
%&\times&
\int d^2q_3 \, e^{-[2q_3^2+2q_3\cdot(k_2+k_3)]/Q_s^2} \nonumber\\
&&\times 
\bigg[ 
\frac{1}{k_2^2k_3^2}\frac{1}{q_3^2} 
+\bigg(\frac{1}{k_2^2}+\frac{1}{k_3^2}\bigg)\frac{1}{q_3^4} 
%\nonumber\\
%&&
+\; 2 \frac{(k_2^i+k_3^i)}{k_2^2\, k_3^2} \frac{q_3^i}{q_3^4}
%\nonumber\\
%&&
%\bigg(\frac{k_2^i}{k_2^2k_3^2}+\frac{k_3^i}{k_2^2k_3^2}\bigg)\frac{q_3^i}{q_3^2} 
+2\bigg(\frac{k_2^i}{k_2^2}+\frac{k_3^i}{k_3^2}\bigg)\frac{q_3^i}{q_3^2}\frac{1}{q_3^4}
+4\frac{k_2^i}{k_2^2}\frac{k_3^j}{k_3^2}\frac{q_3^i}{q_3^2}\frac{q_3^j}{q_3^2}\frac{1}{q_3^2}
+\frac{1}{q_3^6}\bigg],
\eeq
which can be organized as 
\beq
\label{barX4_int2}
\bar X_4 &=& 4\, \alpha_s^3(4\pi)^3(N_c^2-1)\, \mu^6\, (2\pi)^2\, S_\perp\, \frac{1}{(2\pi)^4} \, \bigg(\frac{4\pi}{Q_s^2}\bigg)^3\, (\pi Q_s^2)\, Q_s^2\, e^{-(k_2^2+k_3^2)/Q_s^2}
\nonumber\\
&\times&
\bigg\{ 
\frac{1}{k_2^2k_3^2}\, T_{0,2}+\bigg( \frac{1}{k_2^2}+\frac{1}{k_3^2}\bigg) T_{0,4} +T_{0,6} + 2\frac{(k_2+k_3)^i}{k_2^2\, k_3^2} T^i_{1,4} 
+2\bigg( \frac{k_2^i}{k_2^2}+\frac{k_3^i}{k_3^2}\bigg)T^i_{1,6}+ 4\frac{k_2^i}{k_2^2}\frac{k_3^i}{k_3^2} T_{2,6}^{ij} \bigg\}.
\eeq
Apart from $T^{ij}_{2,6}$, all integrals have been defined and computed previously ($T_{0,2}$ in Eqs. \eqref{T02_def} and \eqref{T02_soln},  $T_{0,4}$ in Eqs. \eqref{T_04_def} and \eqref{T_04_soln}, $T_{0,6}$ in Eqs. \eqref{T06_def} and \eqref{T06_soln}, $T^i_{1,4}$ in Eqs. \eqref{T14_def} and \eqref{T14_soln}, and$T^i_{1,6}$ in Eqs. \eqref{T16_def} and \eqref{T16_soln}). The new type of integral that appears in Eq. \eqref{barX4_int2} is $T^{ij}_{2,6}$ which is defined as 
 \beq
%T_{0,2}&=&\int d^2q_3 \, e^{-[2q_3^2+2q_3\cdot(k_2+k_3)]/Q_s^2} \frac{1}{q_3^2} \\
%T_{0,4}&=&\int d^2q_3 \, e^{-[2q_3^2+2q_3\cdot(k_2+k_3)]/Q_s^2} \frac{1}{q_3^4} \\
%T_{0,6}&=&\int d^2q_3 \, e^{-[2q_3^2+2q_3\cdot(k_2+k_3)]/Q_s^2} \frac{1}{q_3^6} \\
%T^i_{1,4}&=&\int d^2q_3 \, e^{-[2q_3^2+2q_3\cdot(k_2+k_3)]/Q_s^2} \frac{q_3^i}{q_3^4} \\
%T^i_{1,6}&=&\int d^2q_3 \, e^{-[2q_3^2+2q_3\cdot(k_2+k_3)]/Q_s^2} \frac{q_3^i}{q_3^6} \\
T^{ij}_{2,6}&=&\int d^2q_3 \, e^{-[2q_3^2+2q_3\cdot(k_2+k_3)]/Q_s^2} \frac{q_3^i\, q_3^j}{q_3^6}\ ,
\eeq
and can be computed in the same way as the others. Its exponentially enhanced contribution reads
%Note that apart from $T^{ij}_{2,6}$, other integrals were computed previously. The results read 
%
\beq
%T_{0,2}&\approx& \pi Q_s^2\; e^{(k_2+k_3)^2/2Q_s^2}\, \frac{2^2}{(k_2+k_3)^2}\bigg[ \frac{1}{2}+\frac{1}{4} \frac{2^2\, Q_s^2}{(k_2+k_3)^2}+\frac{1}{4}\frac{2^4\, Q_s^4}{(k_2+k_3)^4}\bigg] \\
%T_{0,4}&\approx& \pi Q_s^2\; e^{(k_2+k_3)^2/2Q_s^2}\, \frac{2^4}{(k_2+k_3)^4} \bigg[ \frac{1}{2}+  \frac{2^2\, Q_s^2}{(k_2+k_3)^2}+ \frac{9}{4}\,  \frac{ 2^4\, Q_s^4}{(k_2+k_3)^4} \bigg]\\
%T_{0,6}&\approx& \pi Q_s^2\; e^{(k_2+k_3)^2/2Q_s^2}\, \frac{2^6}{(k_2+k_3)^6} \bigg[ \frac{1}{2}+ \frac{9}{4}\, \frac{2^2\, Q_s^2}{(k_2+k_3)^2}+9\,  \frac{ 2^4\, Q_s^4}{(k_2+k_3)^4} \bigg]\\
%T^i_{1,4}&\approx& - \pi Q_s^2\; e^{(k_2+k_3)^2/2Q_s^2}\, \frac{(k_2+k_3)^i}{(k_2+k_3)^2}\, \frac{2^2}{(k_2+k_3)^2}\bigg[ 1+\frac{2^2\, Q_s^2}{(k_2+k_3)^2}+\frac{3}{2}\, \frac{2^4\, Q_s^4}{(k_2+k_3)^4}\bigg] \\
%T^i_{1,6}&\approx&-\;  \pi Q_s^2\; e^{(k_2+k_3)^2/2Q_s^2}\, \frac{(k_2+k_3)^i}{(k_2+k_3)^2}\; \frac{2^4}{(k_2+k_3)^4}\bigg[ 1+ 3\; \frac{2^2\, Q_s^2}{(k_2+k_3)^2}+9\; \frac{2^4\, Q_s^4}{(k_2+k_3)^4}\bigg]\\
T^{ij}_{2,6}&\approx&  \pi Q_s^2\; e^{(k_2+k_3)^2/2Q_s^2}\, \frac{2^4}{(k_2+k_3)^4} \bigg[ \frac{1}{2}\frac{(k_2+k_3)^i(k_2+k_3)^j}{(k_2+k_3)^2}
+ \frac{2^2\, Q_s^2}{(k_2+k_3)^2}\bigg( \frac{3}{4}\frac{(k_2+k_3)^i(k_2+k_3)^j}{(k_2+k_3)^2}+\frac{\delta^{ij}}{8}\bigg)\nonumber\\
&&
\hspace{7.2cm} + \frac{2^4\, Q_s^4}{(k_2+k_3)^4}\bigg( \frac{3}{2}\frac{(k_2+k_3)^i(k_2+k_3)^j}{(k_2+k_3)^2}+3\frac{\delta^{ij}}{8}\bigg)\bigg].
\eeq
Finally, putting everything together,  $\bar X_4$ can be written as in Eq. \eqref{barX4_final}. 

\end{document}